\documentclass[gmd, manuscript]{copernicus_v}

\begin{document}

\title{PALACE v1.0: Paranal Airglow Line And Continuum Emission model}


\Author[1,2,3][st@noll-x.de]{Stefan}{Noll} 
\Author[2]{Carsten}{Schmidt}
\Author[2]{Patrick}{Hannawald}
\Author[4]{Wolfgang}{Kausch}
\Author[4,5]{Stefan}{Kimeswenger}

\affil[1]{German Space Operations Center, Deutsches Zentrum f\"ur
  Luft- und Raumfahrt, Oberpfaffenhofen, Germany}
\affil[2]{Deutsches Fernerkundungsdatenzentrum, Deutsches Zentrum f\"ur
  Luft- und Raumfahrt, Oberpfaffenhofen, Germany}
\affil[3]{Institut für Physik, Universit\"at Augsburg, Augsburg, Germany}
\affil[4]{Universit\"at Innsbruck, Institut f\"ur Astro- und Teilchenphysik,
  Innsbruck, Austria}
\affil[5]{Universidad Cat\'olica del Norte, Instituto de Astronom\'ia,
  Antofagasta, Chile}




\runningtitle{PALACE v1.0}

\runningauthor{S. Noll et al.}

\received{}
\pubdiscuss{} 
\revised{}
\accepted{}
\published{}


\firstpage{1}

\maketitle

\begin{abstract}
  Below about 2.3\,\unit{\mu{}m}, the nighttime emission of the Earth's
  atmosphere is dominated by non-thermal radiation. Excluding aurorae, the
  emission is caused by chemical reaction chains which are driven by the
  daytime photolysis and photoionisation of constituents of the middle and
  upper atmosphere by hard ultraviolet photons from the Sun. As this airglow
  can even outshine scattered moonlight in the near-infrared regime, the
  understanding of the Earth's night-sky brightness requires good knowledge of
  the complex airglow emission spectrum and its variability. However,
  airglow modelling is very challenging as it would require atomic and
  molecular parameters, rate coefficients for chemical reactions, and
  knowledge of the complex dynamics at the emission heights with a level of
  detail that is difficult to achieve. In part, even the chemical reaction
  pathways remain unclear. Hence, the comprehensive characterisation of
  airglow emission requires large data sets of empirical data. For fixed
  locations, this can be best achieved by archived spectra of large
  astronomical telescopes with a wide wavelength coverage, high spectral
  resolving power, and good temporal sampling. Using 10 years of data from the
  \mbox{X-shooter} echelle spectrograph in the wavelength range from 0.3 to
  2.5\,\unit{\mu{}m} and additional data from the Ultraviolet and Visual
  Echelle Spectrograph at the Very Large Telescope at Cerro Paranal in Chile,
  we have succeeded to build a comprehensive spectroscopic airglow model for
  this low-latitude site under consideration of theoretical data from the
  HITRAN database for molecules and from different sources for atoms. The
  Paranal Airglow Line And Continuum Emission (PALACE) model comprises 9
  chemical species, 26,541 emission lines, and 3 unresolved continuum
  components. Moreover, there are climatologies of relative intensity, solar
  cycle effect, and residual variability with respect to local time and day of
  year for 23 variability classes. Spectra can be calculated with a
  stand-alone code for different conditions, also including optional
  atmospheric absorption and scattering. In comparison to the observed
  \mbox{X-shooter} spectra, PALACE shows convincing agreement and is
  significantly better than the previous, widely used airglow model for Cerro
  Paranal.
\end{abstract}


\introduction[Introduction]  
\label{sec:intro}

Understanding the radiation spectrum of the Earth's night sky at wavelengths
shorter than the thermal emission regime of trace gases such as water vapour
requires the knowledge of non-thermal perpetual radiation processes known as
airglow or nightglow. Different atoms and molecules radiate in the Earth's
mesopause region between about 75 and 105\,km and in the case of atoms also in
the upper thermosphere mainly above about 200\,km at nighttime due to chemical
reactions. The source of the complex chemistry is usually the energy input of
destructive ultraviolet (UV) photons from the Sun, which lead to photolysis
and photoionisation in the middle and upper atmosphere at daytime. The
resulting radicals and ions (mainly involving oxygen) trigger various
reactions and constitute an energy reservoir that is still important at
nighttime. Hence, reactions that produce excited states that can be
deactivated by photon emission are also present during the night, when
scattered sunlight does not disturb.

The different emission processes produce a highly structured spectrum
consisting of various emission lines, bands, and unresolved (pseudo-)continua
\citep[e.g.,][]
{osterbrock96,rousselot00,cosby06,khomich08,vonsavigny17,noll24}. The
ro-vibrational bands of the electronic ground state of the hydroxyl
(\chem{OH}) radical are particularly prominent from the visual to about
2.3\,\unit{\mu{}m} with the highest intensities between about 1.5 and
1.8\,\unit{\mu{}m}. There are also several important bands of molecular oxygen
(\chem{O_2}) in a similar range (0.762, 0.865, 1.27, and 1.58\,\unit{\mu{}m}).
Weak \chem{O_2} bands related to high electronic excitation are present in the
near-UV and blue range. Various weak bands of iron monoxide (\chem{FeO}) form
a broad emission structure near 600\,nm, whereas the near-infrared (near-IR)
(pseudo-)continuum with a prominent peak near 1.51\,\unit{\mu{}m} appears to
be dominated by the hydroperoxyl (\chem{HO_2}) radical. The most crucial
atomic airglow lines are located between about 500 and 800\,\unit{nm}
including prominent atomic oxygen (\chem{O}) emission at 558, 630, 636, and
777\,\unit{nm}, the sodium \chem{Na}\,D doublet at 589\,nm, and the nitrogen
(\chem{N}) doublet at 520\,nm. Most emissions originate in the mesopause
region in a relatively narrow height range. Exceptions are the ionospheric
high-altitude lines at 520, 630, 636, and 777\,\unit{nm}.   

The variability of airglow is also very complex as there are sources of
perturbations with a wide range of time scales and the various emission lines
show individual responses depending on the involved chemical species, relevant
atomic or molecular parameters, and the vertical emission distribution. The
underlying atmospheric dynamics is strongly driven by wave-like variations
such as solar tides with preferred periods of 12 and 24\,\unit{h}, gravity
waves with periods of minutes to hours, and planetary waves with periods from
days to weeks \citep[e.g.,][]{forbes95,fritts03,smith12}. The interaction of
these waves, additional instabilities, and the impact of winds lead to diurnal
variability patterns that depend on the season and the observing site. With
the effect of airglow chemistry, the resulting airglow climatologies can
significantly differ for the investigated emission processes
\citep[e.g.,][]
{takahashi98,shepherd06,gao11,shepherd16,reid17,hart19,noll23,noll24}.
Airglow radiation also shows a clear response to the varying solar activity,
which especially affects the influx of hard UV photons. The activity cycle
of about 11 years can therefore be well recognised with an amplitude depending
on the chemical species and excited state
\citep[e.g.,][]{reid14,gao16,hart19,perminov21,schmidt23,noll23,noll24}.

The modelling of airglow emissions is challenging. Global dynamical models
with included airglow chemistry
\citep[e.g.,][]{yee97,gelinas08,grygalashvyly14,plane15,noll24} or kinetic
models for specific emission processes
\citep[e.g.,][]{dodd94,funke12,vonsavigny12,panka17,noll18b,haider22} rely on
the knowledge of the chemical composition at all relevant heights, geographic
locations, and times, although the implemented dynamics and chemistry can have
significant uncertainties. In particular, the impact of gravity waves with
their relatively small spatial scales on the global dynamics is difficult to
model and the rate coefficients of many chemical reactions and collisional
relaxation processes show significant uncertainties. Sometimes there are only
rough guesses as suitable data do not exist. Even measured profiles of
important species can be relatively uncertain if the retrieval also depends on
an airglow model as in the case of the crucial \chem{O} concentration
\citep[e.g.,][]{mlynczak18,panka18,zhu18}. Theoretical models are important
for a better understanding of airglow emission processes. However, there are
too many uncertainties to calculate comprehensive airglow spectra with
realistic fluxes. Hence, the characterisation of airglow emission has to be
mainly empirical and therefore requires large amounts of spectroscopic
measurements. 

An important application of spectroscopic airglow models is the derivation of
the wavelength-dependent sky brightness, where airglow is a major component
\citep[e.g.,][]{leinert98}. In particular, this is relevant for the
efficient scheduling of observations at large astronomical facilities, the
design of astronomical instruments, and data processing removing atmospheric
signatures. Such a model was developed for the Very Large Telescope (VLT) of
the European Southern Observatory (ESO) at Cerro Paranal (24.6$^{\circ}$\,S,
70.4$^{\circ}$\,W) in Chile \citep{noll12,jones13,noll14}. Until now, this
``Sky Model'' is still the most popular model of this kind. \citet{masana21}
released an alternative sky brightness model but without internal airglow
calculations for different conditions.

The airglow component of the ESO Sky Model consists of a list of 4,764
emission lines in the wavelength range from 0.3 to 2.5\,\unit{\mu{}m}. Up to
0.925\,\unit{\mu{}m}, the list is based on the line identifications of
\citet{cosby06} in the catalogue of \citet{hanuschik03} derived from line
measurements in composite spectra of the high-resolution Ultraviolet and
Visual Echelle Spectrograph \citep[UVES;][]{dekker00} at the VLT. At longer
wavelengths, the simple \chem{OH} level population model of
\citet{rousselot00} described by only two (pseudo-)temperatures (190 and
9,000\,\unit{K}) was used and scaled to the \citet{cosby06} intensities in the
overlapping wavelength region. The population model was combined with
Einstein-$A$ coefficients for photon emission from the HITRAN2008 database
\citep{rothman09}. The latter was also used for obtaining line lists for the
\chem{O_2} bands at 1.27 and 1.58\,\unit{\mu{}m} assuming a temperature of
200\,\unit{K}. The intensities of the \chem{O_2} bands were scaled to be
consistent with those of neighbouring \chem{OH} bands using a small sample of
near-IR spectra of the VLT \mbox{X-shooter} echelle spectrograph
\citep{vernet11} and considering atmospheric absorption. The airglow emissions
were classified using the green \chem{O} line at 558\,\unit{nm}, the red
\chem{O} lines at 630 and 636\,\unit{nm}, the \chem{Na}\,D doublet at
589\,\unit{nm}, the \chem{OH} bands in the range from 642 to 858\,\unit{nm},
and the \chem{O_2} band at 865\,\unit{nm} as references. The intensities of
all lines belonging to each of the five classes were multiplied by correction
factors to achieve that the intensities of the corresponding reference
features match values that were derived from measurements in a sample of 1,186
low-resolution long-slit spectra taken with the VLT FOcal Reducer and low
dispersion Spectrograph~1 \citep[FORS\,1;][]{appenzeller98} and processed by
\citet{patat08}. The spectra were taken between April 1999 and February 2005
and cover parts of the maximum wavelength range from 365 to 890\,\unit{nm}.   
The resulting standard values for the reference features correspond to the
mean intensities for a solar radio flux at 10.7\,\unit{cm} \citep{tapping13}
of 129 solar flux units (\unit{sfu}) or 1.29\,\unit{MJy}, where the unit
jansky (\unit{Jy}) equals \unit{10^{-26}\,W\,m^{-2}\,Hz^{-1}}. Based on a
linear regression analysis for each class, the model intensities can be
adapted to arbitrary solar radio fluxes (although only 95 to 228\,\unit{sfu}
were covered by the data). Moreover, variability is considered by a
climatological grid that comprises six double months (starting with
December/January) and three nightime bins (dividing the night in ranges of
equal length). Apart from these 18 scaling factors, the model also provides
the residual variability for each bin. The ESO Sky Model also considers the
increase of intensity with increasing zenith angle due to the change in the
projected emission layer width along the line of sight \citep{vanrhijn21}.

Apart from five line emission classes, there is also one class related to the
unresolved residual continuum after the subtraction of other radiation
components such as zodiacal light or scattered light from the Moon and stars.
The reference spectrum for this airglow-related component was derived from
the sample of FORS\,1 spectra and the small number of \mbox{X-shooter} spectra
in windows without strong line emission. The uncertainties are relatively
high. For the variability model, only the variation at 543\,\unit{nm} in the
FORS\,1 data was considered. By means of other components of the Sky Model,
the airglow line and continuum emission is also corrected for absorption and
scattering (mainly in the lower atmosphere) depending on the zenith angle and
the season (or the amount of water vapour).

Despite the described complexity, the airglow component of the ESO Sky Model
shows clear limitations. The variability model is only based on about $10^3$
spectra with varying wavelength coverage in the range from 365 to
890\,\unit{nm}. The line list is an unsatisfactory mixture of measurements and
simple models from different sources. The continuum determination suffered
from the low resolution of the FORS\,1 spectra and the calibration
uncertainties related to the few early \mbox{X-shooter} spectra that were
used. Hence, a significant improvement is possible. In the meantime, the
airglow emissions above Cerro Paranal were studied in more detail based on
consistent spectroscopic data sets. UVES spectra covering 15 years (starting
in April 2000) in the wavelength range from 0.57 to 1.04\,\unit{\mu{}m} were
used to investigate long-term \chem{OH} variations \citep{noll17}, the
\chem{OH} ro-vibrational level populations \citep{noll18b,noll20}, and the
variability of the potassium (\chem{K}) emission at 770\,\unit{nm}
\citep{noll19}. The amount and quality of \mbox{X-shooter} spectra covering
the full wavelength range from 0.3 to 2.5\,\unit{\mu{}m} increased in the
course of the years. Starting with a few hundred spectra to study \chem{OH}
and \chem{O_2} emissions \citep{noll15,noll16}, then using a few thousand
spectra to investigate the \chem{FeO}-related pseudo-continuum
\citep{unterguggenberger17}, finally resulted in studies of \chem{OH}
climatologies \citep{noll23} and the airglow continuum \citep{noll24} based on
the order of $10^5$ spectra covering 10 years from October 2009 to September
2019. 

We used the large \mbox{X-shooter} data set to measure the variations of all
remaining significant airglow emissions in order to characterise the full
airglow spectrum. Combined with theoretical data such as level energies,
Einstein-$A$ coefficients, and recombination coefficients as well as fits of
level populations, we could create a list of 26,541 lines with attributed
climatologies. Based on the work of \citet{noll24}, we also added three
continuum components with the corresponding variations. In total, reference
climatologies that cover nocturnal changes, seasonal variations, the response
to solar activity, and residual variations were derived for 23 variability
classes. This Paranal Airglow Line And Continuum Emission (PALACE) model
\citep{noll24ds}, which can be used to calculate airglow spectra for different
conditions by means of an accompanying Python code, will be discussed in the
following. The basic structure and algorithm is explained in
Sect.~\ref{sec:overview}. Section~\ref{sec:dataset} provides details on the
\mbox{X-shooter} and UVES spectrographs and the related data sets that
were used for the development of PALACE. Section~\ref{sec:lines} discusses
the creation of the line lists with reference intensities for the different
chemical species. The pseudo-continuum template spectra are the topic of
Sect.~\ref{sec:continuum}. The reference climatologies are discussed in
Sect.~\ref{sec:variability}. The performance of PALACE is analysed in
Sect.~\ref{sec:eval}, also in comparison with the ESO Sky Model. Finally, we
draw our conclusions in Sect.~\ref{sec:conclusions}.

\section{Model overview}
\label{sec:overview}

\begin{figure*}[t]
\includegraphics[width=15.2cm]{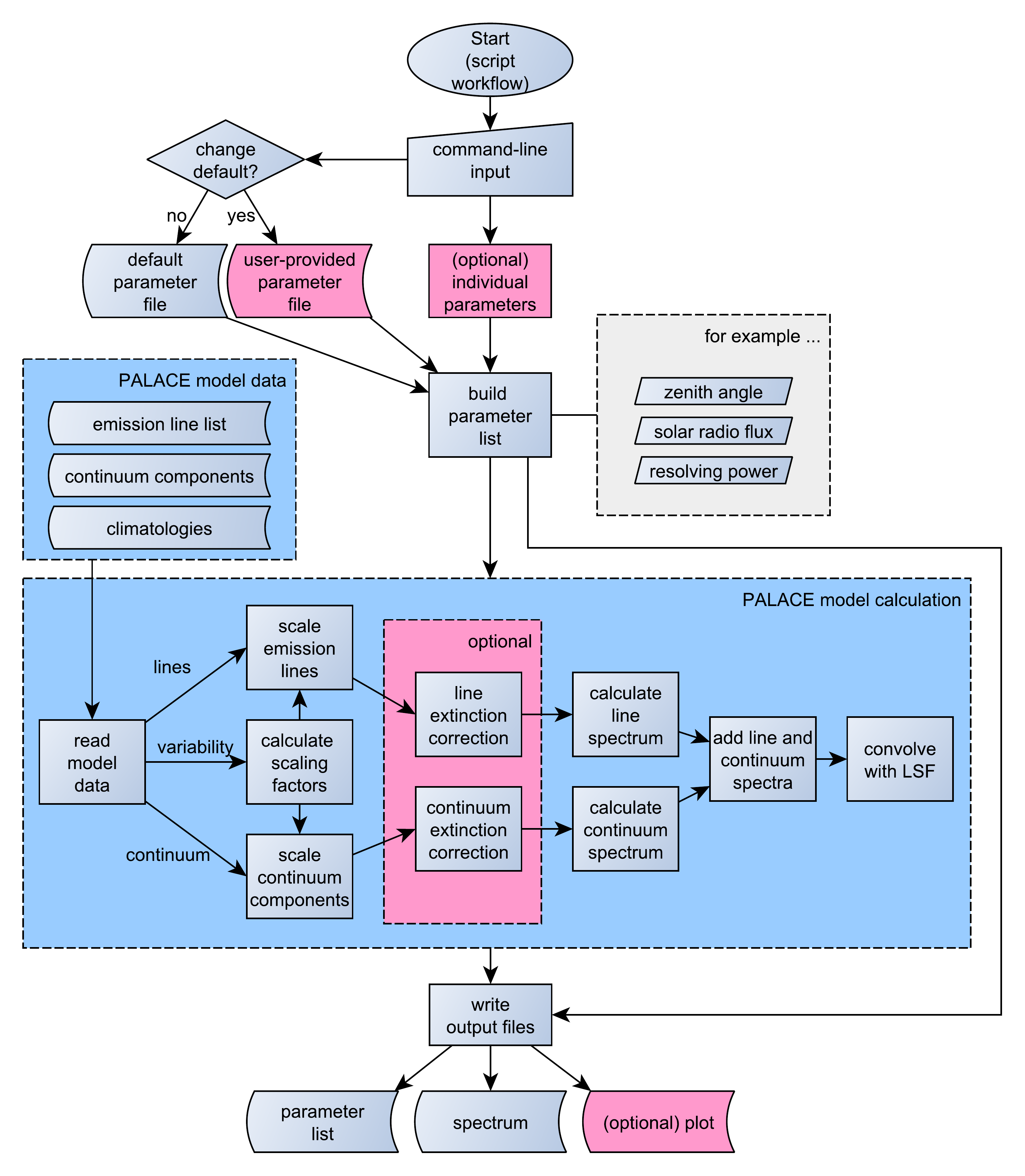}
\caption{Flowchart of PALACE executed as command-line script. By loading the
  Python module `palace', the displayed functions can be called directly. The
  details of the code structure are discussed in the text.}
\label{fig:flowchart}
\end{figure*}

\begin{table*}[t]
\caption{PALACE model parameters and their default values}
\begin{tabular}{llllll}
\tophline
Name & Description & Unit & Type & Constraints & Default \\ 
\middlehline
\texttt{species} & specific molecule/atom or all & -- & str &
  all, OH, O2, HO2, FeO, Na, K, O, N, H & all \\
\texttt{z} & zenith angle & \unit{deg} & float & $\ge 0.$ and $< 90.$ & 0. \\
\texttt{mbin} & month number$^\mathrm{a}$ or 0 for all & -- & int & [0, 12] &
  0 \\
\texttt{tbin} &
  local time$^\mathrm{b}$ bin$^\mathrm{c}$  or 0 for all & -- & int &
  [0, 12] & 0 \\
  & local time$^\mathrm{b}$ & h & float &
  [-6., 6.] or ([0., 6.] and [18., 24.]) & \\
\texttt{srf} & solar radio flux at 10.7\,\unit{cm} & \unit{sfu}$^\mathrm{d}$ &
  float & $\ge 0.$ & 100. \\
\texttt{isair} & wavelength in standard air? & -- & bool &
  True (air) or False (vacuum) & True \\
\texttt{isatm} & absorption and scattering? & -- & bool &
  True (with effects) or False & True \\
\texttt{pwv} & precipitable water vapour$^\mathrm{e}$ & \unit{mm} & float &
  $\ge 0.$ & 2.5 \\
\texttt{lammin} & minimum wavelength & \unit{\mu{}m} & float & $\ge 0.3$ &
  0.3 \\
\texttt{lammax} & maximum wavelength & \unit{\mu{}m} & float & $\le 2.5$ &
  2.5 \\
\texttt{dlam} & constant wavelength step & \unit{\mu{}m} & float &
  $\ge 10^{-7}$ (1e-7) & 1e-4 \\
\texttt{resol} & constant resolving power$^\mathrm{f}$ & -- & float & $> 0.$ &
  1e3 \\
\texttt{outdir} & output directory$^\mathrm{g}$ & -- & str &
  write access required & test/ \\ 
\texttt{outname} & output file name$^\mathrm{h}$ & -- & str &
  write access required & palace\_test \\ 
\texttt{specsuffix} & suffix of spectrum file & -- & str &
  fits for FITS, free suffix for ASCII format & fits \\
\texttt{showplot} & plot of spectrum on screen? & -- & bool &
  True (opens plot window) or False & False \\
\bottomhline
\end{tabular}
\belowtable{}
\begin{list}{}{}
\item[$^\mathrm{a}$] 1 for January and 12 for December.
\item[$^\mathrm{b}$] Solar mean time at Cerro Paranal, i.e. 70.4$^{\circ}$\,W.
\item[$^\mathrm{c}$] 1 for 18--19\,\unit{h} and 12 for 5--6\,\unit{h}.
\item[$^\mathrm{d}$] 1\,\unit{sfu} = $10^4$\,\unit{Jy}.
\item[$^\mathrm{e}$] Only relevant if \texttt{isatm} is True.
\item[$^\mathrm{f}$] Fixed ratio of FWHM of Gaussian line-spread function and
  wavelength.  
\item[$^\mathrm{g}$] Will be created if required.
\item[$^\mathrm{h}$] Supplemented by suffixes .par for a parameter file, .fits
  for a FITS table, or anything else for an ASCII table.
\end{list}
\label{tab:parameters}
\end{table*}

The core of PALACE are three tables in Flexible Image Transport System (FITS)
format which provide the reference line intensities, the continuum component
spectra, and the climatologies describing the variability. The three tables
are linked via the defined airglow variability classes. For certain
applications of the model, these data might already be sufficient. However,
the calculation of airglow spectra for different conditions is not trivial.
Therefore, we wrote a dedicated code that consists of the Python module
`palace' and an optional script for execution (`palace\_run.py'). A detailed
flowchart is shown in Fig.~\ref{fig:flowchart}. For better performance, we
also included optional Cython types, which requires the compilation of the
module if desired. For more details on the installation and execution of
PALACE, we refer to the README file \citep{noll24ds}.

The model output depends on 16 parameters which are listed in
Table~\ref{tab:parameters}. Default values of these parameters are provided
by a standard configuration file. They are also listed in the table. The
model parameters can be modified by providing a different parameter file
and/or changing individual values via command line parameters in the case of
the script (see Fig.~\ref{fig:flowchart}) or manipulating a Python dictionary
in the case of the use of a Python shell or similar environments. The model
writes output files in the \texttt{outdir} directory, which is created
recursively if it does not exist. Relative paths are relative to the working
directory. Usually, two files are produced that are named as \texttt{outname}
plus different suffixes. First, a list of the selected parameter values is
written. This file is marked by `.par'. Second, the airglow spectrum is
written. If \texttt{specsuffix} is set to `fits' (default), a FITS table with
a `.fits' suffix is created. In all other cases, an ASCII file with the given
suffix is produced. In any case, the output spectrum has the three columns
`lam' for the wavelength in micrometres, `flux' for the photon flux in
rayleighs per nanometre (\unit{R\,nm^{-1}}), and `dflux' for the residual
variability not explained by the model with the same unit. In comparison, the
ESO Sky Model uses \unit{photons\,s^{-1} m^{-2} \mu{}m^{-1} arcsec^{-2}} as
radiance unit \citep{noll12}. This unit can be obtained from \unit{R\,nm^{-1}}
by the multiplication of 18.704 as 1\,\unit{R} equals
\unit{10^{10}/(4\pi)\,photons\,\,s^{-1} m^{-2} sr^{-1}}. The wavelength grid of
the output spectrum is defined by the parameters \texttt{lammin},
\texttt{lammax}, and \texttt{dlam}. All values have to be in micrometres.
Wavelengths lower than 0.3\,\unit{\mu{}m} or larger than 2.5\,\unit{\mu{}m}
would be outside the valid model range. Moreover, wavelength steps shorter
than 0.1\,\unit{pm} are not accepted. Such values would be distinctly lower
than the typical natural width of airglow lines of a few picometres. Apart
from writing the spectrum to a file, there is also the option to show it in a
Python plot window. In order to enable this option, \texttt{showplot} needs to
be set to `True'.

In order to adapt the reference line intensities and continuum fluxes to
specific conditions, PALACE calculates scaling factors that depend on 23
different variability classes (see Sect.~\ref{sec:variability}), month, local
time, solar activity, and zenith angle. The month is provided via the input
parameter \texttt{mbin}, where `1' refers to January and `12' to December. It
is also possible to request an annual mean by setting \texttt{mbin} to `0'. In
a similar way, the local time (LT) parameter \texttt{tbin} can be set to
values of `0' for the entire night (for a certain month or the entire year
depending on \texttt{mbin}) or `1' for 18:00 to 19:00~LT to '12' for 05:00 to
06:00~LT. The times refer to the solar mean time at Cerro Paranal, i.e. the
Universal Time is corrected by a fixed amount determined by the longitude of
70.4$^{\circ}$\,W. If floating point numbers are provided as \texttt{tbin},
they are directly interpreted as LTs in hours and the corresponding LT bin is
chosen. In this case, the valid maximum range is either 18.0 to 24.0 and 0.0
to 6.0 or $-6.0$ to 6.0. In fact, the night length constrained by a minimum
solar zenith angle of 100$^{\circ}$ is usually shorter, especially during
austral summer. For month centres, the latest start and the earliest end of
the night are in January ($-4.36$) and December (4.31), respectively. If a
specific LT or a whole bin does not have nighttime contribution, a warning
message is returned. In this case, the model still works but only uses
unreliable extrapolated values.

The parameters \texttt{mbin} and \texttt{tbin} determine the climatological
correction factor $f_0$ in comparison to the annual nocturnal mean intensity
for each variability class. However, these correction factors are only valid
for a fixed solar radio flux of 100\,\unit{sfu}. For other levels of solar
activity given by \texttt{srf}, the factors are adapted using the results of a
linear regression analysis depending on \texttt{mbin} and \texttt{tbin}. The
final climatological correction factors $f$ are then calculated by
\begin{equation}\label{eq:climfac}
  f(\mathtt{mbin},\mathtt{tbin},\mathtt{srf}) =
  f_0(\mathtt{mbin},\mathtt{tbin})\,
  (1 + 0.01\, m_\mathrm{SCE}(\mathtt{mbin},\mathtt{tbin})\,
  (\mathtt{srf} - 100)),
\end{equation}
where $m_\mathrm{SCE}$ is the regression slope relative to 100\,\unit{sfu}.
For the \texttt{srf} parameter, daily and monthly solar radio fluxes at
10.7\,\unit{cm} can be obtained from \texttt{https://spaceweather.gc.ca/}.
Note that the analysis for PALACE was based on centred 27-day averages
\citep[see][for a discussion]{noll17}. Further details on the solar cycle
effect and its analysis are provided in Sect.~\ref{sec:variability}. The
climatological correction factors $\sigma_f$ for the standard deviation of the
residual variability that is given in the `dflux' column in the output
spectrum are calculated from the reference values $\sigma_{f,0}$ relative to
the bin-specific $f_0$ by
\begin{equation}\label{eq:resvarfac}
  \sigma_f(\mathtt{mbin},\mathtt{tbin}) =
  f_0(\mathtt{mbin},\mathtt{tbin})\, \sigma_{f,0}(\mathtt{mbin},\mathtt{tbin}),
\end{equation}
i.e. there is no explicit solar activity term. The reason is the calculation
of the residual fluxes in the analysed data set by the subtraction of the
climatological model represented by Eq.~(\ref{eq:climfac}), which already
contains this effect.

\begin{table*}[t]
\caption{Summary on chemical species in PALACE}
\begin{tabular}{cccccc}
\tophline
Species & Mol mass$^\mathrm{a}$ & Lines$^\mathrm{b}$ & Continua$^\mathrm{c}$ &
Layer height$^\mathrm{d}$ & Temperature$^\mathrm{a}$ \\
& \unit{kg\,mol^{-1}} & & & \unit{km} & \unit{K} \\ 
\middlehline
\chem{OH} & 0.017 & 22,058 & 0 & 87 & 190 \\
& 0.019 & 47 & 0 & & \\
\chem{O_2} & 0.032 & 3,398 & 1 & 89 (a-X), 94 (other) & 190 \\
& 0.033 & 13 & 0 & & \\
& 0.034 & 678 & 0 & & \\
\chem{HO_2} & 0.033 & 0 & 1 & 81 & 190 \\
\chem{FeO}$^\mathrm{e}$ & 0.072 & 0 & 1 & 88 & 190 \\
\chem{Na} & 0.023 & 2 & 0 & 92 & 190 \\
\chem{K} & 0.039 & 2 & 0 & 89 & 190 \\
O & 0.016 & 313 & 0 & 97 (558\,\unit{nm}), 250 (630/636\,\unit{nm}), &
190 (558\,\unit{nm}), \\
& & & & 300 (other) & 1,000 (other) \\
\chem{N} & 0.014 & 2 & 0 & 250 & 1,000 \\
\chem{H} & 0.001 & 28 & 0 & $-1$ & 1,000 \\
all & & 26,541 & 3 & & \\
\bottomhline
\end{tabular}
\belowtable{}
\begin{list}{}{}
\item[$^\mathrm{a}$] For calculation of line width by Doppler broadening, not
  relevant for continuum components.
\item[$^\mathrm{b}$] Number of resolved lines in model, 0 for unresolved
  continua.
\item[$^\mathrm{c}$] Number of continuum components. 
\item[$^\mathrm{d}$] For calculation of emission increase with increasing
  zenith angle (van Rhijn effect), not applicable to \chem{H} (-1).
\item[$^\mathrm{e}$] Additional molecules probably contribute to the
  corresponding continuum component.
\end{list}
\label{tab:species}
\end{table*}

By default, all intensities and fluxes are given for zenith. This can be
changed by means of the input parameter \texttt{z}, which provides the zenith
angle of the line of sight in degrees. An increase of \texttt{z} also
increases the projected layer width and, hence, the intensity or flux. For a
thin layer, the reciprocal correction factor can be calculated by
\begin{equation}\label{eq:vanrhijn}
  f_\mathrm{vR}^{-1} = \sqrt{1 - \left(\frac{R \sin(\texttt{z})}{R + h}\right)^2}
\end{equation}
\citep{vanrhijn21}, where $R$ is the Earth's radius and $h$ corresponds to the
height of the layer above the ground. For $R$, we use the mean radius of
6371\,\unit{km}. The effective layer heights depend on the emission process.
Rough reference values are given in Table~\ref{tab:species}. The heights are
not constant and can easily vary by several per cent. However, this is not
crucial as the resulting $f_\mathrm{vR}$ are very similar for a wide height
range as long as the line of sight is not too close to the horizon. Results
for zenith angles of 60$^{\circ}$ or 70$^{\circ}$ are still quite robust.
Table~\ref{tab:species} shows a wide range of reference layer heights from
81\,\unit{km} for \chem{HO_2} to 300\,\unit{km} for most \chem{O} lines. More
information on these values will be given in Sects.~\ref{sec:lines} and
\ref{sec:continuum}. Note that the van Rhijn effect is not considered in the
case of atomic hydrogen (\chem{H}) as the emissions are caused by
fluorescence, which does not produce a well-defined layer (see
Sect.~\ref{sec:H}).

At the high altitudes of the airglow layers, the atmospheric density is very
low, which results in optically thin environments for most emission lines.
However, for ground-based observations at Cerro Paranal, the airglow photons
have to pass much denser atmospheric layers, which can lead to significant
absorption and scattering. PALACE considers these effects if \texttt{isatm}
is `True', which is the default setting (see Fig.~\ref{fig:flowchart}). The
level of molecular absorption is very different in the model wavelength range.
The transmission $T$ ranges from nearly 1 close to 400\,\unit{nm} to almost 0,
especially in parts of the strong near-IR water vapour absorption bands at
about 1.4 and 1.9\,\unit{\mu{}m} \citep[e.g.,][]{smette15}. For the
consideration of atmospheric absorption, PALACE applies a similar approach as
was also used for the data analysis, but in the reverse direction. The method
is essentially described by \citet{noll15}. The PALACE data for the emission
lines and continuum components include reference transmission values
$T_\mathrm{ref}$ for zenith and typical conditions at Cerro Paranal. In fact,
the latter equal the standard conditions of the ESO Sky Model \citep{noll12}
with a fixed amount of precipitable water vapour (\texttt{pwv}) of
2.5\,\unit{mm}, which is close to the long-term median \citep{holzloehner21}.
The airglow-specific $T_\mathrm{ref}$ values were derived from a transmission
spectrum calculated with the Line-By-Line Radiative Transfer Model
\citep[LBLRTM;][]{clough05} with maximum resolving power of $4\times10^6$ in
order to resolve individual airglow lines. PALACE adapts the reference values
depending on the input parameters \texttt{z} and \texttt{pwv} in
Table~\ref{tab:parameters}. The amount of water vapour can be varied
independently as it is the most important absorber in the model range and is
also highly variable. According to \citet{noll15}, the $T$ values are
approximated from $T_\mathrm{ref}$ by
\begin{equation}\label{eq:transcorr}
  T = T_\mathrm{ref}^{
  \,\left(1 + (r_\mathrm{pwv} - 1)\, f_\mathrm{H_2O} \right)\,X},
\end{equation}
where $r_\mathrm{pwv}$ is the ratio of the selected \texttt{pwv} and the
reference of 2.5\,\unit{mm}, $f_\mathrm{H_2O}$ corresponds to the fraction of
water vapour with respect to the total optical depth, and the airmass $X$ is
calculated by
\begin{equation}\label{eq:airmass}
X = \left(\cos(\texttt{z}) + 0.025 \,\mathrm{e}^{-11 \cos(\texttt{z})}\right)^{-1}
\end{equation}
\citep{rozenberg66}. The values of $f_\mathrm{H_2O}$ are also provided by the
PALACE data files. They were derived by means of the comparison of
transmission spectra for 1 and 5\,\unit{mm}, which enclose a major fraction of
the \texttt{pwv} values for clear sky conditions at Cerro Paranal
\citep{holzloehner21}. Note that the ESO Sky Model uses a large library
of transmission data instead of the discussed approximation. This is more
accurate but only for \texttt{z} and \texttt{pwv} with data in the library.

Scattering of photons at molecules (Rayleigh scattering) and aerosol particles
(Mie theory) also changes the airglow brightness. As airglow is present at the
entire sky, photons can be scattered out of and into the line of sight, which
decreases the effective exinction compared to the case of point sources.
\citet{noll12} modelled this effect for Cerro Paranal and derived recipes that
depend on \texttt{z}. We also use these results combined with the same
Rayleigh and aerosol extinction curves, which describe the wavelength
dependence of the extinction for point sources. For aerosol scattering, the
basis is the standard curve for Cerro Paranal from \citet{patat11}. The change
of airglow emission by scattering is usually small (especially in the
near-IR). However, at near-UV and blue wavelengths combined with high zenith
angles, impacts larger than 10\% are possible. Close to the zenith, scattering
slightly increases the airglow brightness for clear sky conditions as the
intensity minimum at zenith is partly filled by photons from brighter regions
at higher \texttt{z} (van Rhijn effect).

The PALACE data files contain wavelengths in vacuum. However, spectroscopic
data are often given for standard air. Hence, PALACE supports both options. If
the input parameter \texttt{isair} is set to the default `True', air
wavelengths $\lambda_\mathrm{air}$ are provided by calculating 
\begin{equation}\label{eq:vacair}
\lambda_\mathrm{air} = \frac{\lambda_\mathrm{vac}}{n},
\end{equation}
where $n$ is the refractive index. It is approximated by
\begin{equation}\label{eq:edlen}
n = 1 + 10^{-8} \left(8342.13 + \frac{2406030}{130 - \lambda^{-2}} +
\frac{15997}{38.9 - \lambda^{-2}}\right)
\end{equation}
with $\lambda$ in micrometres \citep{edlen66}. This approach is consistent
with \citet{noll12}.

After the application of the scaling factors from the climatologies, van Rhijn
effect, and atmospheric extinction (absorption and scattering) to the
reference data, line intensities need to be converted to line fluxes in order
to obtain spectra (Fig.~\ref{fig:flowchart}). For this purpose, we assume that
the natural line shape is similar to a Gaussian function. This is a good
approximation as thermal Doppler broadening predominates due to the very low
air densites at the emission heights \citep[see also][]{noll12}. In this way,
the line-specific width of the Gaussian $\sigma_\lambda$, which is of the order
of a picometre (except for \chem{H}), can be derived by
\begin{equation}\label{eq:dopplerwidth}
  \sigma_\lambda = \lambda_0
  \sqrt{\frac{k_\mathrm{B} N_\mathrm{A} T_\mathrm{kin}}{M\, c^2}},
\end{equation}
where $k_\mathrm{B}$, $N_\mathrm{A}$, and $c$ are the Boltzmann constant,
Avogadro constant, and the speed of light, respectively. Moreover, the
line-dependent central wavelength $\lambda_0$, the kinetic temperature
$T_\mathrm{kin}$, and the molecular weight $M$ need to be provided. These
parameters are listed in the line table of the model. Molecular weight and
temperature are also summarised in Table~\ref{tab:species}. In the case of
$T_\mathrm{kin}$ only two values are used. For the mesopause region, we take
190\,\unit{K}, which agrees well with the mean temperature profile at Cerro
Paranal \citep{noll16} measured with the satellite-based Sounding of the
Atmosphere using Broadband Emission Radiometry (SABER) instrument
\citep{russell99}. For the upper thermosphere, 1,000\,\unit{K} is a reasonable
assumption for the neutral and ion temperatures \citep[e.g.,][]{bilitza22}.
The temperatures can significantly vary in reality. However, this is not
crucial due to the square root in Eq.~(\ref{eq:dopplerwidth}) and the
visibility of natural line widths only in spectra with extremely high
resolution.

\begin{figure*}[t]
\includegraphics[width=14.2cm]{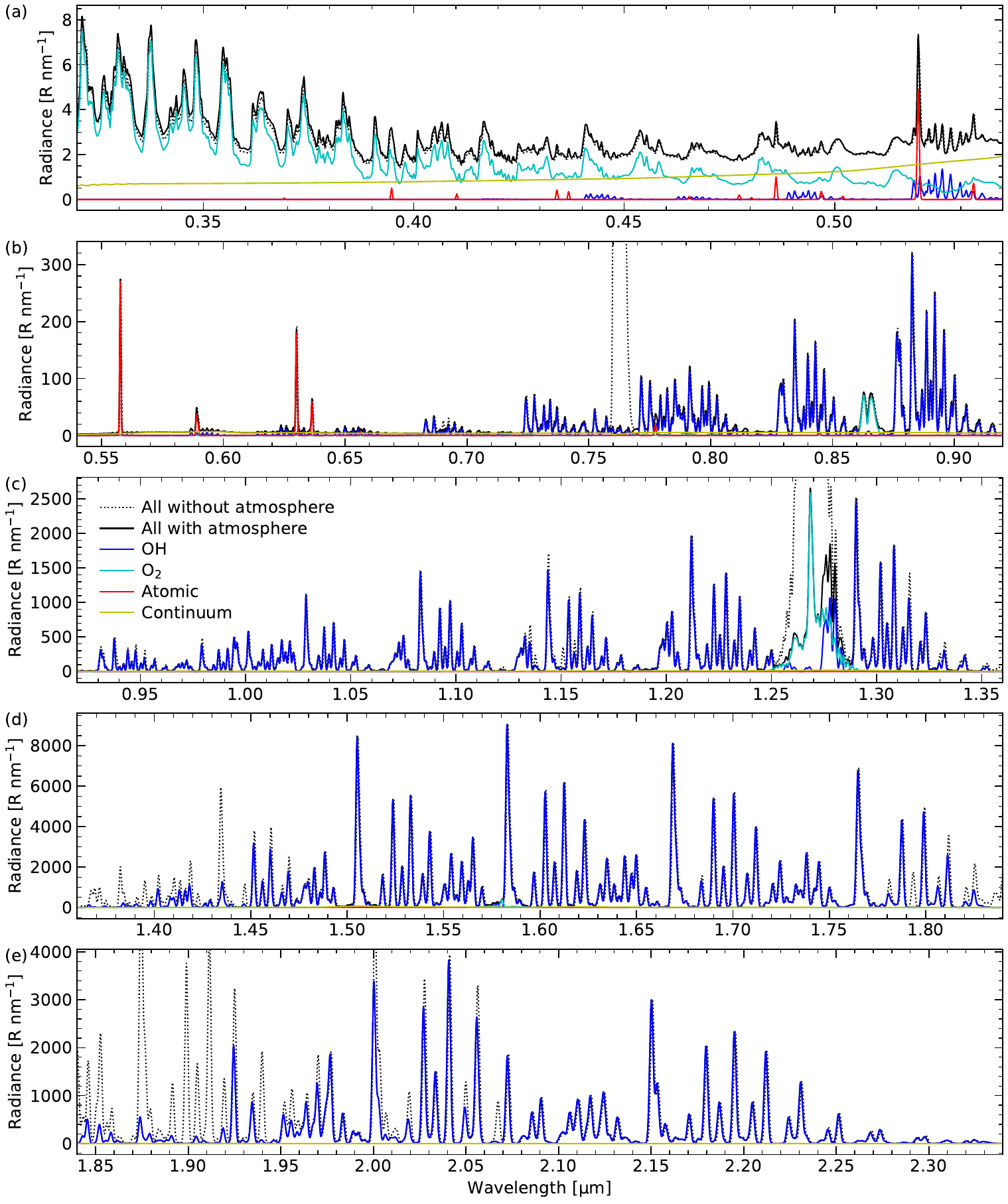}
\caption{Reference PALACE spectrum for default model parameters provided in
  Table~\ref{tab:parameters} in the wavelength range from 0.32 to
  2.34\,\unit{\mu{}m} divided into five sections with different radiance
  ranges. The combined spectrum for all components (black solid lines) is also
  provided without atmospheric absorption and scattering (black dotted lines).
  Moreover, the contributions of \chem{OH} (blue), \chem{O_2} (cyan), atoms
  (red), and the pseudo-continuum related to \chem{HO_2}, \chem{FeO}, and
  other molecules (yellow) are shown.}
\label{fig:refspec}
\end{figure*}

The Gaussian for each scaled line is calculated for the desired wavelength
grid given by \texttt{lammin}, \texttt{lammax}, and \texttt{dlam} and under
consideration of \texttt{isair}. Then, all Gaussians are summed up.
Furthermore, the scaled continuum components are added and the sum spectrum
is mapped to the selected wavelength grid. The combination of the resulting
line and continuum spectra constitutes the total emission
(Fig.~\ref{fig:flowchart}). The spectrum for the residual variability is
calculated in the same way. In principle, the simple summation would require
that overlapping emissions show variations which are fully correlated. In
reality, this can be different (see Sect.~\ref{sec:variability}). Emissions
might even be partly anti-correlated, which would lead to a decreased standard
deviation compared to the individual variations. Hence, the output variability
represents a maximum. Nevertheless, this is only a minor issue as adjacent
lines are often of similar type and most wavelength ranges are dominated by a
single variability class. Moreover, it would be very challenging to derive
effective correlations as they can depend on the specific perturbation.

The final step of the spectrum calculation in PALACE considers the spectral
resolving power $\lambda/\Delta\lambda$. It is a constant for the entire
wavelength range set by the parameter \texttt{resol}. In principle,
\texttt{resol} is independent of the wavelength grid with the step size
\texttt{dlam} but real spectra usually have several pixels per resolution
element. For the default values of \texttt{resol} and \texttt{dlam} in
Table~\ref{tab:parameters} of 1,000 and 0.1\,nm, the number of pixels is 3
at 0.3\,\unit{\mu{}m} and 25 at 2.5\,\unit{\mu{}m}. The limited resolution is
simulated by the convolution of the combined line and continuum spectrum with
wavelength-dependent Gaussians where $\Delta\lambda$ corresponds to the full
width at half maximum (FWHM), which corresponds to
$\sqrt{8\ln{2}}\,\sigma_\lambda$. The half size of the kernel is fixed to
$5\,\sigma_\lambda$. As the convolution near \texttt{lammin} and
\texttt{lammax} requires additional airglow data, the previous steps of the
calculation of the output spectrum are performed with an extended wavelength
grid. Note that the use of only Gaussians for the simulation of the
line-spread function (LSF) is simpler than in the case of the ESO Sky Model,
which also offers boxcar and Lorentzian components. The reproduction of
complicated instrumental LSFs is not in the focus of PALACE.

The final spectrum for the default parameters as listed in
Table~\ref{tab:parameters} is plotted for the wavelength range from 0.32 to
2.34\,\unit{\mu{}m} in Fig.~\ref{fig:refspec}. The most relevant emission
features have already been discussed in Sect.~\ref{sec:intro}. In order to
better distinguish the contributions from the different chemical species, the
figure also shows the specific spectra of \chem{OH}, \chem{O_2}, atomic lines,
and unresolved continua (including \chem{HO_2} and \chem{FeO}-like emissions).
Spectra for each chemical species can also be calculated by PALACE. This is
possible by setting the parameter \texttt{species} to the corresponding
empirical formula. By default, \texttt{species} is set to `all'.
Figure~\ref{fig:refspec} also illustrates the impact of \texttt{isatm}. For
`False', the dotted curve shows the spectrum without atmospheric absorption
and scattering, which indicates a particular high emission increase for the
ranges of the already mentioned water vapour bands near 1.4 and
1.9\,\unit{\mu{}m} as well as the \chem{O_2} bands at 0.762 and
1.27\,\unit{\mu{}m}. Without atmospheric extinction, the total emission of
the reference spectrum amounts to 897\,\unit{kR}, otherwise 603\,\unit{kR}.

\section{Data set}
\label{sec:dataset}

The reference data set for the build-up of the semi-empirical PALACE model
consists of 10 years of \mbox{X-shooter} \citep{vernet11} echelle spectra
taken for astronomical projects at Cerro Paranal between October 2009 and
September 2019. The properties and airglow-opimised processing of this
outstanding data set from the ESO Science Archive Facility were already
described in detail by \citet{noll22,noll23,noll24}. Therefore, we provide
here only a brief overview. \mbox{X-shooter} consists of three so-called arms
with optimum wavelength ranges (after the evaluation of the overlapping
regions) of 0.30 to 0.56\,\unit{\mu{}m} (UVB), 0.55 to 1.02\,\unit{\mu{}m}
(VIS), and 1.02 to 2.48\,\unit{\mu{}m} (NIR). These arms are usually operated
in parallel, although the exposure times of each arm can be set independently.
In any case, \mbox{X-shooter} allows the comparison of the variability of a
high number of airglow emission lines of different origin, which makes its
spectra valuable for combined studies of airglow physics and atmospheric
dynamics. For the standard width of the entrance slits of each arm, the
resulting resolving power is 5,400 (UVB), 8,900 (VIS), and 5,600 (NIR),
respectively. This is sufficiently high to separate many strong and weak
airglow emissions. On the other hand, line multiplets such as \chem{OH}
$\Lambda$ doublets are rarely resolved. However, this is not an issue as the
variations of the components are expected to be similar. The slit widths can
be set to different values, which causes a varying spectral resolving power in
the data set (e.g. from 3,500 to 12,000 in the NIR arm). However, a much wider
range of possibilities is related to the exposure time, which can vary by
several orders of magnitude. Hence, the data selection depends on the
signal-to-noise ratio requirements for the airglow emission features in focus.
This is a drawback of archived astronomical spectra that were taken for very
different targets and purposes. In the same way, residual contaminations of
the astronomical targets in the extracted airglow spectra can be very
different in strength and wavelength dependence. Therefore, each airglow
feature was investigated with an optimum set of spectra. This is not an issue
as long as the resulting samples are still large enough to derive
climatologies with sufficient quality. We only considered those emissions
where this was possible. The maximum useful data set consists of about 56,000
UVB, 64,000 VIS, and 91,000 NIR spectra. The different numbers are mainly due
to a different splitting of exposures in part of the observations. The
thorough flux calibration of the resulting one-dimensional spectra based on
time-dependent sets of master response curves resulted in a scatter (standard
deviation) of only a few per cent in most of the wavelengh range if
flux-calibrated standard star spectra are compared. There is also a possible
wavelength-dependent systematic offset of up to several per cent due to
uncertainties in the theoretical star spectra that were used for the
derivation of the response curves.

The wavelength coverage and spectral resolving power of \mbox{X-shooter}
allows the measurement of many airglow emission lines. However, the
measurement of faint lines in crowded wavelength regions can be too
challenging. Hence, we also used published airglow data from the UVES echelle
spectrograph \citep{dekker00} at the VLT. UVES covers a narrower wavelength
range than \mbox{X-shooter} of 0.31 to 1.04\,\unit{\mu{}m} in maximum if
set-ups with different central wavelengths are combined. However, the
resolving power is much higher. Depending on the width of the entrance slit,
values between about 20,000 and 110,000 are possible. \citet{hanuschik03}
measured 2,810 emission features in composite spectra of different set-ups
covering the maximum wavelength range. The spectra were created from a small
number of long-term exposures taken between June and August 2001. As only
observations with standard slits were included, the resulting resolving power
is between 43,000 and 45,000. \citet{cosby06} identified the airglow lines
related to the measured features. For \citet{noll12}, this catalogue was an
important data source for the ESO Sky Model. For PALACE, we focused on faint
\chem{OH} and \chem{O_2} lines without \mbox{X-shooter} measurements. Another
UVES-related data set was prepared by \citet{noll17}. It consists of about
10,400 spectra taken between April 2000 and March 2015 with the two reddest
set-ups with central wavelengths of 760 and 860\,\unit{nm} covering a maximum
range from 0.57 to 1.04\,\unit{\mu{}m}. A subsample of 2,299 spectra was then
used by \citet{noll19} to study the intensity and variations of the faint
\chem{K} emission line at 770\,\unit{nm}. Moreover, 533 long-term exposures
were selected by \citet{noll20} to calculate a high-quality composite
spectrum, which was used to study \chem{OH} populations. The line measurements
of both publications were also considered for PALACE.

\section{Line emission}
\label{sec:lines}

\begin{figure*}[t]
\includegraphics[width=17cm]{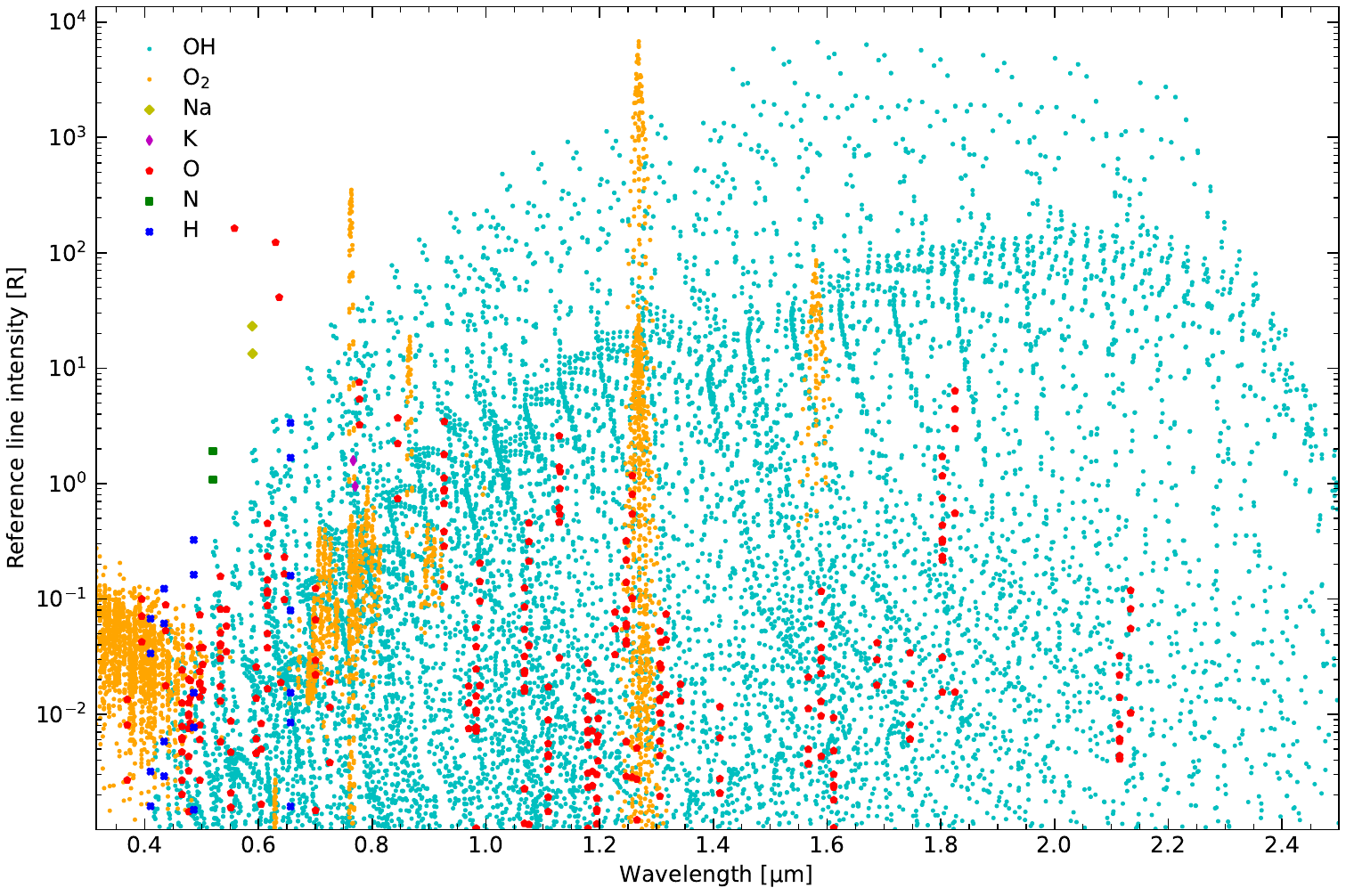}
\caption{Reference line intensities of PALACE model, i.e. annual nocturnal
  mean values for a solar radio flux of 100\,\unit{sfu}, as a function of
  wavelength. Atmospheric absorption and scattering is not applied. The 16,047
  lines of \chem{OH}, \chem{O_2}, \chem{Na}, \chem{K}, \chem{O}, \chem{N}, and
  \chem{H} with a minimum intensity of 1\,mR are marked by different symbols
  and colours (see legend).}
\label{fig:reflineint}
\end{figure*}

\begin{figure*}[t]
\includegraphics[width=17cm]{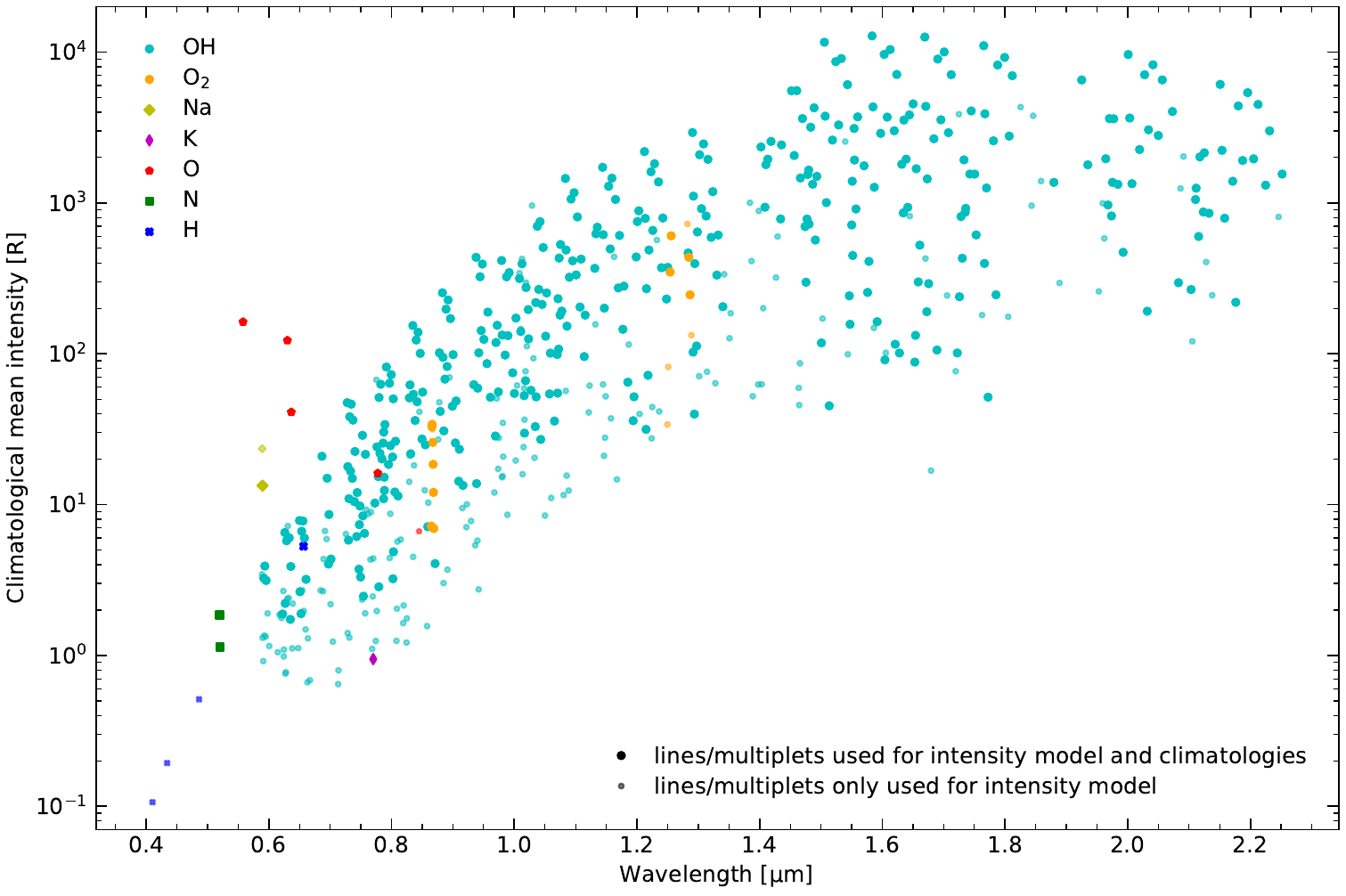}
\caption{Individual lines or multiplets with measured climatologies that were
  used for the development of the mean intensity model as shown in
  Fig.~\ref{fig:reflineint} (574 cases) and the derivation of the reference
  climatologies as listed in Table~\ref{tab:varclasses} (392 cases, marked by
  big symbols, see legend). Except for the UVES-based \chem{K}\,D$_1$ data
  from \citet{noll19}, all data originate from \mbox{X-shooter} measurements.
  Each line or multiplet is identified by the chemical species (different
  symbols and colours, see legend), annual nocturnal mean intensity for
  100\,\unit{sfu}, and wavelength.}
\label{fig:climlines}
\end{figure*}

The list of reference line intensities is an important ingredient of PALACE.
It is illustrated by Fig.~\ref{fig:reflineint}, which shows intensity vs.
wavelength for all lines with a strength of at least 1\,\unit{mR}. The
included chemical species are marked by different symbols and colours. The
total number of lines for each chemical species (or isotopologue if relevant)
is provided in Table~\ref{tab:species}. Figure and table clearly show the
predominance of \chem{OH} lines (22,105 of 26,541 lines), followed by
\chem{O_2} and \chem{O} emissions. In contrast, the atoms \chem{Na}, \chem{K},
and \chem{N} are only present by a doublet. The wavelength distribution of
the line intensites can be compared to the spectrum without atmospheric
extinction in Fig.~\ref{fig:refspec}. A small subset of strong lines is
decisive for the visible structures in most wavelength ranges. An exception
is the near-UV and blue wavelength regime, where only weak lines are present. 

The derivation of reference intensities for 26,541 lines was only possible by
the use of theoretical data such as Einstein-$A$ coefficients, level
energies, and recombination coefficients. Hence, PALACE is a semi-empirical
model. As demonstrated by Fig.~\ref{fig:climlines}, the data set with derived
climatologies is distinctly smaller. It only consists of 574 intensity values,
which are mostly related to \chem{OH} (544 data points). Except for the
UVES-based data for the \chem{K} line at 770\,\unit{nm} \citep{noll19}, all
measurements are related to \mbox{X-shooter}. Line intensities from composite
spectra without variability information \citep{cosby06,noll20} are not
displayed. Even if a climatology was measured and used to derive a reference
intensity, it was not always good enough to be taken for the variability model
(Sect.~\ref{sec:variability}). Hence, the latter was derived from only 392
climatologies (see big symbols in Fig.~\ref{fig:climlines}). Note that the
intensities were often measured for unresolved multiplets. In the case of
\chem{OH} lines, the plotted intensities are related to $\Lambda$ doublets.
Line blends are also relevant for the \chem{H} lines, the \chem{O} 777 and
845\,\unit{nm} emissions, and the \chem{O_2} band at 865\,\unit{nm}.
Therefore, these blends also contribute to the difference of data points in
Figs.~\ref{fig:reflineint} and \ref{fig:climlines}.

Except for the \chem{OH} lines in the NIR arm (Sect.~\ref{sec:dataset})
discussed by \citet{noll23}, the \mbox{X-shooter} intensity time series used
for PALACE are new measurements. For lines without existing data set, we used
the same approach as described by \citet{noll22,noll23,noll24}. First, the
line emission was separated from the underlying continuum, which consists of
unresolved airglow emission (see Sect.~\ref{sec:continuum}) as well as other
contributions to the night-sky brightness such as zodiacal light or scattered
moonlight. The separation was achieved by the application of percentile
filters with varying percentile (50 to 20\,\% for increasing density of strong
lines) and window width relative to the wavelength (0.008 to 0.04). After the
subtraction of the filtered continuum, fluxes were integrated in wavelength
intervals around the target line positions where the width relative to the
wavelength depended on the spectral resolution (defined by the
\mbox{X-shooter} arm and the width of the entrance slit) and wavelength
differences of line components in the case of measured multiplets. For
molecules, the line positions originate from the 2020 version of the HITRAN
database \citep{gordon22} with revised \chem{OH}-related calculations based on
the measurements of \citet{noll20}. For atoms, the current version of the
National Institute of Standards and Technology (NIST) database
\citep{NIST_ASD} was used.

The resulting line intensities were corrected for the van Rhijn effect as well
as atmospheric absorption and scattering. This worked in the same way as
discussed in Sect.~\ref{sec:overview}, with the only difference that the
reverse factors were applied. While the zenith angle in the middle of an
exposure is well determined, the effective \texttt{pwv} value of each spectrum
needs a measurement. As described by \citet{noll22}, the \texttt{pwv} value
was estimated from \chem{OH} line pairs with very different absorption. The
calibration of the relations was achieved by regular water vapour measurements
with a microwave radiometer at Cerro Paranal \citep{kerber12}. In the case of
unresolved multiplets, the effective absorption was derived by means of
component weights depending on Einstein-$A$ coefficients from HITRAN or NIST.
As a final step of the preparation of the time series, the whole period was
divided into intervals with a length of 30\,\unit{min}. All data with the
middle of the exposure in such an interval were averaged with the individual
exposure time as weight. The goal of this procedure was the reduction of the
impact of the large variations in the exposure time. For good data quality
in the entire time series, a minimum summed exposure time of 10\,\unit{min}
was required for each selected interval. As a consequence, intensity
uncertainties are mainly determined by systematic effects such as the quality
of the separation of line and continuum, the possible contamination by other
airglow lines or remnants of the astronomical targets, the quality of the
extinction correction, and uncertainties in the flux calibration.

The procedure for the calculation of climatologies based on the binned time
series is described in Sect.~\ref{sec:climatologies}. In this section, we only
focus on the nocturnal annual mean intensity for a reference solar radio flux
of 100\,\unit{sfu}, which is derived from the climatological grid data
weighted by the nighttime contribution. The climatology-based reference
intensities as shown in Figs.~\ref{fig:reflineint} and \ref{fig:climlines}
differ from the mean values of the binned time series by less than 10\,\%. For
most lines from the mesopause region, the deviation is even of the order of
only 1\,\%. The relatively small deviations are related to the fact that the
mean solar radio flux of the different time series is already close to
100\,\unit{sfu}.

The UVES-based data sets were also adapted to minimise systematic deviations
from the \mbox{X-shooter} data. For the \chem{K} 770\,\unit{nm} time series
from \citet{noll19}, we added the small atmospheric scattering correction and
performed the binning of the data based on 30\,\unit{min} intervals. Despite
the different approach to calculate climatologies (see
Sect.~\ref{sec:climatologies}), the reference intensity did not significantly
change. There are larger changes with respect to the line catalogue of
\citet{hanuschik03} and \citet{cosby06}, which was already used for the ESO
Sky Model. While \citet{noll12} included each emission feature measured by
\citet{hanuschik03} as a single entry in the resulting line list, we used the
individual components of each feature as derived by \citet{cosby06} from
theoretical line position and intensity estimates. The latter allowed us to
use relatively accurate model wavelengths for each line in the blend and to
determine weighting factors for the contribution of the calculated lines to
the feature intensity, which was measured by \citet{hanuschik03} based on fits
of Gaussians. The full line list contains 4,167 unique entries with identified
origin. The line intensities were corrected for the van Rhijn effect and
atmospheric extinction with the same recipes as for the \mbox{X-shooter} data
and also taking the data from Table~\ref{tab:species}. In order to derive the
relevant zenith angle for the composite spectrum of each set-up, we used the
mean airmass values from \citet{hanuschik03}. For the \texttt{pwv} value, we
just set the reference value of 2.5\,\unit{mm} (Table~\ref{tab:parameters}).
This is not an issue for our analysis as absorption by water vapour is not
significant for most crucial lines in the list. We also considered that the
flux calibration of \citet{hanuschik03} was obviously performed with an
outdated extinction curve as the curve of \citet{patat11} was not available
yet. Finally, for the combined use of \mbox{X-shooter} and UVES data, we
performed additional scaling operations in order to remove remaining
differences in the data sets. This was also relevant for the \chem{OH} data
from \citet{noll20}. Details are given in subsequent subsections which discuss
the line intensity models for the different chemical species.

\subsection{Hydroxyl}
\label{sec:OH}

The \chem{OH} radical is the most important contributor to the Earth's
nightglow spectrum with the strongest ro-vibrational bands in the near-IR
(Fig.~\ref{fig:refspec}). The reference effective emission height is at about
87\,\unit{km} with an FWHM of about 8\,\unit{km} \citep[e.g.,][]{baker88},
although the effective height can vary depending on the line and wave-driven
perturbations of the layer by amounts similar to the FWHM in extreme cases
\citep{noll22}. Vibrationally excited \chem{OH} in the electronic ground state
is mostly produced by the reaction
\begin{reaction}\label{eq:H+O3}
  \mathrm{H} + \mathrm{O}_3 \rightarrow \mathrm{OH}^{\ast} + \mathrm{O}_2 
\end{reaction}
\citep{bates50}, which preferentially populates the vibrational levels $v = 9$
to 7 with decreasing fractions
\citep[e.g.,][]{charters71,llewellyn78,adler97}. Subsequent photon emissions
and collisions with other atmospheric constituents (\chem{O}, \chem{O_2}, and
\chem{N_2}) then redistribute the level populations and finally lead to a
predominance of low $v$ \citep[e.g.,][]{dodd94,cosby07,noll15}. Although the
exothermicity of Reaction~(\ref{eq:H+O3}) is not sufficient, very faint lines
with an upper vibrational level of $v^{\prime} = 10$ were detected
\citep{osterbrock98,cosby07}, which requires sufficient kinetic energy of the
reactants and/or excited \chem{O_3}. As a consequence, the \chem{OH} emission
spectrum in Fig.~\ref{fig:refspec} contains various ro-vibrational bands with
$v^{\prime}$ between 2 and 10 and lower vibrational levels $v^{\prime\prime}$
between 0 and 8. Bands with $\Delta v = v^{\prime} - v^{\prime\prime} = 1$ are
outside the \mbox{X-shooter} range. The central wavelengths of \chem{OH} bands
increase with decreasing $\Delta v$ and increasing $v^{\prime}$
\citep[e.g.,][]{osterbrock96,rousselot00,noll15}. The strongest emissions are
related to $\Delta v = 2$.

Each ro-vibrational band consists of R, Q, and P branches (sorted by
wavelength). They are characterised by a change of the upper rotational level
$N^{\prime}$ to the lower rotational level $N^{\prime\prime}$ by $-1$, 0, and,
$+1$, respectively. As the lowest $N$ is defined as 1 for \chem{OH}
\citep[e.g.,][]{pendleton93,dodd94,rousselot00}, $N^{\prime} \geq 2$ for R
branches and $N^{\prime\prime} \geq 2$ for P branches. An example for an
\chem{OH} ro-vibrational band is \mbox{(9-7)} at wavelengths longer than about
2.1\,\unit{\mu{}m} in Fig.~\ref{fig:refspec}, which shows that P-branch lines
are best measurable due to their wider separations. In fact, each branch is
divided into two groups related to the two electronic substates X$^2\Pi_{3/2}$
($F = 1$) and X$^2\Pi_{1/2}$ ($F = 2$). Lines with $F^{\prime} = 1$ are
stronger than those with $F^{\prime} = 2$, which is illustrated by the P$_1$
and P$_2$ branches of \mbox{(9-7)}. Note that $F^{\prime\prime} = F^{\prime}$
for the visible lines since the intercombination lines are fainter by several
orders of magnitude. Lines with high $N^{\prime}$ are also relatively faint.
The $N^{\prime} = 7$ emissions near 2.30\,\unit{\mu{}m} are hard to recognise,
whereas P$_1$($N^{\prime} = 2$) is the strongest P-branch line in the
reference spectrum. Finally, each line characterised by the quantum numbers
$v$, $N$, and $F$ is actually a doublet. However, the separation of these
$\Lambda$ doublets is usually too small for \mbox{X-shooter}. 

The modelling of the complex \chem{OH} emission pattern can be simplified by
the change of the focus from line intensities $I_{i^{\prime}i^{\prime\prime}}$ for
state transitions from $i^{\prime}$ to $i^{\prime\prime}$ to level populations
 for the upper states $n_{i^{\prime}}$, which can be derived by
\begin{equation}\label{eq:levpop}
  n_{i^{\prime}} =
  \frac{I_{i^{\prime}i^{\prime\prime}}}{A_{i^{\prime}i^{\prime\prime}}
  g_{i^{\prime}}}
\end{equation}
\citep[e.g.,][]{noll20}. This transformation requires the knowledge of the
corresponding Einstein coefficients $A_{i^{\prime}i^{\prime\prime}}$ in inverse
seconds and the number of degenerate substates contributing to $i^{\prime}$,
i.e. the statistical weight 
\begin{equation}\label{eq:statweight}
  g_{i^{\prime}} \equiv g^{\prime} = 4\,(N^{\prime} - F^{\prime} + 2) = 2\,
  (2 J^{\prime} + 1),
\end{equation}
where $J^{\prime}$ is the quantum number of the total angular momentum of
$i^{\prime}$. If $\Lambda$ doublets are not separated, this can be implemented
by doubling $g^{\prime}$ and averaging the $A$ coefficients of both components
(which are often very similar) under consideration of possible
temperature-dependent population differences.

\begin{figure*}[t]
\includegraphics[width=17cm]{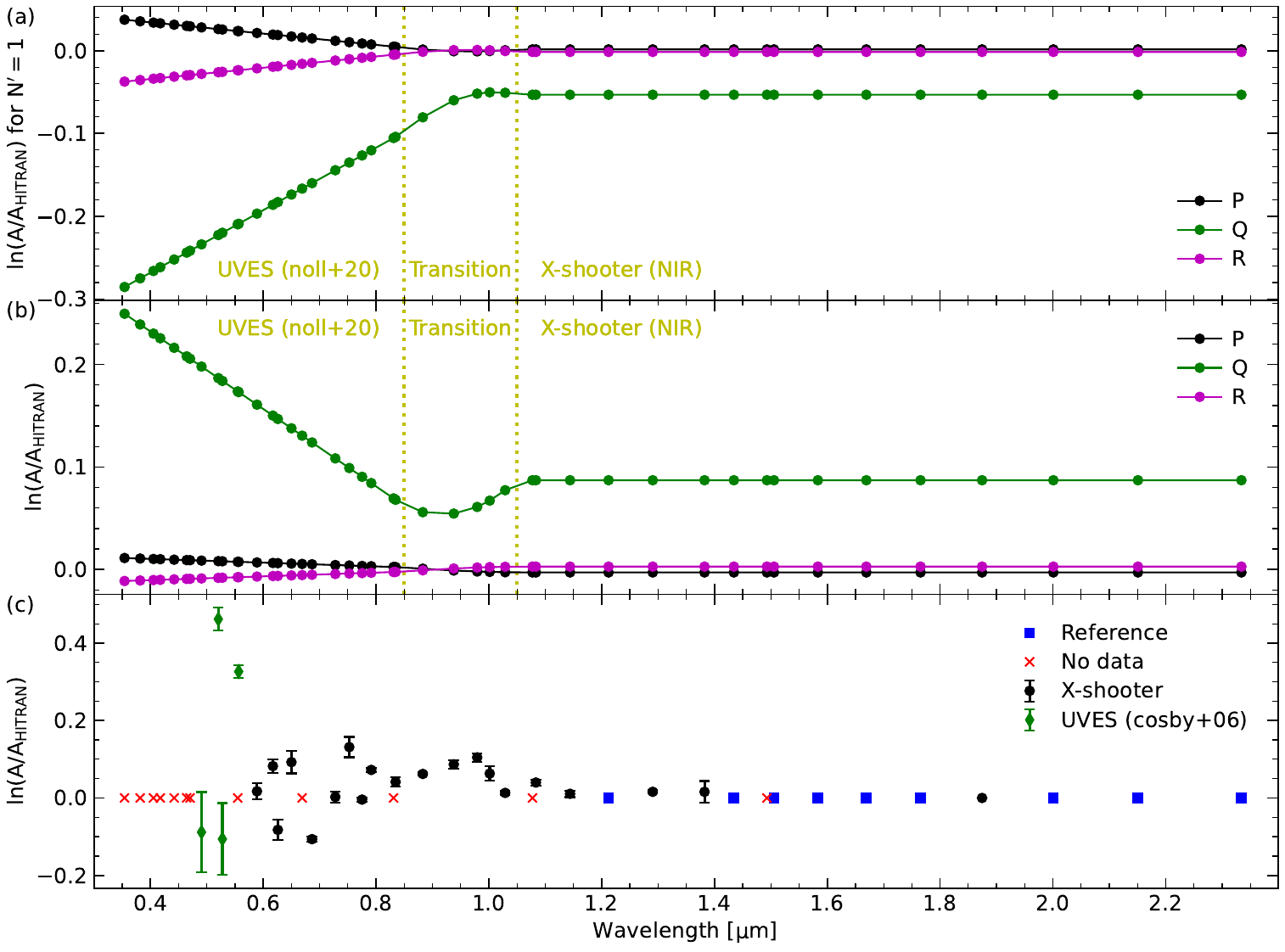}
\caption{Correction factors for \chem{OH} Einstein-$A$ coefficients from
  HITRAN2020 \citep{gordon22}. Each OH band is marked by the wavelength of the
  Q$_1$(1) line. Moreover, the upper two panels distinguish between the
  P (black), Q (green), and R (magenta) branches. In these cases, the
  wavelength-dependent correction functions originate from the UVES-based fits
  of \citet{noll20} and the values for the \mbox{X-shooter} NIR-arm data of
  this study. The wavelength range from 0.85 to 1.05\,\unit{\mu{}m} represents
  a smooth transition between both regimes. In \textbf{(c)}, the band-specific
  correction factors and their uncertainties are either based on
  \mbox{X-shooter} data (black circles) or UVES data from \citet{cosby06}
  (green diamonds). Depending on the upper vibrational level, they are given
  relative to the reference bands marked by blue boxes. \chem{OH} bands
  without data for the derivation of the correction are displayed by red
  crosses. The total correction factor for an OH $\Lambda$ doublet is the
  product of the $A/A_\mathrm{HITRAN}$ of the three panels and depends on
  the band, branch, and upper rotational level $N^{\prime}$. The latter needs
  to be multiplied by the selected $A/A_\mathrm{HITRAN}$ in \textbf{(a)}.}
\label{fig:corrA_OHbands}
\end{figure*}

The quality of the level populations resulting from Eq.~(\ref{eq:levpop})
depends on uncertainties in the $A$ coefficients. There is a long history of
theoretical calculations
\citep[e.g.,][]{mies74,langhoff86,goldman98,vanderloo07,brooke16}, which
showed how challenging the task is for \chem{OH}. \citet{noll20} studied the
quality of different sets of coefficients by comparing populations from lines
with the same upper level. Based on 544 reliable (partly resolved) $\Lambda$
doublets measured in a UVES composite spectrum (see Sect.~\ref{sec:dataset}),
the analysis revealed clear systematic deviations, especially for Q branches
and $\Lambda$ doublet components. In order to improve the quality of the
populations, \citet{noll20} corrected the most recent $A$ coefficients of
\citet{brooke16} based on the results of the population comparisons with
respect to bands, branches, and $\Lambda$ doublets. As \citet{brooke16} is
included in the HITRAN2020 database \citep{gordon22}, the resulting recipes of
\citet{noll20} are also relevant for our population modelling. However, the
UVES-based data do not cover the \chem{OH} bands in the \mbox{X-shooter} NIR
arm. Therefore, we had to extend and revise the correction of the coefficients
of \citet{brooke16}.

The \mbox{X-shooter}-based sample of \chem{OH} climatological mean intensities
(see Fig.~\ref{fig:climlines}) comprises 219 $\Lambda$ doublets in the VIS arm
and 325 double lines in the NIR arm. The latter were already measured by
\citet{noll23}. We used the NIR data for population comparisons between the R
and P as well as the Q and P branches. The number of suitable pairs with the
same upper level was 60 and 37, respectively. The relatively small numbers are
partly explained by the underrepresentation of R- and Q-branch lines in the
full sample (about 40\,\%) due to more severe line blending compared to
P-branch lines. Similar to \citet{noll20}, we performed a linear regression
for both branch comparisons with respect to the change of the logarithmic
population difference $\Delta y$ as a function of the upper rotational level
$N^{\prime}$. The latter was in maximum 8 for R vs. P and 3 for Q vs. P. The
resulting robust slopes and intercepts were the basis for the correction of
the $A$ coefficients. For this, we assumed in agreement with \citet{noll20}
that $\ln(A/A_\mathrm{HITRAN})$ corresponds to $-0.5\,\Delta y_\mathrm{R-P}$
for P, $\Delta y_\mathrm{Q-P} - 0.5\,\Delta y_\mathrm{R-P}$ for Q, and
$+0.5\,\Delta y_\mathrm{R-P}$ for R. The correction of the $A$ coefficients for
the \chem{OH} bands in the NIR arm parameterised by slope and intercept is
shown in Figs.~\ref{fig:corrA_OHbands}a and \ref{fig:corrA_OHbands}b as
constant lines for the three branches. While the corrections for the P
and R branches are very small, the $A$ coefficients for the Q branch are
significantly modified. Note that the values in Fig.~\ref{fig:corrA_OHbands}a
need to be multiplied by $N^{\prime}$ and then added to the values in
Fig.~\ref{fig:corrA_OHbands}b to obtain the total branch effect.

In the \mbox{X-shooter} VIS-arm range, we can rely on the UVES-based fits of
\citet{noll20}. Thanks to the very high resolving power of UVES, larger data
sets can be used for the comparison of the branch populations. In fact, 101
pairs for R vs. P and 67 pairs for Q vs. P were sufficient to even perform a
linear regression analysis for 12 individual \chem{OH} bands. As the
regression slopes and intercepts indicated changes depending on the central
wavelength of the band (defined by Q$_1$($N^{\prime} = 1$)), a
wavelength-based second regression analysis was carried out by \citet{noll20}.
The resulting tilted fit lines are also plotted in
Figs.~\ref{fig:corrA_OHbands}a and \ref{fig:corrA_OHbands}b. Such a clear
wavelength dependence (especially for the Q branch) was not seen in the
NIR-arm data. Hence, we prefer the band-independent but more robust fit there.
Moreover, the tilted lines from the UVES-based fit model get close to the
constants from the \mbox{X-shooter}-based fits for the reddest bands covered
by UVES. In order to remove the remaining discrepancies, we applied a gradual
linear transition between both models in the wavelength range from 0.85 to
1.05\,\unit{\mu{}m} (marked in the figure), which caused changes in the $A$
coefficient corrections for the \mbox{(7-3)}, \mbox{(8-4)}, and \mbox{(3-0)}
bands with central wavelengths below 1\,\unit{\mu{}m} that were fitted
by \citet{noll20}. Except for the intercept for the Q branch, where a local
minimum with a depth of a few per cent is produced, the results are
convincing.

Another kind of correction is possible with respect to the population
differences for bands with the same upper vibrational level $v^{\prime}$ but
different lower vibrational levels $v^{\prime\prime}$. For this correction, we
jointly used the \mbox{X-shooter} VIS- and NIR-arm data in order to obtain
a consistent model for a wide wavelength range. As reference bands for each
$v^{\prime}$, we selected the strong ones with $\Delta v = 2$, which should
also be more reliable with respect to the theoretical $A$ coefficients than
bands with higher $\Delta v$. The only exception is our reference band for
$v^{\prime} = 7$, where we took \mbox{(7-4)} instead of \mbox{(7-5)} because
of the strong water vapour absorption, which only allowed us to use two lines.
In principle, we could also consider the results from \citet{noll20} for this
comparison. However, the narrower wavelength range of UVES without access to
the $\Delta v = 2$ bands and the fact that all bands measured by
\citet{noll20} were also covered by the \mbox{X-shooter} data (although with a
lower number of selected lines per band on average: about 7 vs. about 11), let
us prefer the \mbox{X-shooter} measurements. For each band, we only considered
the relatively strong lines with $N^{\prime} \leq 4$ for the P and R branches
plus the Q$_1$(1) line in maximum. As the corrections of the systematic
effects related to branch and $N^{\prime}$ were already performed before,
differences in the selected line samples for the individual bands should not
have a significant impact.

The results for the band-specific correction of the $A$ coefficients are shown
in Fig.~\ref{fig:corrA_OHbands}c. For the \mbox{X-shooter}-related
$\ln(A/A_\mathrm{HITRAN})$, there is a complex pattern with increasing
discrepancies towards shorter wavelengths. Most corrections are positive with
a maximum of $+0.13$ for \mbox{(4-0)}, whereas only \mbox{(9-3)} ($-0.08$) and
\mbox{(7-2)} ($-0.11$) are clearly negative. It is not easy to understand the
data pattern but the uncertainties of the mean values are relatively small.
Moreover, we compared our correction factors to the UVES-based results and
found a very good agreement for non-reference bands below 1\,\unit{\mu{}m}
with differences of the order of 1\,\%. For the reference bands of
\citet{noll20}, we discovered unexpected non-zero offsets of $-0.16$ for
\mbox{(4-0)} and \mbox{(5-1)} and shifts of $0.00$ to $+0.06$ for
\mbox{(9-4)}, \mbox{(6-2)}, \mbox{(7-3)}, and \mbox{(8-3)}, which point to
issues with the data release. Consequently, the modified $A$ coefficients of
the new analysis should be used in any case.

The already discussed UVES-based line catalogue of \citet{cosby06} that is
based on measurements of \citet{hanuschik03} includes 756 OH $\Lambda$
doublets (40\,\% of them resolved) and 136 cases where only one doublet
component was measured. As this line list contains \chem{OH} bands that were
too weak for \mbox{X-shooter} and were not measured by \citet{noll20}, it is
possible to extend the population analysis. First, we had to level out
possible intensity differences compared to our \mbox{X-shooter} data due to
differences in the sample properties and the flux calibration. For this, we
compared the lines in the catalogue to those in the \mbox{X-shooter}-based
line list in order to obtain specific mean corrections for the UVES
set-ups centred on 580 and 860\,\unit{nm} that were considered by
\citet{cosby06}. As the spectra of the latter (which show a gap in the centre
due to the use of two detector chips denoted by L and U) were taken at
different times and the systematic uncertainties in the flux calibration may
differ, this division of the wavelength range is reasonable. We found that the
UVES-based intensities were higher by similar mean factors of 1.19 (860U, 0.86
to 1.04\,\unit{\mu{}m}), 1.15 (860L, 0.67 to 0.86\,\unit{\mu{}m}), and 1.18
(580U, 0.58 to 0.68\,\unit{\mu{}m}) with uncertainties of about 0.02. The main
reason for the deviation from 1.0 is probably the relatively high solar
activity in 2001 when the UVES data were taken (Sect.~\ref{sec:dataset}). We
corrected the discrepancies using the given factors, also considering the
value for 580U for 580L (0.48 to 0.58\,\unit{\mu{}m}). Then, we also
calculated band-specific population differences for the \citet{cosby06} data.
Overall, the pattern is similar to the \mbox{X-shooter}-based results but with
some scatter. Most deviations were smaller than 0.05 but cases up to 0.11 were
also found. As the 580L range also covers four \chem{OH} bands not measured in
the \mbox{X-shooter} spectra, we were able to extend the correction of the $A$
coefficients in Fig.~\ref{fig:corrA_OHbands}c. While the
$\ln(A/A_\mathrm{HITRAN})$ of \mbox{(9-2)} and \mbox{(7-1)} are strongly
positive but with low uncertainties, the values of the very faint bands
\mbox{(8-1)} and \mbox{(6-0)} with line intensities much smaller than
1\,\unit{R} are negative with high uncertainties. For the remaining bands
without data, we kept $\ln(A/A_\mathrm{HITRAN})$ at 0 as in the case of the
reference bands.

\citet{noll20} also derived corrections for the ratio of the $A$ coefficients
of the $\Lambda$ doublet components, which is usually considered to be unity
in theoretical calculations. However, especially doublets with high
$N^{\prime}$ can show significant deviations. This correction does not affect
the subsequent population analysis, which is only based on combined
$\Lambda$ doublets, but was relevant for the calculation of the \chem{OH} line
intensities based on the HITRAN database for the PALACE line model.

\begin{figure*}[t]
\includegraphics[width=17cm]{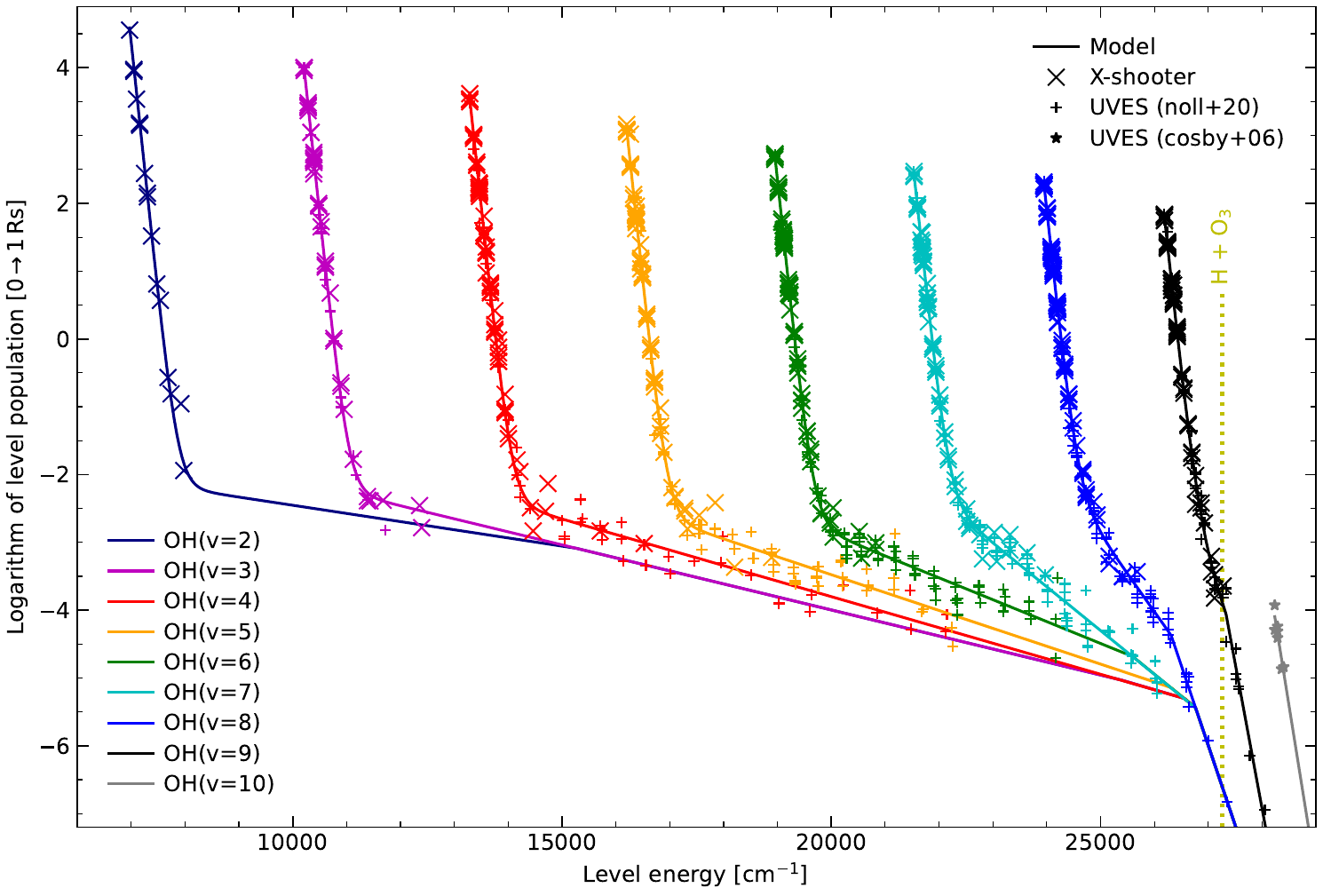}
\caption{\chem{OH} population model for the electronic ground state and
  vibrational levels $v = 2$ to 10 in logarithmic units and depending on the
  level energy. For $v \leq 7$, 8 to 9, and 10, the model (solid lines)
  consists of two, three, and one fitted components, respectively. The
  corresponding parameters are provided in Table~\ref{tab:popfits}. The basic
  data for the $v \leq 9$ model originate from \mbox{X-shooter} measurements
  (crosses) and the UVES-related data from \citet{noll20} (plus signs). For
  $v = 10$, UVES-based data from \citet{cosby06} (stars) were used. The energy
  provided by the \chem{OH}-producing reaction of \chem{H} and \chem{O_3} is
  marked by a vertical dotted line.}
\label{fig:popmodel_OH}
\end{figure*}

\begin{table*}[t]
\caption{Fit parameters and their uncertainties for cold and hot \chem{OH} and
\chem{O_2} vibrational level populations}
\begin{tabular}{ccccccc@{ }cc@{ }cc@{ }cc@{ }c}
\tophline
Mol. & $e$ & $v$ & Data$^\mathrm{a}$ & $N_\mathrm{sel}$$^\mathrm{b}$ &
$E_0$$^\mathrm{c}$ &
\multicolumn{2}{c}{$y_{0,\mathrm{cold}}$$^\mathrm{d}$} &
\multicolumn{2}{c}{$T_\mathrm{cold}$$^\mathrm{e}$} &
\multicolumn{2}{c}{$y_{0,\mathrm{hot}}$$^\mathrm{d}$} &
\multicolumn{2}{c}{$T_\mathrm{hot}$$^\mathrm{e}$} \\
& & & & & \unit{cm^{-1}} & \multicolumn{2}{c}{ln(\unit{R\,s})} &
\multicolumn{2}{c}{\unit{K}} & \multicolumn{2}{c}{ln(\unit{R\,s})} &
\multicolumn{2}{c}{\unit{K}} \\
\middlehline
\chem{OH} & X & 2 & xuo & 18 & 6,971.37 & 4.591 & 0.108 & 191.0 & fixed &
$-2.090$ & fixed & 12,000. & fixed \\
\chem{OH} & X & 3 & xuo & 75 & 10,210.57 & 3.994 & 0.022 & 191.0 & fixed &
$-2.118$ & fixed & 7,500. & fixed \\
\chem{OH} & X & 4 & xu & 155 & 13,287.18 & 3.538 & 0.026 & 191.0 & fixed &
$-2.265$ & 0.045 & 6,294. & 285. \\
\chem{OH} & X & 5 & xu & 166 & 16,201.32 & 3.088 & 0.027 & 191.0 & fixed &
$-2.475$ & 0.055 & 5,455. & 341. \\
\chem{OH} & X & 6 & xu & 210 & 18,951.87 & 2.701 & 0.019 & 191.0 & fixed &
$-2.540$ & 0.032 & 4,469. & 158. \\
\chem{OH} & X & 7 & xu & 185 & 21,536.21 & 2.415 & 0.023 & 191.0 & fixed &
$-2.067$ & 0.035 & 2,233. & 48. \\
\chem{OH} & X & 8 & xu & 219 & 23,949.83 & 2.262 & 0.015 & 191.0 & fixed &
$-1.732$ & 0.048 & 1,288. & 35. \\
& & & xu & 13 & 26,309.32 & & & & & $-4.369$ & 0.062 & 611. & 33. \\
\chem{OH} & X & 9 & xu & 165 & 26,185.80 & 1.780 & 0.016 & 191.0 & fixed &
$-1.937$ & 0.286 & 760. & 108. \\
& & & xu & 12 & 27,321.20 & & & & & $-4.020$ & 0.105 & 336. & 24. \\
\chem{OH} & X & 10 & c & 7 & 28,234.08 & & & & & $-4.090$ & 0.085 & 290. &
58. \\
\chem{O_2} & a & 0 & x & 8 & 7,892.02 & 15.747 & 0.015 & 193.1 & 1.4 & & &
& \\
\chem{O_2} & a & 1 & x & 0 & 9,375.37 & 4.693 & fixed & 193.1 & fixed & & &
& \\
\chem{O_2} & b & 0 & x & 8 & 13,122.01 & 7.359 & 0.010 & 187.3 & 1.2 & & &
& \\
& & & c & 30 & 13,122.01 & (7.572) & (0.045) & (195.0) & (2.9) &
$-1.963$ & 0.519 & 2,167. & 759. \\
\chem{O_2} & b & 1 & c & 44 & 14,526.74 & 0.406 & 0.137 & (195.0) & fixed &
$-1.679$ & 0.504 & 6,181. & high \\
\chem{O_2} & b & 2 & c & 24 & 15,903.50 & 1.100 & high & (195.0) &
fixed & $-1.807$ & high & 6,181. & fixed \\
\bottomhline
\end{tabular}
\belowtable{}
\begin{list}{}{}
\item[$^\mathrm{a}$] x = \mbox{X-shooter} data (NIR-arm OH intensities already
  used by \citet{noll23}), u = UVES data from \citet{noll20}, c = UVES data
  from \citet{cosby06} (O$_2$-related fit values in parentheses replaced by
  \mbox{X-shooter}-based results for the final model), o = fit parameters for
  OH low-$v$ hot populations from \citet{oliva15}.
\item[$^\mathrm{b}$] Number of line measurements used for the fit.
\item[$^\mathrm{c}$] Minimum level energy according to HITRAN2020
  \citep{gordon22}.
\item[$^\mathrm{d}$] Logarithm of population of individual level at $E_0$.
\item[$^\mathrm{e}$] Effective (pseudo-)temperature of population
  distribution.
\item[] Uncertainties: values given for uncertainties below 100\,\%
  (otherwise `high'), `fixed' for parameters without a fit.
\end{list}
\label{tab:popfits}
\end{table*}

With the corrected $A$ coefficients, the populations resulting from
Eq.~(\ref{eq:levpop}) are more consistent. The natural logarithm of the
selected 544 \mbox{X-shooter}-based populations in rayleigh seconds, $y$, is
plotted in Fig.~\ref{fig:popmodel_OH} as a function of the level energy $E$
in inverse centimetres. As the UVES-based line list of \citet{noll20} also
contains relatively faint lines with higher $N^{\prime}$ (7.2 vs. 3.9 on
average), we also included the corresponding 664 populations in our analysis
in order to achieve more complete population distributions. For this purpose,
we had to correct possible systematic discrepancies caused by the sample
differences and the flux calibration. From the comparison of 135 lines, we
found that the UVES-based populations had to be increased by $\Delta y$ of
0.096 for vibrational level $v = 3$, 0.063 for $v = 4$, 0.042 for $v = 5$, and
between 0.030 and 0.034 for the higher $v$. The corrected populations are
also plotted in Fig.~\ref{fig:popmodel_OH}. There are multiple population
measurements for most levels depending on the number of measured lines with
different lower levels and the coverage of a level by the two samples.
Finally, we also used the previously corrected data from \citet{cosby06}.
Here, we only focused on lines with $v^{\prime} = 10$, which are not present in
the two other catalogues. The list of safe detections is short. It only
contains one \mbox{(10-4)} and six \mbox{(10-5)} $\Lambda$ doublets. Hence,
possible differences in the HITRAN-related $A$ coefficients for these two
bands could not be corrected (see Fig.~\ref{fig:corrA_OHbands}c). The
resulting populations are also displayed in Fig.~\ref{fig:popmodel_OH}. 

The distribution of the plotted level populations shows a steep decrease for
low rotational levels $N$ for each $v$. For high $N$ (if available), the
drop is distinctly weaker, especially for low $v$. Moreover, the populations
tend to decrease with increasing $v$ with a major decrease between $v = 9$ and
10, which is obviously related to the exothermicity limit of
Reaction~(\ref{eq:H+O3}) of about 27,300\,\unit{cm^{-1}}. The structure of the
population pattern is well known
\citep[e.g.,][]{pendleton93,cosby07,oliva15,noll15,noll20}. For its
characterisation, it can be exploited that the $v$-specific logarithmic
populations for low and high $N$ are located along relatively straight lines.
The slopes $\frac{\mathrm{d}y}{\mathrm{d}E}$ of these lines can be converted
into pseudo-temperatures $T$ by
\begin{equation}\label{eq:Trot}
  T = - \frac{1}{k_\mathrm{B} \frac{\mathrm{d}y}{\mathrm{d}E}} 
\end{equation}
\citep{mies74,noll18b}, where $k_\mathrm{B}$ is the Boltzmann constant. The
steep population decrease for low $N$ is therefore related to low $T$, which
should be close to the ambient temperature in the emission layer, whereas
high-$N$ populations are characterised by high pseudo-temperatures, which
reflect a lack of thermalisation of the nascent population distribution from
Reaction~(\ref{eq:H+O3}) by collisional processes
\citep[e.g.,][]{pendleton89,pendleton93,dodd94,cosby07,kalogerakis18,noll18b}.
This bimodal distribution for fixed $v$ can be best fitted by a
two-temperature model \citep{oliva15,kalogerakis18,kalogerakis19a,noll20}. We
therefore fitted the populations for $v$ from 2 to 9 using
\begin{equation}\label{eq:2Tfit}
  n(E) = n_{0,\mathrm{cold}} e^{-(E-E_0)/(k_\mathrm{B}T_\mathrm{cold})} +
         n_{0,\mathrm{hot}} e^{-(E-E_0)/(k_\mathrm{B}T_\mathrm{hot})},
\end{equation}
where $n_0$ refers to the population at the lowest energy $E_0$ of a given
$v$.

Fits with unconstrained parameters were carried out for $v$ of 4 to 9. The
results agree within the uncertainties with the results of \citet{noll20}.
Hence, the inclusion of the \mbox{X-shooter} data and the slightly modified
$A$ coefficients for the UVES data only had a minor impact on the best fit
parameters. However, the increase of the sample size reduced the
uncertainties. The fits showed very similar values for $T_\mathrm{cold}$ with
a mean of $191.6 \pm 0.7$\,\unit{K} for $v$ between 5 and 9 and a slight
outlier of about 196\,\unit{K} for $v = 4$. This suggests to fix
$T_\mathrm{cold}$ and set it to the ambient temperature. SABER-based
temperature profiles for the region around Cerro Paranal indicate that the
mean profile is fairly constant in most of the altitude range relevant for
\chem{OH} with changes of the order of only 1\,\unit{K} \citep{noll16}. The
SABER data set from 2002 to 2015 collected by \citet{noll17} indicates a mean
temperature of about 191\,K consistent with the population fits. Hence, we
repeated the fits with this value as $T_\mathrm{cold}$. The resulting fit
parameters (with $y_0 = \ln(n_0)$) are shown in Table~\ref{tab:popfits}. With
increasing $v$, $y_{0,\mathrm{cold}}$ and $T_\mathrm{hot}$ decrease (from about
6,300 to 800\,\unit{K} for the latter), whereas there is no clear trend for
$y_{0,\mathrm{hot}}$. For $v$ of 2 and 3, we needed to set additional constraints
as the hot populations are not sufficiently covered. Fortunately,
\citet{oliva15} were able to perform two-component fits for these low $v$,
based on line measurements between 0.95 and 2.4\,\unit{\mu{}m} in a
high-resolution spectrum of the GIANO echelle spectrograph at the island of La
Palma (Spain). For $v = 2$, we directly took their results for the hot
population, i.e. $T_\mathrm{hot} = 12,000$\,\unit{K} and a ratio of
$n_{0,\mathrm{hot}}$ and $n_{0,\mathrm{cold}}$ of 0.14\,\%. For $v = 3$ with
at least some data points in the transition region between cold and hot
population, we used the values of \citet{oliva15} (7,000\,\unit{K} and
0.23\,\%) as a starting point for our own fits, which then showed the most
convincing results (also with respect to the higher $v$) for fixed values of
7,500\,\unit{K} and 0.22\,\%. The changes should not be larger than the fit
uncertainties for the GIANO data \citep[see][]{kalogerakis18}.

Close to the exothermicity limit of about 27,300\,\unit{cm^{-1}}, the
two-component fits do not work as the populations decrease faster than given
by $T_\mathrm{hot}$. For this reason, the fits of $v = 8$ and 9 in
Table~\ref{tab:popfits} were performed with upper energy limits of
26,300\,\unit{cm^{-1}} for $v = 8$ ($N \leq 13$) and 27,330\,\unit{cm^{-1}} for
$v = 9$ ($N \leq 9$). For the populations above these limits, fits with a
single temperature turned out to be promising. In a plot of $v$ vs. $E$ as
Fig.~\ref{fig:popmodel_OH}, this is just a linear regression. We performed the
fits including the data points with the highest $N$ that were used for the
two-component fits, i.e. 13 and 9, respectively. However, we only applied the
results for levels with higher $N$. In Table~\ref{tab:popfits}, the listed
$E_0$ refer to the intersection of the one- and two-component fits. The
resulting temperatures are only about half the $T_\mathrm{hot}$ from the
two-component model but still distinctly higher than $T_\mathrm{cold}$. The
small sample of $v = 10$ population measurements was also fitted with a
one-component model. The resulting temperature is about 300\,\unit{K}. It
might be somewhat lower than the corresponding fit for $v = 9$. The
$y_{\,0}$ are very similar for both $v$.

The complete \chem{OH} population model is shown in
Fig.~\ref{fig:popmodel_OH}. For very high $N$ not covered by data, we
introduced the rule that curves for different $v$ never cross. Instead, the
curve of the lower $v$ then follows the path of the next higher one. This
recipe considers that the measured high-$N$ populations of the different $v$
tend to converge. This interpretation is also supported by similar effective
emission heights \citep{noll22} and variability patterns \citep{noll23} of
high-$N^{\prime}$ lines. The population model shows good agreement with the
measured populations. The mean population ratio is very close to 1.00 and the
standard deviation is 0.10 for the \mbox{X-shooter} and 0.21 for the
\citet{noll20} UVES data. In the case of levels with $N \leq 4$ neglecting
Q-branch lines, the scatter is only 0.05 for both data sets. The main reason
for the deviations should be measurement uncertainties, but higher-order
population features not covered by the model may also contribute. For the
PALACE line list, the \chem{OH} population model was applied to 11,029
$\Lambda$ doublets from HITRAN2020. Using the corrected $A$ coefficients, the
resulting total reference intensity (without atmospheric extinction) amounts
to 715\,\unit{kR} in the full model range from 0.3 to 2.5\,\unit{\mu{}m}. The
strongest band is \mbox{(4-2)} at about 1.6\,\unit{\mu{}m} with 92\,\unit{kR}
for all band-related lines. For $\Delta v = 3$, \mbox{(8-5)} at about
1.3\,\unit{\mu{}m} indicates the highest reference intensity of 24\,\unit{kR}.
For $\Delta v$ from 4 to 7, where also measured data exist, the bands with
$v^{\prime} = 9$ are the most intense ones showing 3.8\,\unit{kR},
570\,\unit{R}, 61\,\unit{R}, and 5.1\,\unit{R} for $v^{\prime\prime}$ from 5 to
2. The weakest band of this list, i.e. \mbox{(9-2)}, emits near
520\,\unit{nm}. The total reference emission of all $v^{\prime} = 10$ bands
amounts to about 330\,\unit{R}. The integrated \chem{OH} emission between 642
and 858\,\unit{nm} is about 3.1\,\unit{kR}, which is very close to
3.2\,\unit{kR} as reported by \citet{noll12} for the ESO sky model, although
for a solar radio flux of about 129\,\unit{sfu}.

So far, the discussion has focused on the main isotopologue \chem{^{16}OH}.
However, other isotopologues can also emit. \citet{cosby06} succeeded to
identify 26 $\Lambda$ doublets of \chem{^{18}OH} (covering only one component
in five cases). The lines were very faint with intensities below 1\,\unit{R}
and a total emission of 8.4\,\unit{R}. Compared to the model intensities of
the same lines for \chem{^{16}OH}, we found a mean ratio of
$0.00241 \pm 0.00018$ for the 15 most reliable $\Lambda$ doublets. This result
is consistent with $0.00235 \pm 0.00007$ obtained by \citet{osterbrock98}
based on a different data set. The ratio is higher than the standard abundance
ratio of 0.00201 from the HITRAN database. Deviations can be related to
differences in the energy levels and $A$ coefficients. The discrepancy may
also be caused by an enrichment of heavier \chem{O} isotopes in \chem{O_3}
\citep{osterbrock98}. As HITRAN does not contain \chem{^{18}OH} lines in the
PALACE wavelength range, we only included the lines identified by
\citet{cosby06} in our line model. This is certainly highly incomplete.
However, the small number of detected lines also shows that the contributions
of heavier \chem{OH} isotopologues to the airglow emission spectrum are
mostly negligible. For the model, we took the measured intensities (with the
general corrections discussed before) and the modelled wavelengths from
\citet{cosby06}. For \chem{^{16}OH}, the latter are in good agreement with the
wavelengths from the HITRAN database. The relative deviations are only of the
order of $10^{-7}$.

\subsection{Molecular oxygen}
\label{sec:O2}

Excited \chem{O_2} has a significant impact on the shape of the nocturnal
airglow (nightglow) spectrum (see Figs.~\ref{fig:refspec} and
\ref{fig:reflineint}). In contrast to \chem{OH} where the emission is only
related to ro-vibrational bands of the electronic ground state X$^2\Pi$, the
relevant \chem{O_2} bands always involve electronic transitions including the
states X$^3\Sigma^-_\mathrm{g}$, a$^1\Delta_\mathrm{g}$,
b$^1\Sigma^+_\mathrm{g}$, c$^1\Sigma^-_\mathrm{u}$,
A$^{\prime\,3}\Delta_\mathrm{u}$, and A$^3\Sigma^+_\mathrm{u}$, which are
listed with increasing energy. Relevant band systems are \mbox{(A-X)},
\mbox{(A$^{\prime}$-a)}, \mbox{(c-b)}, \mbox{(c-X)}, \mbox{(b-X)}, and
\mbox{(a-X)}
\citep[e.g.,][]{rousselot00,slanger03,cosby06,noll16,vonsavigny17}. Systems
related to the higher electronic levels cause relatively weak bands in the
near-UV to visual wavelength regime. The \mbox{(b-X)(0-0)} band at
762\,\unit{nm} is very strong but not visible from the ground as the lower
level is X$^3\Sigma^-_\mathrm{g}$($v = 0$), which is strongly affected by
atmospheric \chem{O_2} absorption (see Fig.~\ref{fig:refspec}). The
\mbox{(b-X)(0-1)} band at 865\,\unit{nm} is significantly fainter but not
affected by self-absorption. There are other but weak bands of this system
with $v^{\prime} \leq 15$. The strongest nightglow band at all is
\mbox{(a-X)(0-0)} at 1.27\,\unit{\mu{}m}. Although also strongly affected by
self-absorption, it remains a relatively bright band for observations at the
ground. It is also clearly brighter than \mbox{(a-X)(0-1)} at
1.58\,\unit{\mu{}m}, the only other known band of the system.

The classical nighttime production mechanism for excited \chem{O_2} is atomic
oxygen recombination 
\begin{reaction}\label{eq:O+O+M}
  \mathrm{O} + \mathrm{O} + \mathrm{M} \rightarrow \mathrm{O}_2^{\ast} +
  \mathrm{M}
\end{reaction}
\citep[e.g.,][]{slanger03}, where \chem{M} can be any atmospheric
constituent. The reaction preferentially produces populations in the higher
electronic states. Then, the lower states are essentially populated by
level-changing collisions of the excited molecules. However, there appear to
be other important processes. \citet{kalogerakis19b} proposes that
b$^1\Sigma^+_\mathrm{g}$ could be excited up to $v = 1$ by collisions with
\chem{O} in the excited $^1$D state, which could be produced by collisions of
\chem{O} in the ground state $^3$P with excited \chem{OH}. Concerning
a$^1\Delta_\mathrm{g}$, \citet{noll24} discussed pathways such as reactions of
\chem{OH} and \chem{O} or reactions involving \chem{HO_2}. Nevertheless, there
is no satisfying explanation, so far. In any case, it needs to be explained
why the nighttime \mbox{(a-X)} emission peaks several kilometres below the
emissions of the other band systems. For \mbox{(b-X)(0-0)}, 94\,\unit{km} is
a typical peak height \citep[e.g.,][]{yee97}, whereas it is about
90\,\unit{km} for \mbox{(a-X)(0-0)} as measured at Cerro Paranal using SABER
profiles from the second half of the night \citep{noll16}. At the beginning of
the night, most emission originates from significantly lower heights with
contributions even from the lower mesosphere. This change in the emission
profile is caused by populations produced by \chem{O_3} photolysis at daytime,
which can produce excited \chem{O_2}(a$^1\Delta_\mathrm{g}$). As this state has
a long effective lifetime of about 1~hour in the mesopause region, this
pathway still matters in the first few hours of the night indicated by an
exponential intensity decrease \citep{lopez89,mulligan95,noll16}. For the van
Rhijn correction, we chose 89\,\unit{km} as a representative centroid height
for the major fraction of the night (see Table~\ref{tab:species}).

\begin{figure*}[t]
\includegraphics[width=17cm]{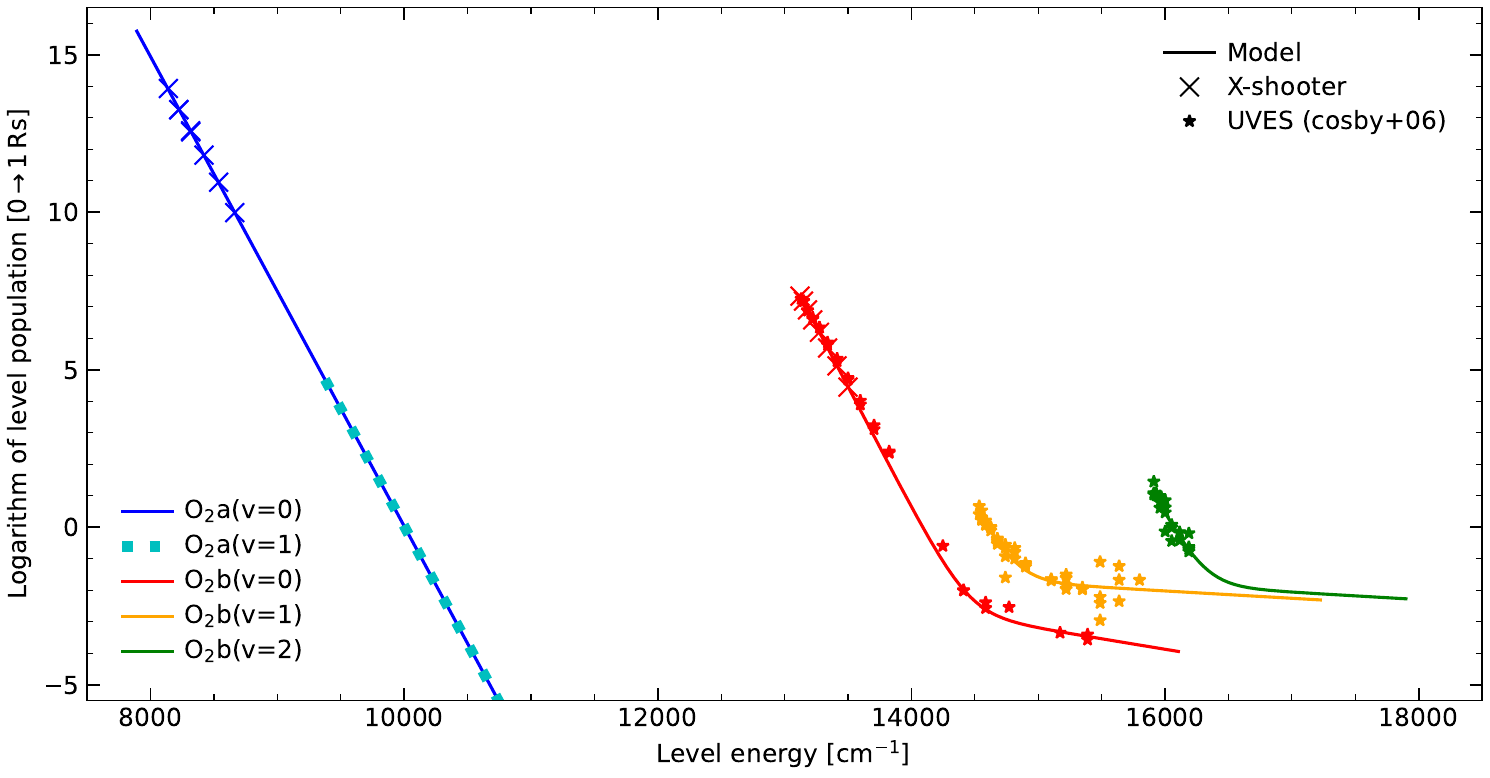}
\caption{\chem{O_2} population model for the electronic states
  $\mathrm{a}^1\Delta_{\mathrm{g}}$ for vibrational levels $v \leq 1$ and
  $\mathrm{b}^1\Sigma_{\mathrm{g}}^{+}$ for $v \leq 2$ in logarithmic units and
  depending on the level energy. The model (solid or dotted lines) consists of
  a linear fit for both $v$ of $\mathrm{a}^1\Delta_{\mathrm{g}}$ or two fitted
  components for each $v$ of $\mathrm{b}^1\Sigma_{\mathrm{g}}^{+}$. The
  corresponding parameters are provided in Table~\ref{tab:popfits}. The basic
  data for the \chem{O_2} population model originate from \mbox{X-shooter}
  measurements (crosses) and UVES-based data from \citet{cosby06} (stars).}
\label{fig:popmodel_O2}
\end{figure*}

In order to model the mean a$^1\Delta_\mathrm{g}$($v = 0$) population, we
focused on the strong \mbox{(a-X)(0-0)} band. As this band consisting of nine
branches \citep[e.g.,][]{rousselot00} shows high line density especially
in the band centre and is strongly affected by self-absorption, only a small
number of lines is suitable for measurements at \mbox{X-shooter} resolution.
With a small data set of \mbox{X-shooter} data, such measurements were already
performed by \citet{noll16}. We carried out a similar line selection only
considering lines with relatively high $N^{\prime}$ of the branches
$^\mathrm{S}\mathrm{R}$ and $^\mathrm{O}\mathrm{P}$, which are related to an
$N$ change of $-2$ and $+2$ (denoted by the letters S and O), respectively,
and an opposite spin change by 1 leading to $-1$ (R) and $+1$ (P) in terms of
the total angular momentum $J$. Our final sample consists of the four
$^\mathrm{S}\mathrm{R}$ lines with $N^{\prime}$ of 15, 17, 21, and 23 at
wavelengths between 1.249 and 1.256\,\unit{\mu{}m} and the four
$^\mathrm{O}\mathrm{P}$ lines with $N^{\prime}$ of 13, 15, 17, and 19 at
wavelengths between 1.282 and 1.289\,\unit{\mu{}m} (see
Fig.~\ref{fig:climlines}). These lines are relatively far from the crowded
band centre and only experience moderate self-absorption with zenith
transmissions between 0.22 and 0.87, which can be corrected in a reliable way.
The reference intensities from the derived climatologies were then converted
into populations using Eq.~(\ref{eq:levpop}) and \chem{O_2} $A$ coefficients
from HITRAN2020 \citep{gordon22}. The latter are more robust than in the case
of \chem{OH} and do not need to be optimised. We fitted the data with a single
temperature (see Fig.~\ref{fig:popmodel_O2}) and obtained
$193.1 \pm 1.4$\,\unit{K} as shown by Table~\ref{tab:popfits}. This is very
close to the expected ambient temperature and therefore reliable. In the view
of the mentioned long lifetime of a$^1\Delta_\mathrm{g}$($v = 0$), the
rotational population should be well thermalised. Hence, the resulting
$T_\mathrm{cold}$ should be representative for all rotational levels. We
therefore applied the fit results to all levels in order to calculate
reference intensities for the \mbox{(0-0)} band with a total strength of
155\,\unit{kR} without self-absorption (otherwise only about 17\,\unit{kR})
and the much weaker \mbox{(0-1)} band with 2.0\,\unit{kR}. 

The HITRAN \chem{O_2} line list also includes the \mbox{(a-X)(1-0)} band near
1.07\,\unit{\mu{}m}. \citet{noll24} speculated whether a feature in the
residual airglow continuum (see Sect.~\ref{sec:continuum}) could be related
to this band. We tested this by calculating the \mbox{(a-X)} model with
different factors for the ratio of the $v = 1$ and $v = 0$ populations of
a$^1\Delta_\mathrm{g}$ and comparing the resulting spectrum to an
\mbox{X-shooter} NIR-arm mean spectrum. However, we found that the calculated
and measured structures do not match suggesting only a very small contribution
of \mbox{(a-X)(1-0)}. We therefore assumed that the vibrational temperature
equals the rotational temperature (see Fig.~\ref{fig:popmodel_O2} and
Table~\ref{tab:popfits}), which resulted in a band intensity of only
0.012\,\unit{R}. The calculated strength of the also available \mbox{(1-1)}
band of 2.6\,\unit{R} is much higher, but the band is located close to the
strong \mbox{(0-0)} emission. For the latter, the HITRAN database also
includes 322 lines of the \chem{^{16}O^{18}O} isotopologue. For the scaling of
these lines compared to those of \chem{^{16}O_2}, we applied the standard
abundance ratio of 0.0040 from HITRAN, which resulted in a band strength of
1.3\,\unit{kR}. The whole \mbox{(a-X)} model comprises 1,196 lines.

The only sufficiently strong \mbox{(b-X)} band that can be measured with
\mbox{X-shooter} is \mbox{(0-1)} at about 865\,\unit{nm}. It is not affected
by self-absorption. In agreement with the \mbox{X-shooter}-based measurements
of \citet{noll16}, we focused on the P branch, where the lines are stronger
and better separated than in the R branch. Nevertheless, the two subbranches
$^\mathrm{P}\mathrm{Q}$ and $^\mathrm{P}\mathrm{P}$, which are related to the
multiplicity of the electronic ground state X$^3\Sigma^-_\mathrm{g}$, are not
separated for the same $N^{\prime}$. Only in the case of the minimum
$N^{\prime}$ of 0, it is a single line as the $^\mathrm{P}\mathrm{Q}$ component
is forbidden. We selected this single line and seven doublets with
$N^{\prime}$ from 4 to 16 (odd numbers are not allowed) to study the population
of b$^1\Sigma^+_\mathrm{g}$($v = 0$). A linear fit of the logarithmic
populations calculated by means of the climatological mean intensities and the
corresponding HITRAN $A$ coefficients returned $187.3 \pm 1.2$\,\unit{K}
as convincing single temperature (see Table~\ref{tab:popfits} and
Fig.~\ref{fig:popmodel_O2}). As the \mbox{(b-X)(0-1)} emission heights are
closer to the temperature minimum of the mesopause, the resulting
$T_\mathrm{cold}$ should be slightly lower than in the case of
\mbox{(a-X)(0-0)} and the different \chem{OH} bands \citep{noll16}. 

The fit model works very well for the covered range of $N^{\prime}$. However,
there might be discrepancies for distinctly higher rotational levels. The
lifetime of b$^1\Sigma^+_\mathrm{g}$($v = 0$) is only about 11\,\unit{s}
\citep[e.g.,][]{zhu19} (compared to more than an hour for
a$^1\Delta_\mathrm{g}$) and the chemistry also seems to differ as discussed
above. Fortunately, we could investigate this by using the UVES-based line
catalogue of \citet{cosby06}, where we selected 21 resolved \mbox{(b-X)(0-1)}
P-branch lines with $N^{\prime}$ between 2 and 22 (0 had calibration issues)
and nine \mbox{(b-X)(0-0)} P-branch lines with $N^{\prime}$ between 28 and 40.
The latter were measurable as self-absorption becomes weaker with increasing
$N^{\prime}$. The reference transmission values were between 0.15 and 1.00 for
the selected lines. Before the study of the populations, we had to scale the
intensities to be consistent with the \mbox{X-shooter} data. For this, we
compared \mbox{(b-X)(0-1)} lines which are present in both data sets. The
logarithmic line ratios were linearly fitted depending on $N^{\prime}$ in order
to obtain a robust intensity ratio for $N^{\prime} = 0$. The result was
$1.55 \pm 0.03$, i.e. the \citet{cosby06} data were significantly brighter.
The factor is most probably caused by the high solar activity for the UVES
data. In June 2001, the monthly mean was 174\,\unit{sfu}. As discussed in
Sect.~\ref{sec:comparison}, \chem{O_2} lines show a relatively strong
correlation with the solar radio flux.

After the correction of the lines intensities, which we applied to all
\chem{O_2} lines of \citet{cosby06}, we calculated the level populations using
the HITRAN database and then fitted the population distribution using the
two-component model in Eq.~(\ref{eq:2Tfit}). As indicated by
Fig.~\ref{fig:popmodel_O2} and Table~\ref{tab:popfits}, the fit revealed a
weak hot population with roughly $2,200 \pm 800$\,\unit{K} that is only
visible in the highest $N$. Moreover, $T_\mathrm{cold}$ resulted in
$195.0 \pm 2.9$\,\unit{K}, which is distinctly higher than in the case of the
\mbox{X-shooter} data. The most likely explanation is also the high solar
activity during the UVES observations. As the reference intensities should be
valid for a solar radio flux of 100\,\unit{sfu}, we constructed our final
b$^1\Sigma^+_\mathrm{g}$($v = 0$) model by using the \mbox{X-shooter}-based fit
for the cold population and only the hot population from the UVES-based fit.
For this approach, it was important to have a scaling factor for the
UVES-based intensities optimised for $N^{\prime} = 0$ as described above. The
resulting reference intensity for the whole \mbox{(b-X)(0-0)} band (150 lines)
is 6.5\,\unit{kR} without self-absorption, which is in good agreement with
satellite-based measurements \citep[e.g.,][]{yee97}. The corresponding
intensity for \mbox{(b-X)(0-1)} amounts to 340\,\unit{R}. The 49 lines for
this band include two $N^{\prime} = 24$ lines, where the intensities (about
0.5\,\unit{R} in sum) were directly taken from the corrected \citet{cosby06}
data set since these lines are not included in HITRAN2020. The derived
reference band intensity is very close to the one for the ESO Sky Model of
320\,\unit{R} \citep{noll12}. However, the latter would be only about
275\,\unit{R} for a solar radio flux of 100 instead of 129\,\unit{sfu}.

\citet{slanger00b} discussed the detection of various \mbox{(b-X)} bands
with $v^{\prime}$ between 1 and 15 in spectra of the High Resolution Echelle
Spectrometer (HIRES) from Mauna Kea (Hawai'i). We could also extract more than
2,000 of such lines from the line list of \citet{cosby06}. The HITRAN database
only covers a small number of these bands, essentially only those that matter
for absorption calculations such as \mbox{(1-0)}, \mbox{(1-1)}, \mbox{(2-0)},
and \mbox{(2-1)}. Hence, our approach was to directly use the calculated
wavelengths (with a tested systematic uncertainty significantly below
$10^{-6}$) and the scaled intensities from \citet{cosby06} as primary input
for the model. However, the latter can have relatively large measurement
uncertainties, especially for the relatively faint $v^{\prime} \geq 1$ lines.
Moreover, the catalogue is incomplete due to line blending and detection
limits, and there is a possible issue with respect to the high solar radio
flux during the UVES observations as discussed before. Therefore, we performed
population fits for lines with UVES and HITRAN data in order to improve the
model, although the covered $v^{\prime} = 1$ and 2 bands appear to be
relatively faint compared to the integrated emission for higher $v^{\prime}$
such as 3, 4, 5, and even 12 \citep{slanger00b}.

Using 44 \mbox{(1-1)} lines, a two-component fit (see
Fig.~\ref{fig:popmodel_O2}) with $T_\mathrm{cold} = 195.0$\,\unit{K} from the
\mbox{(0-1)} fit resulted in a hot population with an uncertain temperature
of about 6,000\,\unit{K} (see Table~\ref{tab:popfits}). For the fit of 24
\mbox{(2-1)} lines, $T_\mathrm{hot}$ was also fixed (using the previous result)
as the measured lines do not cover the hot population
(Fig.~\ref{fig:popmodel_O2}). In a similar way as for the
b$^1\Sigma^+_\mathrm{g}$($v = 0$) model, we then replaced $T_\mathrm{cold}$ in
both $v^{\prime}$ cases by 187.3\,\unit{K} from the \mbox{X-shooter}-based fit
of the \mbox{(0-1)} band. The resulting summed reference intensity for the
272 lines from the HITRAN database with $v^{\prime} = 1$ and 2 and
$v^{\prime\prime} \leq 1$ amounts to 14\,\unit{R}. For the remaining
$v^{\prime} \geq 1$ bands, the integrated intensity from 460 measured lines
results in 83\,\unit{R}. With additional three weak lines from the
\mbox{(0-2)} band, the total line number of the \mbox{(b-X)} model for
\chem{^{16}O_2} is 934. These lines emit 6.9\,\unit{kR} under reference
conditions.

HITRAN2020 also includes lines related to b$^1\Sigma^+_\mathrm{g}$($v = 0$)
for the \chem{^{16}O^{18}O} isotopologue. As the UVES-based data set also
contains such measurements, we could scale the population model for $v = 0$
by means of a comparison of level populations from the measured intensities
of \mbox{(0-0)} for \chem{^{16}O^{18}O} and \mbox{(0-1)} for \chem{^{16}O_2}.
The resulting reference intensity model for the former comprises 140 lines
with a summed emission of 20\,\unit{R}, which corresponds to 0.0031 times the
radiation from the \mbox{(0-0)} band of \chem{^{16}O_2}. This is lower than
the standard abundance ratio of 0.0040 but consistent with
\citet{slanger00b} based on the HIRES data set. HITRAN also contains the
\mbox{(1-0)} and \mbox{(2-0)} bands for \chem{^{16}O^{18}O}, which could also
be modelled using the corresponding population fits. However, the 216 very
weak lines have a negligible impact on the total emission for
\chem{^{16}O^{18}O}. For the \mbox{(b-X)} model, we also considered 13
detected lines of the \mbox{(0-0)} band of \chem{^{16}O^{17}O} from the
\citet{cosby06} data set without population calculations. The total emission
(which represents only a minor fraction of the lines of this band) is just
1.3\,\unit{R}.

Lines of the high-energy c$^1\Sigma^-_\mathrm{u}$,
A$^{\prime\,3}\Delta_\mathrm{u}$, and A$^3\Sigma^+_\mathrm{u}$ states are not part
of HITRAN2020 in the PALACE model range. Therefore, we directly used 1,590
corresponding lines in the wavelength range from 314 to 550\,\unit{nm}
identified by \citet{cosby06}. We also applied the intensity scaling factor
of $1 / 1.55$ as derived for \mbox{(b-X)(0-1)} and $N^{\prime} = 0$. However,
the difference in the states and the UVES set-ups
\citep[mainly 346 and 437 instead of 860;][]{hanuschik03} could cause a
different factor. Therefore, we compared the integrated intensities of
several clear emission features in the range from 330 to 400\,\unit{nm} (see
Fig.~\ref{fig:refspec}) as derived from a preliminary model and an
\mbox{X-shooter} UVB-arm mean spectrum, which revealed that our model was
still too bright by a factor of about $1.1 \pm 0.1$. We therefore divided the
intensities by 1.1 to obtain our final model for the high-energy states. The
total reference emission amounts to about 62\,\unit{R}. The contributions of
the \mbox{(A-X)}, \mbox{(A$^{\prime}$-a)}, \mbox{(c-b)}, and \mbox{(c-X)} band
systems are 35.7, 22.6, 3.4, and 0.3\,\unit{R}, respectively. As the bands of
the different systems strongly overlap and the individual lines are very weak
($< 0.3$\,\unit{R}), the line list is highly incomplete. The lack of coverage
is particularly severe for intensities below about 0.01\,\unit{R} as
Fig.~\ref{fig:reflineint} demonstrates. Therefore, the PALACE model of the
\chem{O_2} emission related to the c$^1\Sigma^-_\mathrm{u}$,
A$^{\prime\,3}\Delta_\mathrm{u}$, and A$^3\Sigma^+_\mathrm{u}$ states also
consists of an unresolved continuum component, which is discussed in 
Sect.~\ref{sec:O2cont}.

\subsection{Sodium}
\label{sec:Na}

The alkali metal \chem{Na} produces a prominent doublet at air wavelengths of
589.0 and 589.6\,\unit{nm} (see Fig.~\ref{fig:refspec}). Accurate wavelengths
can be taken from the NIST atomic line database \citep{NIST_ASD}. These lines
are usually called D$_2$ and D$_1$ and are related to the transitions from the
excited $^2\mathrm{P}_{3/2}$ and $^2\mathrm{P}_{1/2}$ levels to the ground state
$^2\mathrm{S}_{1/2}$. The excitation mechanism is a multi-step process starting
with
\begin{reaction}\label{eq:Na+O3}
  \mathrm{Na} + \mathrm{O}_3 \rightarrow \mathrm{NaO}^{\ast} + \mathrm{O}_2
\end{reaction}
and ending with
\begin{reaction}\label{eq:NaO+O}
  \mathrm{NaO}^{(\ast)} + \mathrm{O} \rightarrow \mathrm{Na}^{\ast} +
  \mathrm{O}_2.
\end{reaction}
The version of Reaction~(\ref{eq:NaO+O}) with \chem{NaO} in excitation (marked
by an asterisk) is the classical mechanism of \citet{chapman39}. In addition,
laboratory measurements by \citet{slanger05} indicated that \chem{NaO^{\ast}}
from Reaction~(\ref{eq:Na+O3}) can be first deactivated by collisions, most
likely with \chem{O_2}, before Reaction~(\ref{eq:NaO+O}) starts. For both
processes, it is also possible that \chem{Na} is produced in the ground state,
which does not generate airglow emission. For the effective emission height,
we take 92\,\unit{km}, which agrees well with satellite-based measurements
\citep[e.g.,][]{koch22} and is consistent with \citet{noll12}.

Both lines can be well measured in \mbox{X-shooter} VIS-arm spectra. However,
the D$_2$ line is also used at Cerro Paranal for the production of artifical
guide stars via laser-induced fluorescence in the \chem{Na} layer. The amount
of contamination depends on the angular distance of the line of sights of the
two telescopes operating \mbox{X-shooter} and the laser if it is used. Hence,
we had to perform a very careful outlier detection in terms of the D$_2$
intensity and the D$_2$-to-D$_1$ ratio. Strong outliers are easily found but
the identification of small contributions remains uncertain. Due to the
different pathways for the \chem{Na} excitation, the D$_2$-to-D$_1$ ratio is
variable \citep{slanger05}. For a subsample consisting of the most reliable
measurements, we found a mean ratio of 1.73. This is well in the range of
values reported by \citet{slanger05} based on spectra from Mauna Kea. For the
line list of \citet{cosby06}, about 1.70 was obtained. For the calculation of
the D$_1$ climatology, we used a 4 times larger sample (27,942 vs. 6,999). As
the same for the D$_2$ line would have caused contaminations by the laser, we
derived the reference intensity for D$_2$ by just multiplying the result of
13.4\,\unit{R} for D$_1$ by the factor 1.73, which returned 23.1\,\unit{R}.
Both intensities are plotted in Figs.~\ref{fig:reflineint} and
\ref{fig:climlines}. The resulting total \chem{Na} emission amounts to
36.5\,\unit{R}. This is slightly lower than the corresponding value of
43\,\unit{R} in the ESO Sky Model. \citet{unterguggenberger17} obtained from
a sample of 3,662 \mbox{X-shooter} spectra (only covering the period up to
March 2013) about 40\,\unit{R}.

\subsection{Potassium}
\label{sec:K}

The airglow emission of the alkali metal \chem{K} is also characterised by a D
doublet with the same upper and lower states as in the case of \chem{Na}
(Sect.~\ref{sec:Na}). The D$_2$ and D$_1$ lines at air wavelengths of 766.5
and 769.9\,\unit{nm} \citep{NIST_ASD} are also produced by a similar
main excitation mechanism \citep{swider87,noll19}. In
Reactions~(\ref{eq:Na+O3}) and (\ref{eq:NaO+O}), \chem{Na} only needs to be
replaced by \chem{K}. The reference layer height of 89\,\unit{km} in
Table~\ref{tab:species} was taken from emission simulations by \citet{noll19}.

The D$_2$ line is strongly affected by \chem{O_2} absorption. The reference
model transmission is just 0.0026. Satellite-based measurements are not
possible as well due to the very strong airglow emission of the
\mbox{(b-X)(0-0)} band (Sect.~\ref{sec:O2}). Hence, only D$_1$ can be
measured. As this line is weak and located in a region crowded by \chem{OH}
and \chem{O_2} emission, the spectral resolving power of \mbox{X-shooter} is
not sufficient and we therefore used 2,299 UVES-based line measurements from
\citet{noll19} as already mentioned in Sect.~\ref{sec:dataset} and the
beginning of Sect.~\ref{sec:lines}. The reference intensity from the resulting
climatology is 0.95\,\unit{R} (see Figs.~\ref{fig:reflineint} and
\ref{fig:climlines}) in agreement with the result of \citet{noll19}
and also very close to the first published intensity measurement by
\citet{slanger00a} of 1.0\,\unit{R} from a HIRES composite spectrum from Mauna
Kea. As D$_2$ cannot be measured, we used the theoretical D$_2$-to-D$_1$ ratio
of 1.67 from \citet{noll19} to obtain a reference intensity of 1.58\,\unit{R}.
Hence, the total emission of the \chem{K} model is 2.53\,\unit{R}.

\subsection{Atomic oxygen}
\label{sec:O}

There are various \chem{O} lines that contribute to the nightglow emission
spectrum. The most prominent line is certainly the green line at an air
wavelength of 557.7\,\unit{nm} (see Fig.~\ref{fig:refspec}), which is related
to the transition between the two excited states $^1\mathrm{S}$ and
$^1\mathrm{D}$ \citep[see][]{NIST_ASD}. The established production mechanism
of $^1\mathrm{S}$ in the mesopause region \citep{barth61} consists of the
atomic oxygen recombination in Reaction~(\ref{eq:O+O+M}) (Sect.~\ref{sec:O2})
producing excited \chem{O_2} and
\begin{reaction}\label{eq:O2+O}
\mathrm{O}_2^{\ast} + \mathrm{O} \rightarrow \mathrm{O}_2 + \mathrm{O}^{\ast}.
\end{reaction}
The green line emission peaks relatively high compared to the previously
discussed emissions \citep[e.g.,][]{yee97,vonsavigny13}. As reference height,
we use 97\,\unit{km} in agreement with the ESO Sky Model \citep{noll12}.

The 557.7\,\unit{nm} line is located in the overlapping region of the
\mbox{X-shooter} UVB and VIS arms, which receive light in their wavelength
ranges by means of dichroics \citep{vernet11}. As the beam splitting shows
some variability, there are relatively high flux calibration uncertainties for
wavelengths around the green line. On the other hand, it was possible to
measure intensity time series in both arms. The resulting climatology-based
reference intensities are 156\,\unit{R} for the UVB arm and 170\,\unit{R} for
the VIS arm, which reflects the systematic uncertainties. Assuming that the
most reliable value is close to the middle, we averaged the climatologies of
both data sets with equal weight, which returned a reference intensity of
163\,\unit{R} (Figs.~\ref{fig:reflineint} and \ref{fig:climlines}). This is
similar to a value of 153\,\unit{R} that we obtained for the ESO Sky Model
with the given solar activity correction for a change from 129 to
100\,\unit{sfu} and a reference intensity of 190\,\unit{R} \citep{noll12}.

Other prominent \chem{O} emissions are the red lines at air wavelengths of
630.0 and 636.4\,\unit{nm}, which are related to the transition from the
lowest excited state $^1\mathrm{D}$ to the ground state $^3\mathrm{P}$
\citep{NIST_ASD}. These lines are negligible in the mesopause region due to
the long radiative lifetime of about 2\,min compared to the frequency of
relevant collisions. However, they are the most important emission lines in
the PALACE wavelength range that originate (at a distinctly lower air density)
in the upper thermosphere. The major fraction of the emission is usually
emitted between 200 and 300\,\unit{km} and typical peak heights are around
250\,\unit{km} \citep[e.g.,][]{adachi10,haider22}, but there are significant
variations. At nighttime, the main excitation mechanism
\citep[e.g.,][]{link88} is triggered by atomic oxygen ions (\chem{O^+})
produced at daytime. Then, the charge transfer reaction
\begin{reaction}\label{eq:O++O2}
\mathrm{O}^+ + \mathrm{O}_2 \rightarrow \mathrm{O}_2^+ + \mathrm{O}.
\end{reaction}
and dissociative recombination with electrons (\chem{e^-})
\begin{reaction}\label{eq:O2++e-}
\mathrm{O}_2^+ + \mathrm{e}^- \rightarrow \mathrm{O}^{\ast} + \mathrm{O}
\end{reaction}
can happen, which produces \chem{O} atoms in the required $^1\mathrm{D}$
state. In principle, $^1\mathrm{S}$ populations and 557.7\,\unit{nm} emission
can also be produced in this way. However, this ionospheric radiation is
usually mucher weaker than the contribution from the mesopause region. 

The reference intensities for the 630.0 and 636.4\,\unit{nm} lines derived
from the corresponding climatologies are 123 and 41\,\unit{R}, respectively
(Figs.~\ref{fig:reflineint} and \ref{fig:climlines}). This corresponds to an
intensity ratio of 2.99, which is very close to the ratio of the NIST-based
$A$ coefficients of theses lines of 3.09 \citep{NIST_ASD}. We therefore used
these $A$ coefficients to also estimate the reference intensity of the third
line of the triplet at 639.2\,\unit{nm}, which can only be generated by an
electric quadrupole transition due to the necessary angular momentum change of
2. As a consequence, the resulting intensity is only 0.019\,\unit{R}. The
total emission of the red lines amounts to 164\,\unit{R}, which is in good
agreement with a value of 161\,\unit{R} for 100\,\unit{sfu} derived from the
reference intensity of 190\,\unit{R} and solar activity correction of the ESO
Sky Model \citep{noll12}. With an average nocturnal intensity similar to the
557.7\,\unit{nm} line, the combined red lines are relatively strong at Cerro
Paranal compared to observing sites at northern mid-latitudes
\citep[e.g.,][]{hart19,mackovjak21}. Further details are discussed in
Sect.~\ref{sec:comparison}.

Levels with energies higher than $^1\mathrm{S}$ also contribute to the
nightglow emission. They can emit by radiative \chem{O} recombination, which
involves the reaction
\begin{reaction}\label{eq:O++e-}
\mathrm{O}^+ + \mathrm{e}^- \rightarrow \mathrm{O}^{\ast}
\end{reaction}
and subsequent radiative cascades to lower states \citep{meier91,slanger04c}.
As the emission correlates with the electron density squared, it peaks near
the ionisation maximum in the ionospheric F-layer, which is higher than
the peak of the red line emission that also depends on the \chem{O_2} density.
We assume a rough difference of 50\,\unit{km} \citep[see][]{makela01}, which
results in a reference height of 300\,\unit{km} (Table~\ref{tab:species}).
The relevant emission lines can be grouped depending on the spin multiplicity
of the upper level \citep{meier91,slanger04c}. Quintet states are most
important with noteworthy emissions located at 777, 926, 645, and
616\,\unit{nm} (listed in the order of decreasing intensity) in the PALACE
wavelength regime. Each emission feature consists of several components
(e.g., three for 777\,\unit{nm} and nine for 926\,\unit{nm}). Examples for
emissions related to triplet states are the features located at 845 and
700\,\unit{nm}. Both lines consist of three components.

Using \mbox{X-shooter} spectra, we could measure intensity time series for the
features listed above and derive climatologies and reference intensities.
However, safe line detections were only possible for favourable observing
conditions such as high solar activity (see Sect.~\ref{sec:comparison}) in the
case of the weaker lines. Moreover, the relatively strong 777 and
926\,\unit{nm} emissions are affected by blending with \chem{OH} lines. Hence,
we corrected the intensities by taking measurements of a nearby unblended
\chem{OH} line with a similar upper level as the blending line and using the
\chem{OH} model from Sect.~\ref{sec:OH} for the estimate of a correction. For
example, we subtracted 0.13 times the intensity of
\chem{OH}\mbox{(9-4)P$_1$($N^{\prime} = 2$)} from the 777\,\unit{nm} intensity.

As the \chem{O} recombination line measurements were difficult, we built our
line model only based on the strongest quintet and triplet state features,
i.e. the emissions at 777 and 845\,\unit{nm} with resulting reference
intensities of 16.1 and 6.7\,\unit{R} (see Fig.~\ref{fig:climlines}). For the
calculation of the strength of the other lines from the NIST database
\citep{NIST_ASD} in the PALACE wavelength regime, we used theoretical
radiative recombination coefficients from \citet{escalante92}. In accordance
with \citet{slanger04c}, we selected the data for the optically thin case~A
and a temperature of 1000\,\unit{K} for this task. NIST lines without
coefficients were not considered for our model. The resulting line ratios
appear to be realistic as a comparison of the intensities of the measured
features of the same multiplet group indicated. The quintet and triplet groups
need to be treated separately as the latter is directly connected to the
triplet ground state $^3\mathrm{P}$, which causes an increased optical depth.
For this reason, the ratio of the 845 and 777\,\unit{nm} is about 2.8 times
higher than expected from the recombination coefficients for the optically
thin case. However, it is still lower than a factor of 4.8 for the transition
to the case with infinite optical depth \citep{escalante92}, which is in
qualitative agreement with results of \citet{slanger04c} based on line
measurements in spectra from the Echelette Spectrograph and Imager (ESI) and
HIRES at Mauna Kea. As the optical depth can depend on the specific line of
the triplet group, the model uncertainties are higher than in the case of the
quintet transitions. In order to model the strength of the individual
multiplet components, we used the NIST $A$ coefficients. Their quality could
roughly be checked as some multiplet features were partly resolved. The final
\chem{O} recombination line model comprises 309 lines
(Fig.~\ref{fig:reflineint}) and has an integrated reference intensity of
72\,\unit{R}, where the contribution of the 204 quintet transitions is 55\,\%. 

Finally, the full \chem{O} line model comprises 313 lines with a summed
intensity of 399\,\unit{R}.

\subsection{Atomic nitrogen}
\label{sec:N}

The \chem{N} transition from the excited $^2$D level to the ground state $^4$S
generates a doublet at air wavelengths of 519.8 and 520.0\,\unit{nm}
\citep{NIST_ASD}. The chemistry of this ionospheric emission is closely
related to the chemistry of the red \chem{O} lines (see
Reactions~(\ref{eq:O++O2}) and (\ref{eq:O2++e-})). The $^2$D level population
is also produced by dissociative recombination, i.e.
\begin{reaction}\label{eq:NO++e-}
  \mathrm{NO}^+ + \mathrm{e}^- \rightarrow \mathrm{N}^{\ast} + \mathrm{O},
\end{reaction}
where \chem{NO^+} originates from charge transfer reactions of either
\chem{O^+} with \chem{N_2} or \chem{O_2^+} with \chem{N}
\citep[e.g.,][]{khomich08}. We assume the same reference emission height of
250\,\unit{km} as for the red \chem{O} lines (Table~\ref{tab:species}).

The doublet was measured in \mbox{X-shooter} UVB-arm spectra. However, we
only considered spectra for widths of the entrance slit not larger than the
standard value of 1\,\unit{arcsec} in order to avoid blending of both
components. Moreover, the intensity ratio of both lines showed a dependence on
the total intensity. However, \citet{sharpee05} found a relatively constant
ratio of $1.759 \pm 0.014$ based on ESI and HIRES data from Mauna Kea, which
is also consistent with the UVES-based result of 1.760 from the catalogue of
\citet{cosby06}. As only the strongest emissions in the \mbox{X-shooter}-based
sample showed a satisfying agreement, there was obviously an issue with the
separation of the weak line and relative strong underlying emission (see
Fig.~\ref{fig:refspec}). Therefore, we corrected the mean ratio of the time
series using 1.759 as a reference. The total emission of the doublet remained
untouched. From the resulting climatologies, we then obtained reference
intensities of 1.91 and 1.08\,\unit{R} for the lines at 519.8 and
520.0\,\unit{nm}, respectively (Figs.~\ref{fig:reflineint} and
\ref{fig:climlines}). As a result of this approach, the intensity ratio
slightly changed to 1.757. The integrated reference emission of the \chem{N}
line model amounts to 2.99\,\unit{R}.

\subsection{Atomic hydrogen}
\label{sec:H}

Low-latitude nightglow emission is essentially caused by chemiluminescence,
i.e. the excited states are produced by chemical reactions. However, there
is also a small contribution from resonant fluorescence to the nocturnal
radiation. Although this process requires the absorption of solar photons,
such emission is visible if the absorbing atoms are frequent and located
higher than the Earth's shadow. These criteria are usually fulfilled for
\chem{H} in the Earth's geocorona that extends thousands of kilometres in
altitude. In the PALACE wavelength regime, the lowest Balmer lines are most
relevant. The Balmer series is related to the \chem{H} level with a principal
quantum number $n$ of 2. In emission, the first four lines \chem{H\alpha},
\chem{H\beta}, \chem{H\gamma}, and \chem{H\delta} have upper levels
$n^{\prime}$ of 3 to 6 and are located at air wavelengths of 656.3, 486.1,
434.0, and 410.2\,\unit{nm} \citep{NIST_ASD}. Each transition consists of
seven fine structure components. The intensity of the different lines is
affected by the \chem{H} density profile, the altitude of the Earth's shadow,
the solar spectrum, radiative transfer by scattering, excitation factors, and
radiative cascades in the \chem{H} atom \citep{meier95,gardner17}. The
emission strongly depends on the shadow height due to decreasing \chem{H}
density with increasing altitude. Hence, a fixed reference height as for the
chemiluminescent emissions cannot be given. Moreover, the shadow height
depends on zenith distance and azimuth for a fixed time, which requires more
information than the PALACE input parameters (Table~\ref{tab:parameters}) can
provide. As a consequence, we neglected a correction of this effect in the
data analysis and the model.

We measured intensity time series for \chem{H\alpha} to \chem{H\delta} in the
\mbox{X-shooter} data set. Higher Balmer lines were too faint. As
\chem{H\beta} to \chem{H\delta} are weak lines in a wavelength region with
complex emission structure due to various \chem{O_2} bands
(Fig.~\ref{fig:refspec}), systematic intensity uncertainties are likely. We
therefore checked the mean intensities for very high shadow heights above
80,000\,\unit{km} and found negative values between $-0.23$ and
$-0.30$\,\unit{R}, which we interpreted as systematic errors. In the case of
\chem{H\alpha}, we obtained $+1.16$\,\unit{R}, which we did not correct as
non-zero emission is expected due to the distinctly stronger multiple
scattering for \chem{H\alpha} compared to the other Balmer lines
\citep[e.g.,][]{gardner17}. It is not clear whether \chem{H\alpha} emission
from the interstellar medium might contribute. From the calculated
climatologies, we derived reference intensities of 5.29, 0.51, 0.19, and
0.11\,\unit{R} for \chem{H\alpha} to \chem{H\delta} (see
Fig.~\ref{fig:climlines}), i.e. the total \chem{H} model intensity, which is
dominated by \chem{H\alpha}, amounts to 6.10\,\unit{R}. As the shadow height
effect was not considered, these intensities refer to the mixture of lines of
sight and observing times of the astronomical observations in the archive.
Very different samples might cause a significant change in the intensity. For
the final \chem{H} model (see Fig.~\ref{fig:reflineint}), we had to derive
the relative contributions of the seven fine structure components, which
cannot be resolved in the \mbox{X-shooter} spectra. \citet{gardner17} reported
calculated fractions for \chem{H\alpha}. In fact, the two lines related to
the transitions from $^2$P to $^2$S appear to comprise about 95\,\% of the
total emission. Only $^2$P should directly be excited by photons originating
from the solar emission in the far-UV Lyman series line Ly$\beta$ that changes
$n$ from 1 to 3 in \chem{H} \citep{meier95}. The upper levels of the other
fine structure components need to be populated by radiative cascades, which is
less efficient. We used the intensity fractions from \citet{gardner17} for the
other Balmer lines as well. This is a rough extrapolation but the absolute
uncertainties are small for these weak lines.

\subsection{Helium}
\label{sec:He}

Another source of fluorescent emission is \chem{He}. The most interesting
line here is probably at 1,083\,\unit{nm} \citep[e.g.,][]{noto98}. It is
related to metastable ortho-\chem{He} (spin multiplicity of 3), which can be
produced by \chem{He^+} recombination. It was not possible to measure this
line in \mbox{X-shooter} spectra due to insufficient resolution to separate
it from strong \chem{OH} emission. The catalogue of \citet{cosby06} includes
another ortho-\chem{He} line at 389\,\unit{nm} with an intensity of
0.54\unit{R}. However, we did not succeed to detect this line in a difficult
wavelength region with many other emission lines (see Fig.~\ref{fig:refspec}).
As we could not measure any \chem{He} line and derive a climatology, we
neglected this minor contribution in the PALACE line model.

\section{Continuum emission}
\label{sec:continuum}

\begin{figure*}[t]
\includegraphics[width=17cm]{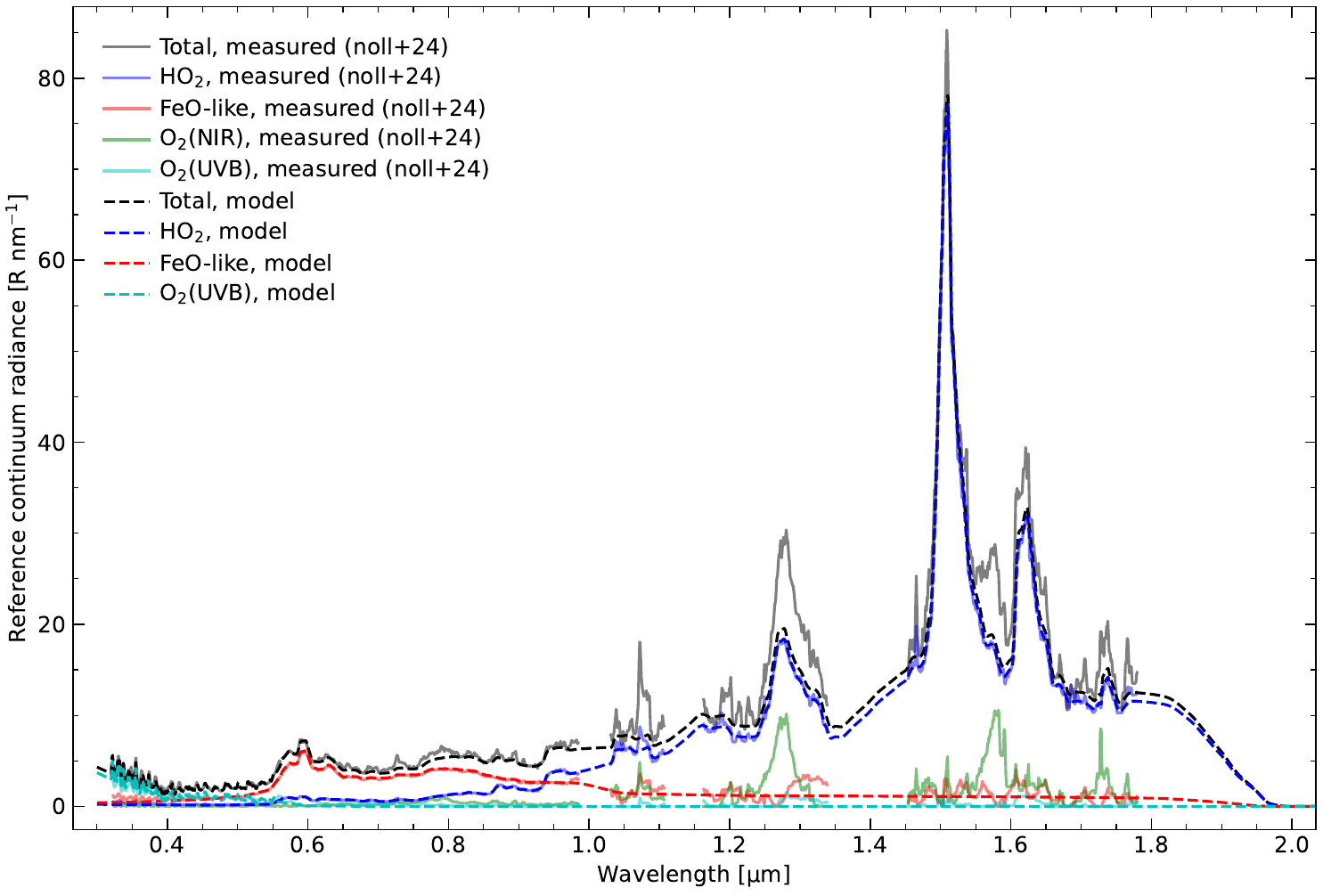}
\caption{PALACE reference continuum model (dashed spectra) compared to
  \mbox{X-shooter}-based components from \citet{noll24} (solid spectra with
  gaps). The total (black), \chem{HO_2} (blue), \chem{FeO}-like (red), and
  \chem{O_2}(UVB) emissions (cyan) are shown in both cases. The
  \chem{O_2}(NIR) residuals (green) are not part of PALACE as they should be
  covered by the line emission model.}
\label{fig:refcont}
\end{figure*}

The line model described in Sect.~\ref{sec:lines} does not cover airglow
emission components without resolved lines in \mbox{X-shooter} spectra. These
pseudo-continua were investigated by \citet{noll24} in all three
\mbox{X-shooter} arms, which required the subtraction of other sky brightness
contributions such as scattered moonlight, scattered starlight, and zodiacal
light as well as disturbing thermal emission from the telescope and instrument
at the longest wavelengths. This correction was performed by means of the ESO
Sky Model \citep{noll12,jones13}. Moreover, the spectra were corrected for
atmospheric absorption and emission in a similar way as discussed in
Sects.~\ref{sec:overview} and \ref{sec:lines}. The only difference was that
the correction was calculated for pixels instead of lines. The separation of
continuum and lines was carried out with the same percentage filters and
filter widths as used for the line extraction described in
Sect.~\ref{sec:lines}. The correction of the van Rhijn effect expressed by
Eq.~(\ref{eq:vanrhijn}) was performed with a reference height of
90\,\unit{km}. For the decomposition of the resulting regridded continuum
spectra in the useful wavelength range between 0.3 and 1.8\,\unit{\mu{}m},
\citet{noll24} selected a sample of 10,633 high-quality spectra. The mean
spectrum is shown in Fig.~\ref{fig:refcont}. There are several gaps without
data due to strong atmospheric absorption or low-quality marginal ranges of
the \mbox{X-shooter} arms. Wavelengths with weak absorption above about
2.0\,\unit{\mu{}m} could not be analysed due to too strong telescope-related
thermal emission for a convincing correction. The continuum decomposition was
carried out by variability-driven non-negative matrix factorisation applied to
the set of spectra. This approach returned an optimum number of four component
spectra (see Fig.~\ref{fig:refcont}) and corresponding time series of scaling
factors that were used for the calculation of climatologies. In the following,
we discuss the different continuum components and the modification of the
measured spectra for the PALACE continuum model, which is also shown in 
Fig.~\ref{fig:refcont}.

\subsection{Hydroperoxyl}
\label{sec:HO2}

The predominating component in the \mbox{X-shooter} NIR-arm range is
characterised by a conspicuous peak at 1.51\,\unit{\mu{}m} and weaker emission
features near 1.27 and 1.62\,\unit{\mu{}m} as well as a continuum with slowly
decreasing flux for shorter wavelengths. This continuum component was first
described and analysed by \citet{noll24}, although the residual continuum of
the ESO Sky Model \citep{noll12} roughly reveals its structure. \citet{noll24}
found that the emission pattern is very likely the result of \chem{HO_2}
emission. The strong 1.51\,\unit{\mu{}m} feature can be identified as the
vibrational \mbox{(200-000)} transition of the electronic ground state
X$^2\mathrm{A}^{\prime\prime}$, which changes the \chem{OO-H} stretching mode
\citep[e.g.,][]{becker74}. The feature near 1.27\,\unit{\mu{}m} would mainly
be related to a vibronic band with the change of the electronic and
vibrational levels from A$^2\mathrm{A}^{\prime}$(001) to
X$^2\mathrm{A}^{\prime\prime}$(000), which involves the \chem{O-OH} stretching
mode. The related \mbox{(000-000)} transition, which was bright in laboratory
experiments \citep[e.g.,][]{becker74,holstein83}, could not be detected in
the \mbox{X-shooter}-based continuum spectrum as the band at
1.43\,\unit{\mu{}m} is in a wavelength range with strong water vapour
absorption. On the other hand, the 1.62\,\unit{\mu{}m} feature in
Fig.~\ref{fig:refcont} has not been identified in laboratory studies, so far.

In order to better understand the origin of the emission, \citet{noll24}
estimated the total radiation, studied the variability in the data, and
compared the results to simulations of the \chem{HO_2}-related chemistry with
an optimised version of the Whole Atmophere Community Climate Model
\citep[WACCM;][]{gettelman19}. Moreover, SABER-based retrievals (especially of
\chem{H} concentrations) were used for the analysis. This investigation
revealed that the main production process of \chem{HO_2} in the mesosphere
\citep[e.g.,][]{makhlouf95}
\begin{reaction}\label{eq:H+O2+M}
  \mathrm{H} + \mathrm{O}_2 + \mathrm{M} \rightarrow \mathrm{HO}_2^{\ast} +
  \mathrm{M}
\end{reaction}
is also the most important reaction for the nightglow emission. In this way,
the spectral distribution of the emission matches much better than in the
case of excitations due to collisions with \chem{O_2} in the
a$^1\Delta_\mathrm{g}$ state \citep{holstein83}, which can only produce a minor
fraction of the total emission. This constraint is also justified by major
differences in the climatological variability patterns. The WACCM simulations
revealed a mean centroid height of the emission related to
Reaction~(\ref{eq:H+O2+M}) of 81\,\unit{km}, which is in good agreement with
rough \mbox{X-shooter}-based estimates derived from the timing of a passing
high-amplitude wave compared to various \chem{OH} lines
\citep[see also][]{noll22}. The use of 81\,\unit{km} in PALACE
(Table~\ref{tab:species}) instead of an height of 90\,\unit{km} that was taken
for the data analysis does not cause a systematic error as the van Rhijn
factors are almost identical for the mean zenith angle of the
continuum-related sample of 34$^{\circ}$. 

The measured \chem{HO_2} pseudo-continuum in Fig.~\ref{fig:refcont} still
shows small-scale structures that could be produced by residuals from other
emissions (especially \chem{OH}). Moreover, the spectrum only covers a part of
the PALACE wavelength range. In order to obtain a smooth model spectrum, we
therefore set suitable data points to remove obvious residuals and to bridge
gaps by means of spline interpolation. As the component spectrum showed
significant contamination by \chem{O_2} emission features at the shortest
wavelengths, we removed any features there and kept the flux almost constant.
The \chem{HO_2} spectrum indicates significant flux at the longest measured
wavelength of 1.78\,\unit{\mu{}m}. As it is not clear how the emission
pattern evolves at longer wavelengths, we slowly reduced the flux up to about
2.0\,\unit{\mu{}m} and then assumed zero emission. The whole smoothing
procedure is somewhat arbitrary, but the highest systematic uncertainties are
related to regions with very strong absorption or strong emission of different
origin, where either the absolute or relative \chem{HO_2} flux is negligible.
Nevertheless, the reference flux without atmospheric extinction could show
significant deviations in wavelength regions without measurements. An
example is the possible emission feature at 1.43\,\unit{\mu{}m}, which is not
present in the model. For the output spectrum in the PALACE wavelength range
from 0.3 to 2.5\,\unit{\mu{}m}, we used a wavelength step of 0.02\,\unit{nm},
which is consistent with the pixel sizes in the UVB- and VIS-arm
\mbox{X-shooter} spectra (0.06\,\unit{nm} in the NIR arm). This grid is finer
than the 0.5\,\unit{nm} used by \citet{noll24} in order to allow for a higher
resolution in the precalculated reference atmospheric transmission
(Sect.~\ref{sec:overview}). In the course of the regridding based on the
spline interpolation, we also changed air to vacuum wavelengths, which are
also the standard for the line model. As last step of the modification of the
spectrum, we considered that the mean intensity of the time series and the
climatological reference intensity can slightly differ. From the climatology
of the component scaling factors, we obtained a correction factor of 1.011,
which we applied. The final reference spectrum has an integrated flux of
13.1\,\unit{kR}. \citet{noll24} reported an intensity of 11.8\,\unit{kR} for
wavelengths between 0.323 and 1.78\,\unit{\mu{}m} and using linear
interpolation for the gaps. For the same wavelength range, the model spectrum
shows a very similar value of 11.7\,\unit{kR}.

\subsection{Iron monoxide and other molecules}
\label{sec:FeO}

The pseudo-continuum in the wavelength range from about 500 to 720\,\unit{nm}
with a main emission peak at about 595\,\unit{nm} (Fig.~\ref{fig:refcont})
was already the topic of several studies. Based on spectra from the
satellite-based Optical Spectrograph and Infrared Imaging System
\citep[OSIRIS;][]{evans10}, ESI at Mauna Kea \citep{saran11}, and also
\mbox{X-shooter} \citep{unterguggenberger17} as well as laboratory data
\citep[e.g.,][]{west75} and theoretical calculations \citep{gattinger11a}, the
emission structures could be identified as emission of the \chem{FeO}
`orange arc' bands, which are related to transitions with the two upper
electronic levels $\mathrm{D}\,^5\Delta_i$ and
$\mathrm{D^{\prime}}\,^5\Delta_i$ and the ground state
$\mathrm{X}\,^5\Delta_i$ as the lower level. The excited states should be
produced by
\begin{reaction}\label{eq:Fe+O3}
  \mathrm{Fe} + \mathrm{O}_3 \rightarrow \mathrm{FeO}^{\ast} + \mathrm{O}_2
\end{reaction}
\citep{evans10}.

The decomposition of the \mbox{X-shooter}-based continuum by \citet{noll24}
revealed that also emissions at shorter and longer wavelengths (with a shallow
maximum near 800\,\unit{nm}) showed a strong correlation with the \chem{FeO}
main feature, which suggested that additional band systems as investigated by
\citet{west75} in the laboratory contribute to the \chem{FeO}-related
component spectrum in Fig.~\ref{fig:refcont}. However, WACCM simulations by
\citet{noll24} indicated that \chem{FeO} emission is probably an order of
magnitude too faint to explain the whole component spectrum. Minor additional
contributions in the visual range could be caused by excited \chem{NiO}
\citep{evans11} or reactions of \chem{NO} and \chem{O} \citep{gattinger10}.
\citet{noll24} also proposed emission from excited \chem{OFeOH} produced by
the reaction of \chem{FeOH} and \chem{O_3}. This possible nightglow radiation
(with uncertain spectral distribution) could even be stronger than \chem{FeO}
emission. Nevertheless, the integrated emission from all proposed processes
could still be too low. Any other suitable mechanism needs to show similar
variations as \chem{FeO} (see Sect.~\ref{sec:comparison}). In order to set
the reference height for the continuum component that dominates the
\mbox{X-shooter} VIS arm, we focused on \chem{FeO} and took the
WACCM-simulated mean centroid height of 88\,\unit{km}
(Table~\ref{tab:species}), which is in good agreement with previous
measurements and simulations \citep{evans10,saran11}.

In a similar way as discussed for \chem{HO_2} in Sect.~\ref{sec:HO2}, we
modified the measured continuum with \chem{FeO} contribution in order to
reduce residuals from other emissions and fill gaps. While in the VIS-arm
range the changes by the spline interpolation are minor and only worked as a
smoothing filter, the UVB-arm range shows an increasing discrepancy with
decreasing wavelength in order to remove clear contaminations from \chem{O_2}
band systems. As the measured NIR-arm spectrum indicates large flux
variations due to decomposition uncertainties, we strongly smoothed it to
obtain slowly declining fluxes up to a maximum wavelength of
2.0\,\unit{\mu{}m}. As some peaks could be related to residuals of other
emissions, we obtained a 12\,\% (16\,\%) lower mean in the range from 1.05 to
1.27\,\unit{\mu{}m} (1.45 to 1.78\,\unit{\mu{}m}). Although the uncertainties
remain high in this regime, non-negligible emissions from \chem{FeO} or
\chem{OFeOH} appear to be possible. In agreement with Sect.~\ref{sec:HO2}, we
also scaled the resulting model spectrum to be consistent with the
climatological mean from the component scaling factors. The correction factor
was 1.021. The integrated flux of the resulting continuum component amounts to
2.8\,\unit{kR}. In the reduced range from 0.323 to 1.78\,\unit{\mu{}m}, where
\citet{noll24} derived 2.9\,\unit{kR}, it would be 2.7\,\unit{kR}. The slight
decrease reflects the attempted reduction of residuals from other emissions.

\subsection{Unresolved molecular oxygen bands}
\label{sec:O2cont}

The continuum decomposition of \citet{noll24} also produced two
\chem{O_2}-related components (Fig.~\ref{fig:refcont}). The component in the
\mbox{X-shooter} NIR arm mainly consists of residuals of the \mbox{(a-X)}
bands at 1.27 and 1.58\,\unit{\mu{}m} (see Sect.~\ref{sec:O2}). The other
contributions also appear to be caused by line residuals, especially by
\chem{OH}. As these emissions should fully be covered by the PALACE line
model (Sect.~\ref{sec:lines}), we did not use this component for the continuum
model. This is different for the continuum in the UVB arm, which should mainly
be produced by \chem{O_2} band systems with upper levels of
c$^1\Sigma^-_\mathrm{u}$, A$^{\prime\,3}\Delta_\mathrm{u}$, and
A$^3\Sigma^+_\mathrm{u}$. As discussed in Sect.~\ref{sec:O2}, the
model-relevant UVES-based line measurements of \citet{hanuschik03} that were
interpreted by \citet{cosby06} are clearly incomplete. Hence, we had to add
a continuum to the model that includes the missing emission from the line
measurements. First, we derived a smooth curve from the component spectrum
of \citet{noll24} using spline interpolation (see Sect.~\ref{sec:HO2}). Here,
we removed any \chem{O_2}-related band features. Moreover, the flux was
assumed to be zero above 680\,\unit{nm}. Then, PALACE was used to calculate
the reference line spectrum as shown in Fig.~\ref{fig:refspec} without
atmospheric extinction. This model spectrum was added to the smoothed
continuum and compared to a representative UVB-arm airglow mean spectrum in
the range from 320 to 556\,\unit{nm} with corrected atmospheric extinction and
van Rhijn effect. The difference was the basis for the correction of the
\chem{O_2} continuum component. In order to avoid impacts of observational
noise, differences in the line-spread function, and wavelength calibration, we
applied a moving averaging filter with a width of 2\,\unit{nm} and subsequent
automatic spline interpolation with gaps between the selected data points of
1\,\unit{nm} to obtain the final spectrum (Fig.~\ref{fig:refcont}) with the
wavelength grid of the PALACE continuum model (see Sect.~\ref{sec:HO2}). The
integrated flux is 405\,\unit{R}. Neglecting the extrapolation range between
300 and 320\,\unit{nm} without measured band structures, it would be
339\,\unit{R}. The latter intensity is still significantly larger than a value
of 62\,\unit{R} from the line model discussed in Sect.~\ref{sec:O2}. The
uncertainties in the \chem{O_2} continuum component are relatively high as
the correction of other sky radiance components and atmospheric extinction is
difficult in the near-UV.

\section{Variability}
\label{sec:variability}

\begin{table*}[tp]
\caption{PALACE variability classes}
\begin{tabular}{llccccclcc}
\tophline
Class & Selection & Sample$^\mathrm{a}$ & Clim.$^\mathrm{b}$ &
Lines$^\mathrm{c}$ & Cont.$^\mathrm{c}$ & $1 - r^2_\mathrm{max}$$^\mathrm{d}$ &
Nearest$^\mathrm{e}$ & $\langle m_\mathrm{SCE} \rangle$$^\mathrm{f}$ &
$\langle \sigma_{f,0} \rangle$$^\mathrm{g}$ \\
\middlehline
OH3a & \chem{OH}, $v^{\prime} = [2, 3]$, $N^{\prime} = [1, 3]$ & 19,372 & 30 &
189 & 0 & 0.028 & OH5a & $+0.116$ & 0.297 \\
OH5a & \chem{OH}, $v^{\prime} = [4, 5]$, $N^{\prime} = [1, 3]$ & 19,461 & 65 &
438 & 0 & 0.010 & OH3b & $+0.131$ & 0.288 \\
OH7a & \chem{OH}, $v^{\prime} = [6, 7]$, $N^{\prime} = [1, 3]$ & 19,296 & 62 &
704 & 0 & 0.022 & OH5a & $+0.143$ & 0.267 \\
OH9a & \chem{OH}, $v^{\prime} = [8, 9]$, $N^{\prime} = [1, 3]$ & 18,593 & 77 &
947 & 0 & 0.025 & OH4c & $+0.160$ & 0.256 \\
OH3b & \chem{OH}, $v^{\prime} = [2, 3]$, $N^{\prime} = [4, 6]$ & 19,492 & 15 &
216 & 0 & 0.010 & OH5a & $+0.179$ & 0.351 \\
OH5b & \chem{OH}, $v^{\prime} = [4, 5]$, $N^{\prime} = [4, 6]$ & 19,486 & 35 &
505 & 0 & 0.024 & OH7a & $+0.183$ & 0.345 \\
OH7b & \chem{OH}, $v^{\prime} = [6, 7]$, $N^{\prime} = [4, 6]$ & 19,254 & 34 &
792 & 0 & 0.021 & OH9b & $+0.189$ & 0.318 \\
OH9b & \chem{OH}, $v^{\prime} = [8, 9]$, $N^{\prime} = [4, 6]$ & 18,320 & 32 &
1,080 & 0 & 0.021 & OH7b & $+0.192$ & 0.288 \\
OH4c & \chem{OH}, $v^{\prime} = [2, 5]$, $N^{\prime} = [7, 8]$ & 19,449 & 7 &
480 & 0 & 0.024 & OH7b & $+0.191$ & 0.286 \\
OH8c & \chem{OH}, $v^{\prime} = [6, 9]$, $N^{\prime} = [7, 8]$ & 19,118 & 9 &
1,246 & 0 & 0.047 & OH9b & $+0.198$ & 0.301 \\
OH6d & \chem{OH}, $v^{\prime} = [2, 9]$, $N^{\prime} \ge 9$ \& & 19,002 & 6 &
15,508 & 0 & 0.117 & OH8c & $+0.187$ & 0.248 \\
& \chem{OH}, $v^{\prime} = 10$, $N^{\prime} \ge 1$ & & & & & & & & \\
O2a & \chem{O_2}\mbox{(a-X)} bands & 17,479 & 4 & 1,196 & 0 & 0.225 & OH3a &
$+0.298$ & 0.346 \\
O2b & \chem{O_2}\mbox{(b-X)} bands & 15,763 & 7 & 1,303 & 0 & 0.079 & O2Ac &
$+0.399$ & 0.355 \\
O2Ac & \chem{O_2} A, A$^\prime$, and c upper states & 7,971 & 1 & 1,590 & 1 &
0.079 & O2b & $+0.395$ & 0.368 \\
& (bands in UVB range) & & & & & & & & \\
HO2 & \chem{HO_2} continuum & 17,482 & 1 & 0 & 1 & 0.427 & O2a & $+0.042$ &
0.346 \\
FeO & \chem{FeO} (+ other) continuum & 7,971 & 1 & 0 & 1 & 0.143 & Na &
$+0.086$ & 0.343 \\
Na & \chem{Na}\,D doublet & 12,935 & 1 & 2 & 0 & 0.143 & FeO & $+0.235$ &
0.418 \\
K & \chem{K}\,D doublet & 2,145 & 1 & 2 & 0 & 0.485 & Na & $-0.108$ & 0.464 \\
Og & \chem{O} green line & 17,525 & 1 & 1 & 0 & 0.232 & O2Ac & $+0.754$ &
0.369 \\
Or & \chem{O} red lines & 17,488 & 2 & 3 & 0 & 0.054 & N & $+1.432$ & 0.676 \\
Orc & \chem{O} recombination lines & 16,450 & 1 & 309 & 0 & 0.522 & OH3a &
$+2.679$ & 0.967 \\
N & \chem{N} 520\,\unit{nm} doublet & 9,161 & 2 & 2 & 0 & 0.054 & Or &
$+1.570$ & 0.579 \\
H & \chem{H} Balmer lines & 8,472 & 1 & 28 & 0 & 0.707 & OH4c & $+0.172$ &
0.346 \\
\bottomhline
\end{tabular}
\belowtable{}
\begin{list}{}{}
\item[$^\mathrm{a}$] Average size of sample of 30\,\unit{min} bins.
\item[$^\mathrm{b}$] Number of measured climatologies used for the derivation
  of the class variability. 
\item[$^\mathrm{c}$] Number of lines or continuum components with the given
  class.
\item[$^\mathrm{d}$] Uniqueness or fraction of unexplained variance with
  respect to the climatologies of relative intensity.  
\item[$^\mathrm{e}$] Class with maximum correlation coefficient
  $r_\mathrm{max}$.
\item[$^\mathrm{f}$] Climatology-averaged relative solar cycle effect
  $m_\mathrm{SCE}$ for a change of 100\,\unit{sfu} as given in
  Eq.~(\ref{eq:climfac}).
\item[$^\mathrm{g}$] Climatology-averaged relative residual variability
  $\sigma_{f,0}$ as given in Eq.~(\ref{eq:resvarfac}).
\end{list}
\label{tab:varclasses}
\end{table*}

\subsection{Reference climatologies}
\label{sec:climatologies}

The reference line and continuum models described in Sect.~\ref{sec:lines} and
\ref{sec:continuum} are only valid for the nocturnal annual average, a solar
radio flux of 100\,\unit{sfu}, and zenith. In order to calculate spectra for
different conditions, a variability model was constructed. As already
discussed in Sect.~\ref{sec:overview}, it consists of two-dimensional
climatologies of relative intensity $f_0(\mathtt{mbin},\mathtt{tbin})$, solar
cycle effect $m_\mathrm{SCE}(\mathtt{mbin},\mathtt{tbin})$, and residual
variability $\sigma_{f,0}(\mathtt{mbin},\mathtt{tbin})$ for 23 different
variability classes, which cover all emission lines and continuum components
of the model. Scaling factors for the reference values can then be calculated
using Eq.~(\ref{eq:climfac}), (\ref{eq:resvarfac}), and (\ref{eq:vanrhijn})
for the given PALACE input parameters \texttt{mbin}, \texttt{tbin},
\texttt{srf}, and \texttt{z}. Table~\ref{tab:varclasses} lists the 23
variability classes. The basis for their definition was the involved chemical
species and upper electronic, vibrational, and rotational levels.

As most measured climatologies are related to \chem{OH}, the most detailed
variability model could be developed for this molecule. In total, 372
individual climatologies were used to derive 11 reference climatologies. As
indicated by Fig.~\ref{fig:climlines}, the sample of lines for this purpose is
smaller than it was for the derivation of the population model (544 doublets,
Sect.~\ref{sec:OH}). The quality requirements are higher for full
climatologies than for mean values. The detailed investigation of \chem{OH}
line climatologies by \citet{noll23} revealed that differences in the
variability mainly depend on the effective emission height and the mixing of
cold and hot rotational populations. Emission heights were investigated by
\citet{noll22} by means of a passing high-amplitude quasi-2-day wave (see also
Sect.~\ref{sec:comparison}) with a vertical wavelength of about 30\,\unit{km}
that could be observed in \mbox{X-shooter} spectra and \chem{OH} emission
profiles from SABER. The resulting reference heights are between about 86 and
94\,\unit{km}. They tend to increase with increasing upper vibrational and
rotational level, i.e. $v^{\prime}$ and $N^{\prime}$, although the dependence on
$v^{\prime}$ gets weaker with increasing $N^{\prime}$. The contribution of cold
and hot populations as visualised in Fig.~\ref{fig:popmodel_OH} for the
different $v^{\prime}$ is also important as both populations vary in a
different way \citep{noll20,noll23}. As small changes in $T_\mathrm{cold}$ can
significantly affect the rotational energy where contributions from the cold
and hot populations are similar, the transition zone between the energy ranges
dominated by the two populations shows increased variability.

In order to cover this pattern, we defined four rotational ranges with
$N^{\prime}$ from 1 to 3 (\texttt{a}), 4 to 6 (\texttt{b}), 7 to 8
(\texttt{c}), and at least 9 (\texttt{d}). Moreover, the height dependence of
the emission layers for different $v^{\prime}$ was also considered. In the
ranges \texttt{a} and \texttt{b}, we built classes for pairs of $v^{\prime}$,
whereas the ranges \texttt{c} and \texttt{d} only contain two or one
vibrational intervals because of a lack of reliable climatologies from faint
high-$N^{\prime}$ lines (only 6 to 9 in Table~\ref{tab:varclasses}). However,
this classification can also be justified by the decreasing deviations between
climatologies of different $v^{\prime}$ for higher $N^{\prime}$. In any case,
the individual climatologies in each class showed very strong correlation. The
reference climatologies were derived from the corresponding set of individual
ones by averaging weighted by the square root of the effective intensity. This
approach considered that climatologies of brighter lines tend to have a better
quality. The square root avoids that only the brightest lines contribute. The
intensity can differ by several orders of magnitude as
Fig.~\ref{fig:climlines} illustrates.

The quality of individual climatologies depends on the size of the time
series. As already discussed in Sect.~\ref{sec:lines}, the climatologies were
calculated based on time bins with a length of 30\,\unit{min} and a mininum
summed exposure time of 10\,\unit{min}, which significantly decreased the
varying quality of exposures of very different length. Moreover, the intervals
lowered the discrepancies in the number of spectra in each \mbox{X-shooter}
arm (see Sect.~\ref{sec:dataset}). Table~\ref{tab:varclasses} shows the
resulting average sizes of the time series for each class. They vary between
18,320 and 19,492. The relatively low numbers for the classes \texttt{OH9a}
and \texttt{OH9b} are related to the fact that a minor fraction of the NIR-arm
spectra is only useful up to 2.1\,\unit{\mu{}m} \citep{vernet11}, which
affects the \mbox{(8-6)} and \mbox{(9-7)} bands. Nevertheless, all samples
were large enough for the derivation of robust climatologies. For this
purpose, subsamples of the intensity time series for each $\mathtt{mbin}$
(month) and $\mathtt{tbin}$ (local time) combination were selected by
including all data points within a circle with a radius of 1 month or 1 hour
around the corresponding climatological grid point \citep[see][]{noll23}.
Thus, the data sets are overlapping, which smoothes the result. Moreover, a
minimum size of the selected subsamples of 400 was required, which is usually
fulfilled. An exception are cells close to twilight or even with daytime
contribution. Here, the sample size was optimised by an increase of the
selection radius in steps of 0.1 (relative to 1 month or 1 hour) until the
threshold size was achieved. As a consequence, the temporal resolution of the
climatologies is lower at the margins of the nighttime period. Nevertheless,
the effective selection radius for the nocturnal climatologies was only about
1.07 as the cells with the largest radii do not contribute to the average
weighted by the nighttime fraction.

The lower limit for the subsample size is important for the calculation of the
solar cycle effect, which relies on linear regression using the centred 27-day
average of the solar radio flux as reference (Sect.~\ref{sec:overview}). As a
regression line is more uncertain than an average, the
$m_\mathrm{SCE}(\mathtt{mbin},\mathtt{tbin})$ climatologies were more critical
for the preselection of suitable \chem{OH} lines than the
$f_0(\mathtt{mbin},\mathtt{tbin})$ ones. Moreover, the
$\sigma_{f,0}(\mathtt{mbin},\mathtt{tbin})$ climatologies can easier be
disturbed. They were calculated from the standard deviation of the difference
between the measured intensities and the intensity model based on
Eq.~(\ref{eq:climfac}). Hence, systematic deviations in the data especially
due to outliers can cause uncertainties. Therefore, the weighted averaging
of individual climatologies further increased the robustness of the measured
variability patterns.

The number of available climatologies for all non-\chem{OH} variability
classes was relatively small because of either the difficulty of measurement
or the existence of only a single multiplet matching the variability class.
The sizes of the data sets used for the calculation of these climatologies
were also smaller than in the case of \chem{OH} (Table~\ref{tab:varclasses}).
The data selection for the few lines or features tended to be more
restrictive. In part, the emissions were relatively faint. In some cases, the
slit width of the spectrograph had to be limited to avoid line blending.
Overall, this leads to higher uncertainties (and lower temporal resolution)
in the reference climatologies, especially with respect to $m_\mathrm{SCE}$ and
$\sigma_{f,0}$. Exceptions are the few relatively strong emissions
(Fig.~\ref{fig:climlines}) and classes with strong variations
(Sect.~\ref{sec:comparison}).

For the \chem{O_2}\mbox{(a-X)} bands, only eight \mbox{X-shooter}-based
climatologies were used for the population study in Sect.~\ref{sec:O2}. For
the derivation of the reference climatology, we further reduced this number to
four, only involving the $^\mathrm{S}\mathrm{R}$ and $^\mathrm{O}\mathrm{P}$
lines of the \mbox{(0-0)} band with $N^{\prime} = 15$ and 17, which showed the
best quality. The averaging of these climatologies weighted by reference
intensity resulted in an effective $N^{\prime}$ of 15.7. As a consequence, the
reference climatology is only an approximation for lines with low $N^{\prime}$
that could not be measured. In the case of the \chem{O_2}\mbox{(b-X)} bands,
climatologies of P-branch lines of the \mbox{(0-1)} band with $N^{\prime}$
between 0 and 16 were derived. While $N^{\prime} = 2$ was already skipped for
the population study, $N^{\prime} = 8$ was also dropped for the variability
analysis due to an outlier-affected $\sigma_{f,0}$ climatology. In the end,
seven levels were used for the intensity-weighted averaging of the
climatologies, which resulted in an effective $N^{\prime}$ of 8.0. As
variations in the ambient temperature change the rotational level population
distribution, the climatologies are expected to deviate for increasing
discrepancy in $N^{\prime}$. This effect is visible for the comparison of
$N^{\prime} = 0$ and 16 (mainly revealing stronger variability for higher
$N^{\prime}$) but it is relatively weak. The correlation coefficent $r$ for
the two intensity climatologies is $+0.975$. Nevertheless, the deviations
could be more significant for very high $N^{\prime}$ dominated by a hot
population (see Fig.~\ref{fig:popmodel_O2}) as well as higher $v^{\prime}$,
where a maximum of 15 was measured \citep{slanger00b}. As the most extreme
lines are very weak, the systematic uncertainties in the resulting airglow
spectrum should be low.

The variability of the \chem{HO_2}, \chem{FeO}-like, and UVB-arm \chem{O_2}
emissions was investigated by \citet{noll24}. Climatologies were derived from
the scaling factors of the mean components (Sect.~\ref{sec:continuum}).
Moreover, there are climatologies based on the intensities of specific
features that are representative of the continuum components. We used the
latter as these are more direct measurements with lower systematic
uncertainties. The intensities were derived by linear interpolation between
two limiting wavelengths in the continuum spectra and subsequent integration
of the flux above the line. In detail, we used the results for the main
features of \chem{HO_2}, \chem{FeO}, and \chem{O_2} in the ranges 1.485 to
1.550\,\unit{\mu{}m}, 584 to 607\,\unit{nm}, and 335 to 388\,\unit{nm},
respectively (Fig.~\ref{fig:refcont}). The latter range includes several
emission features (see Fig.~\ref{fig:refspec}). For \chem{FeO} and
\chem{O_2}, the same sample of 7,971 spectra as for the continuum
decomposition \citep{noll24} was used. As this number is much lower than for
\chem{OH}, this significantly lowered the resolution for the resulting
\chem{O_2} climatology (effective nighttime selection radius of 1.56). In
the case of \chem{FeO}, we decided to set the minimum subsample size from 400
to 200, which increased the noise but kept the climatology at a higher
resolution (radius of 1.15). Such changes were not necessary for the
relatively strong 1.51\,\unit{\mu{}m} feature of \chem{HO_2}, where we could
use a relatively large data set of 17,482 bins that was selected by
\citet{noll24} optimised for the feature.  

Climatologies were measured for both \chem{Na} lines of the D doublet
(Fig.~\ref{fig:climlines}). In principle, some variation in the line ratio is
expected \citep{slanger05}. However, we do not have an unbiased measurement
of this variation as we used the line ratio to identify laser contaminations
of the D$_2$ line at 589.0\,\unit{nm} (see Sect.~\ref{sec:Na}). Hence, we
directly took the unaffected D$_1$ climatology as reference for the \chem{Na}
variability class. The sample size of 12,935 bins caused a mild decrease of
the resolution (effective selection radius of 1.22). By far, the smallest
input sample is related to the weak \chem{K} D$_1$ line at 769.9\,\unit{nm}.
The 2,145 bins were derived from intensity measurements of \citet{noll19} in
UVES data (Sect.~\ref{sec:dataset}). As a consequence, the resolution of the
\chem{K}-related climatology is relatively low (radius of 1.47), although we
only used a minimum subsample size of 100, which increased the noise. Despite
these limitations, the climatological patterns for $f_0$, $m_\mathrm{SCE}$, and
$\sigma_{f,0}$ are still meaningful (see Sect.~\ref{sec:comparison}).

As already discussed in Sect.~\ref{sec:O}, the green \chem{O} line was
measured in the UVB and VIS arm of \mbox{X-shooter}. This resulted in two
time series with similar sizes that were converted into climatologies of good
quality with very similar patterns. For the relative intensity, the
correlation coefficient $r$ was $+0.999$. The only noteworthy discrepancy was
the mean intensity due to issues with the flux calibration (see
Sect.~\ref{sec:O}). As reference climatology for the green line, we use the
arithmetic mean of both climatologies. The treatment of the two red \chem{O}
lines, which were measured in good quality, was relatively simple. As
expected, the resulting climatologies proved to be equal. We therefore
obtained the reference climatology for this small variability class from the
averaging of the individual climatologies weighted by the intensity. All
\chem{O} recombination lines related to Reaction~(\ref{eq:O++e-}) were
assigned to the same variability class based on the climatology of the
strongest emission feature at 777\,\unit{nm}, which consists of three
components. The other measured features (see Sect.~\ref{sec:O}) showed noisy
climatologies. In principle, lines related to quintet and triplet upper states
could indicate some discrepancies in their variability. According to
\citet{slanger04c}, there might be differences at the lowest intensities,
where a fluorescence contribution to the triplet-related states could become
significant. The strongest emission feature of the triplet group is at
845\,\unit{nm}. A comparison of the intensity time series did not show a clear
effect and the correlation coefficient $r$ for the $f_0$ climatology (which is
also robust for the 845\,\unit{nm} emission) turned out to be $+0.998$, which
suggests that the reference climatology based on the 777\,\unit{nm} emission
is also valid for the other recombination lines.

For the two lines of the \chem{N} doublet at 520\,\unit{nm}, individual
climatologies could be derived (Sect.~\ref{sec:N}). A correlation analysis
indicated almost perfect agreement, which confirms the result of
\citet{sharpee05} that the line ratio should be fixed. We therefore produced
a single reference climatology for \chem{N} by intensity-weighted averaging.
As the sample size for the climatology calculation was relatively small
(9,161 bins), the temporal resolution is relatively coarse with an effective
selection radius of 1.45 for nighttime conditions. As the \chem{H\alpha}
reference intensity contributes 87\,\% to the total intensity of the \chem{H}
line model (Sect.~\ref{sec:H}) and the summed intensity for the three other
Balmer lines is less than 1\,\unit{R}, we used the \chem{H\alpha} climatology
as the reference for all \chem{H} lines. This approach neglects small
systematic changes in the variability patterns for $f_0$, $m_\mathrm{SCE}$,
and $\sigma_{f,0}$. For $f_0$, the correlation coefficients for \chem{H\alpha}
compared with \chem{H\beta}, \chem{H\gamma}, and \chem{H\delta} are 0.96,
0.88, and 0.65, respectively. Although the latter value is certainly affected
by noise, there is a clear trend. Such discrepancies are expected as the
impact of radiative cascades and the opacity of the geocorona for scattering
differ, which leads to a different dependence of the intensity on the shadow
height \citep{roesler14,gardner17}. The sample size for the \chem{H\alpha}
climatology amounts to 8,472 time bins. In order to avoid very low temporal
resolution, we calculated the climatology for a minimum subsample size of
300, which resulted in an effective selection radius of 1.34.

\subsection{Comparison}
\label{sec:comparison}

\begin{figure*}[tp]
\includegraphics[width=14.4cm]{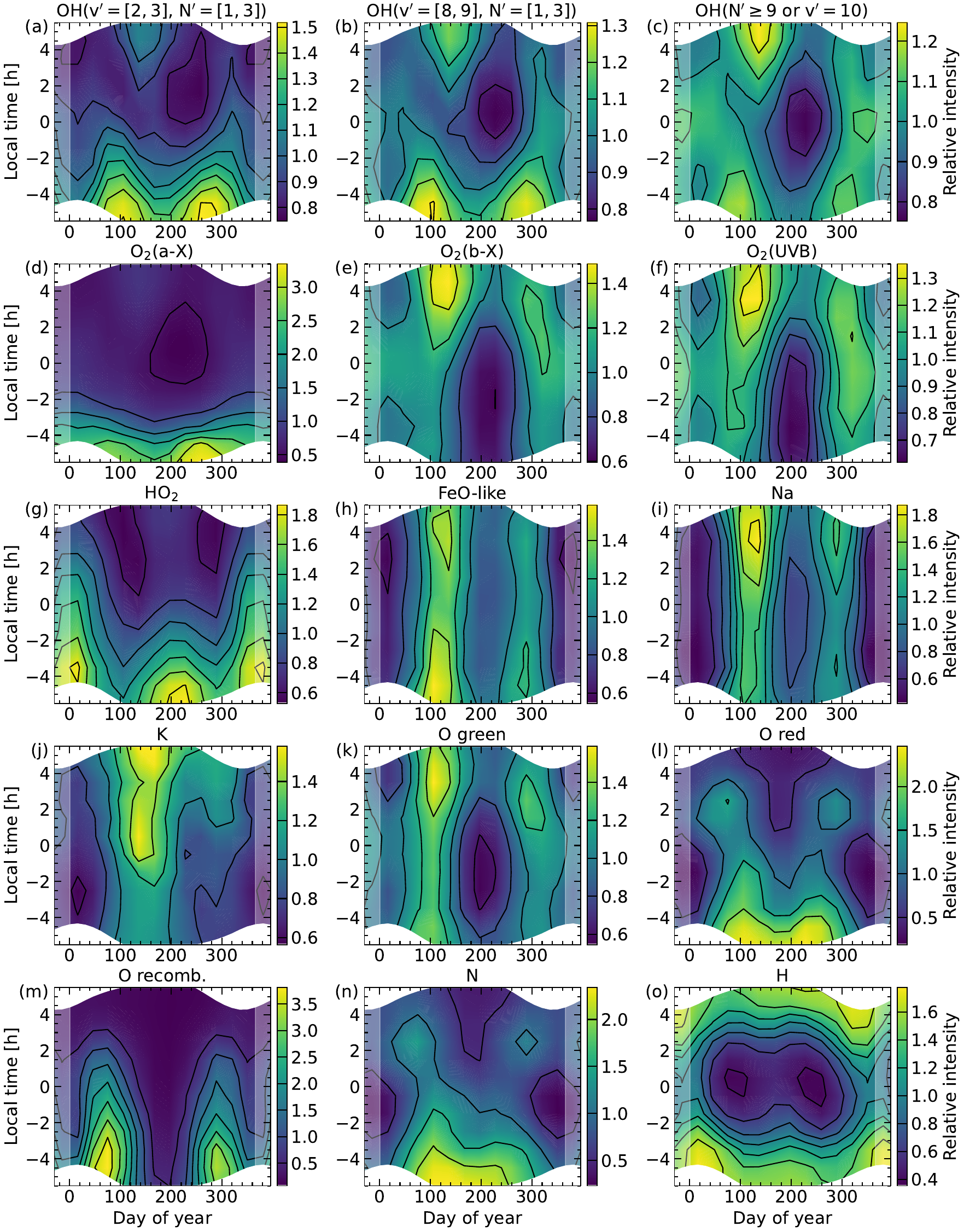}
\caption{Climatologies of intensity relative to the annual mean for a solar
  radio flux of 100\,\unit{sfu}, $f_0$, with respect to local time (mean
  solar time at Cerro Paranal) and day of year for 15 PALACE variability
  classes as defined in Table~\ref{tab:varclasses}. The coloured contours only
  show data with a minimum solar zenith angle of 100$^{\circ}$. The lighter
  colours at the left and right margins mark the repeated patterns of December
  and January.}
\label{fig:refclim_rI}
\end{figure*}

\begin{figure*}[tp]
\includegraphics[width=14.4cm]{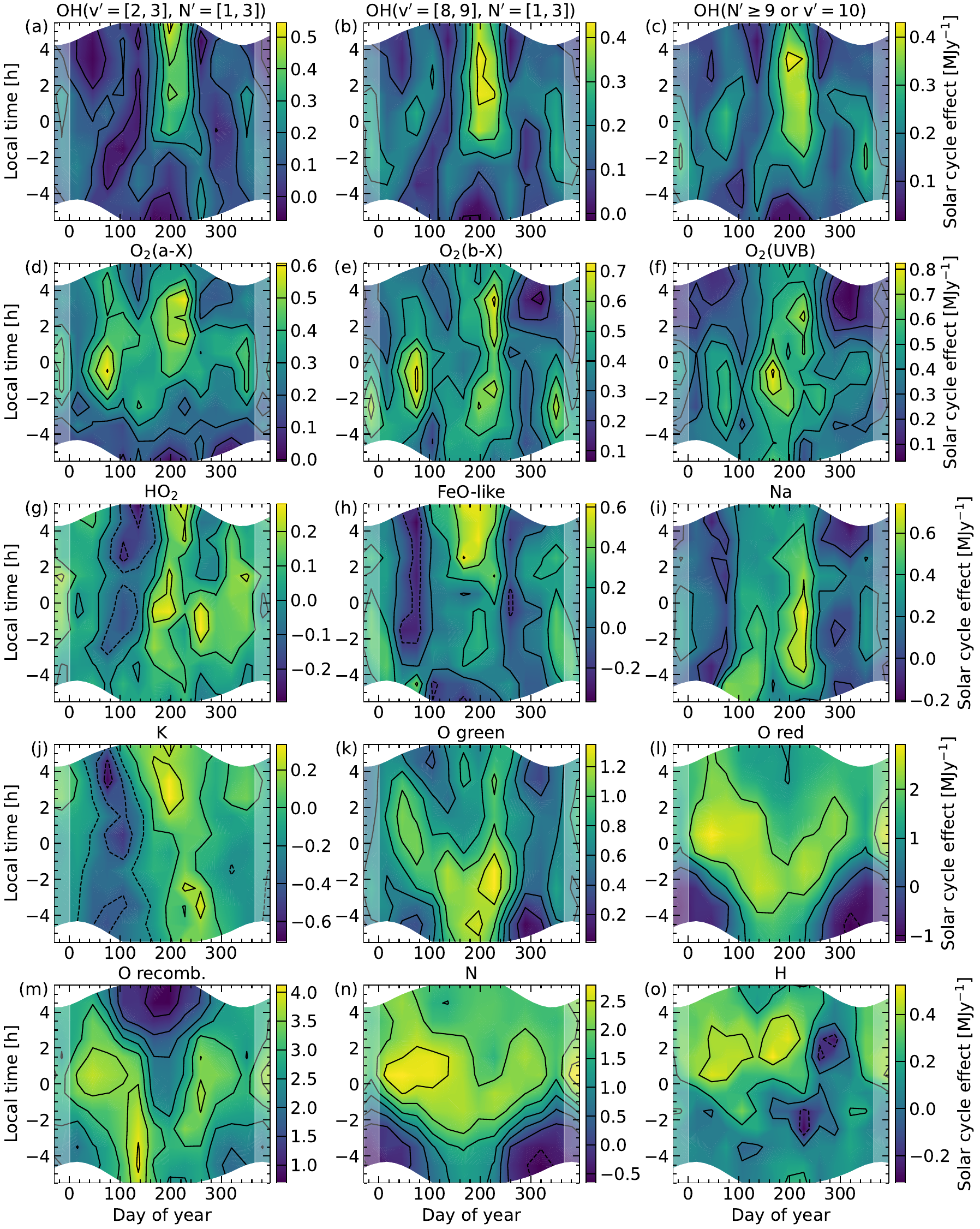}
\caption{Climatologies of the relative intensity change by an increase of the
  solar radio flux by 100\,\unit{sfu} or 1\,\unit{MJy}, $m_\mathrm{SCE}$, with
  respect to local time and day of year for 15 PALACE variability classes as
  defined in Table~\ref{tab:varclasses}. The solar cycle effect is provided
  relative to the corresponding relative intensities $f_0$ in
  Fig.~\ref{fig:refclim_rI}.}
\label{fig:refclim_SCE}
\end{figure*}

\begin{figure*}[tp]
\includegraphics[width=14.4cm]{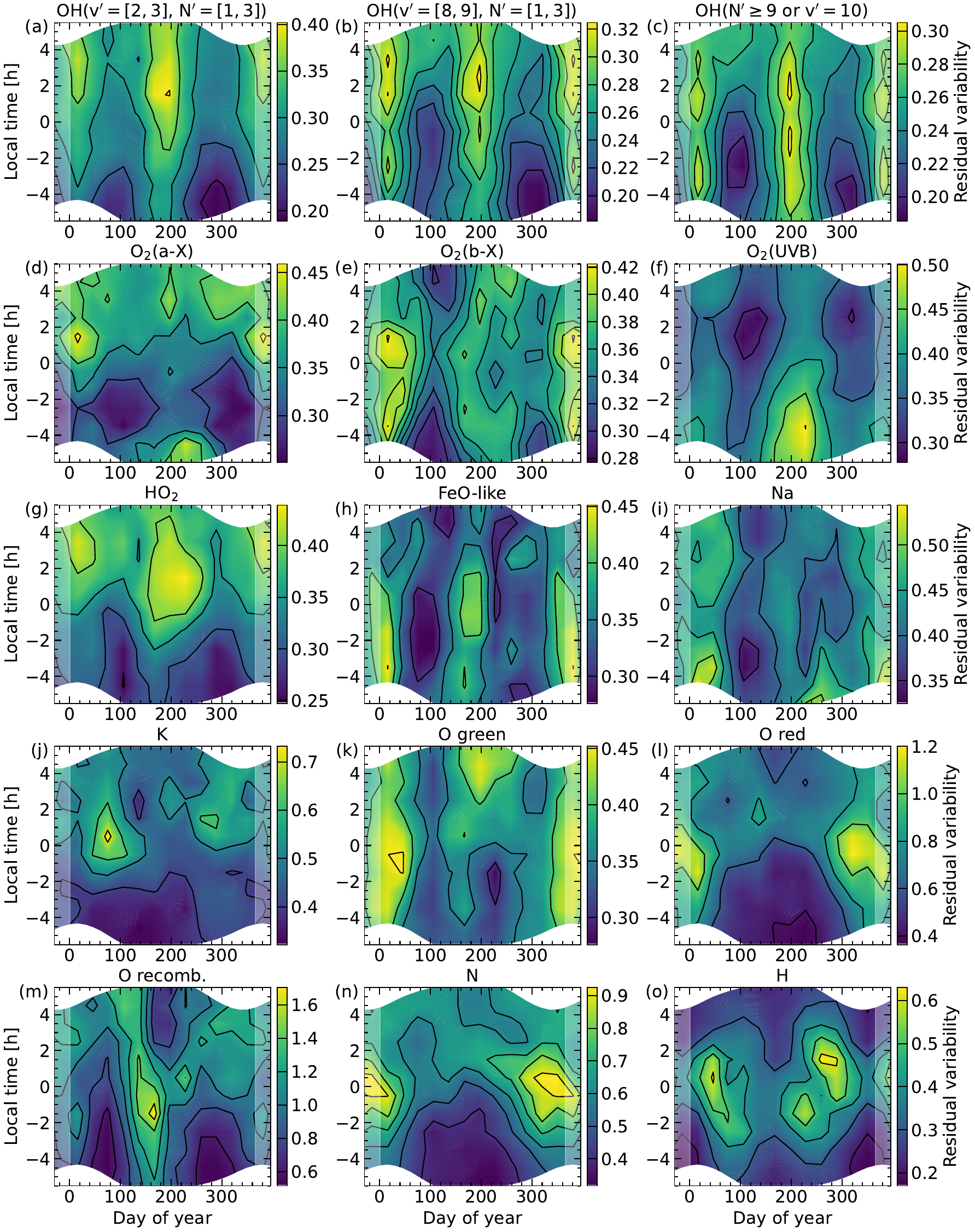}
\caption{Climatologies of the relative residual variability (after
  subtraction of the climatological intensity variations considering the solar
  cycle effect), $\sigma_{f,0}$, with respect to local time and day of year for
  15 PALACE variability classes as defined in Table~\ref{tab:varclasses}. The
  standard deviation is provided relative to the corresponding relative
  intensities $f_0$ in Fig.~\ref{fig:refclim_rI}.}
\label{fig:refclim_rdI}
\end{figure*}

The resulting reference climatologies of 15 variability classes for the
parameters $f_0$, $m_\mathrm{SCE}$, and $\sigma_{f,0}$ are shown in
Figs.~\ref{fig:refclim_rI}, \ref{fig:refclim_SCE}, and \ref{fig:refclim_rdI},
respectively. All non-\chem{OH} climatologies are included. Starting in the
second row of each figure, they are plotted in the same order as listed in
Table~\ref{tab:varclasses}. In the top row, we focus on three examples for 
\chem{OH}, where the variability was discussed by \citet{noll23} in detail.
The class \texttt{OH3a} (a) includes the lines with the lowest effective
emission heights \citep{noll22}. The change in $v^{\prime}$ without changing
$N^{\prime}$ is indicated by \texttt{OH9a} (b), whereas \texttt{OH6d} (c) is
the class for the plain hot population, which comprises the majority of all
lines in the PALACE line model (see Table~\ref{tab:varclasses}). The contours
in each climatology are limited to a ground-based minimum solar zenith angle
of 100$^{\circ}$, which corresponds to complete nighttime up to a height of
about 200\,\unit{km}. As already explained in Sect.~\ref{sec:overview}, the
given local time (LT) refers to the mean solar time at the longitude of Cerro
Paranal.

The climatologies in relative intensity in Fig.~\ref{fig:refclim_rI} show a
wide range of variability patterns but also clear similarities. In order to
quantify this, Table~\ref{tab:varclasses} provides a uniqueness parameter
which was derived from the correlation coefficients $r$ for all combinations
of $f_0$ climatologies. The maximum $r$ for each class was then used to
calculate the parameter by means of $1 - r^2_\mathrm{max}$. The labels of the
best-matching climatologies are also listed in the table. As expected, the
\chem{OH}-related classes show a relatively high similarity with the lowest
uniqueness of 0.010 for the pair \texttt{OH5a} and \texttt{OH3b}. Here, the
increase of $N^{\prime}$ for the latter appears to be almost completely
compensated by the lower $v^{\prime}$. The lines of both classes should have
very similar emission heights \citep[see][]{noll22}.

We start the discussion of the variability patterns in
Fig.~\ref{fig:refclim_rI} with emissions originating in the mesopause region,
which cover the plots from (a) to (k). Here, \chem{HO_2} in (g) indicates a
high uniqueness of 0.427. As already described by \citet{noll24}, the
climatology shows high $f_0$ at the beginning of the night and seasonal maxima
in January and July/August. This pattern correlates well with SABER \chem{H}
density retrievals \citep{mlynczak14}, which confirms
Reaction~(\ref{eq:H+O2+M}) as crucial for the production of excited
\chem{HO_2}. In the reaction, only \chem{H} is a volatile minor species. It is
mainly produced by \chem{H_2O} photolysis at daytime and consumed at nighttime
by different processes, which would explain the nocturnal trend. There is no
other known nightglow process in the mesopause region that predominantly
depends on \chem{H}. In Fig.~\ref{fig:refclim_rI}h, the climatology derived
from the main \chem{FeO} emission feature (Sect.~\ref{sec:climatologies}) is
very different. As also analysed by \citet{noll24}, it is characterised by an
almost opposite seasonal variation with maxima in April/May and October and
the lack of a clear nocturnal trend. The main driver for this variability is
ozone (\chem{O_3}) in Reaction~(\ref{eq:Fe+O3}). Similar variations were found
in SABER-based \chem{O_3} densities for the region around Cerro Paranal
\citep{noll19}. The impact of \chem{Fe} atoms on the climatology is the higher
intensity in austral autumn/winter compared to the other seasons
\citep{feng13,unterguggenberger17}. The source of metals in the mesopause
region is their ablation from heated meteoric dust, which produces permanent
metal layers \citep[e.g.,][]{plane15}. As this also includes \chem{Na} atoms
and the production of excited $^2\mathrm{P}$ states also involves \chem{O_3}
in Reaction~(\ref{eq:Na+O3}), the similarity of the $f_0$ climatologies in (h)
and (i) is expected \citep{unterguggenberger17}. However, the emission of the
other alkali metal \chem{K} in (j) shows clear discrepancies, especially a
remarkable maximum in June at the end of the night that was first observed
by \citet{noll19}. The climatology has a very high uniqueness of 0.485 based
on the relation to \chem{Na}, although the reactions leading to the nightglow
emission are similar. However, the whole reaction networks of both species
show differences \citep{plane14} that could explain the discrepancies. 

The production of excited \chem{OH} requires \chem{H} and \chem{O_3} according
to Reaction~(\ref{eq:H+O3}). As a consequence, the seasonal variability
patterns of both species cancel out in the plots (a) to (c) for most local
times. An exception is the beginning of the night, where a semiannual
variability pattern with maxima near the equinoxes as expected for \chem{O_3}
(see the plots for \chem{FeO} and \chem{Na}) is present, although relatively
weak for the hot population lines in (c). With decreasing effective emission
height \citep{noll22}, this structure becomes stronger. This trend and the
LT-related pattern can be explained by the strong decrease of the mean \chem{O}
concentrations in the lower parts of the \chem{OH} emission layer
\citep[e.g.,][]{smith10}. As \chem{O} radicals are mostly produced by
photolysis of \chem{O_2} at daytime, this causes only short lifetimes of
\chem{O} below about 83\,\unit{km} \citep{marsh06,noll23} due to its nocturnal
consumption by the production of ozone via
\begin{reaction}\label{eq:O+O$_2$}
\mathrm{O} + \mathrm{O}_2 + \mathrm{M} \rightarrow \mathrm{O}_3 + \mathrm{M}
\end{reaction}
and the subsequent creation of excited \chem{OH}. In order to explain the
observed seasonal pattern, the variability of the \chem{O_3} formed near
80\,\unit{km} needs to be stronger than in the case of the \chem{H} density
(see \chem{HO_2} in (g) as a tracer), which peaks at these heights
\citep{mlynczak14,noll24}. As described by \citet{noll23}, all \chem{OH}
climatologies show a local maximum in May close to dawn and a global minimum
in the middle of the night in August/September. These structures are related
to the influence of solar thermal tides with periods of integer fractions of
24\,\unit{h} \citep{forbes95,smith12} on the density, temperature, and
\chem{O} concentration. The effect of tide-induced vertical transport on
\chem{O} is amplified by the strong vertical density gradient. According to
\citet{noll23}, the tidal modes (and/or the interacting perturbations)
relevant for the most conspicuous variability structures should have a
relatively long vertical wavelength as there is no significant change in the
features for different effective emission heights of the \chem{OH} lines.

The tidal features as visible in (c) can also be observed in the
\chem{O_2}-related climatologies in (e) and (f), which also depend on a
reaction involving \chem{O} (\ref{eq:O+O+M}). The differences at the beginning
of the night are probably related to the several kilometres higher emissions
(causing a lower impact of heights with relatively short lifetime of the
\chem{O} radicals). The $f_0$ climatologies of the \texttt{O2b} and
\texttt{O2Ac} classes (the latter based on the \chem{O_2}(UVB) continuum
component in Fig.~\ref{fig:refcont}) show a good agreement with a uniqueness
of 0.079 in Table~\ref{tab:varclasses}. The climatology of the green \chem{O}
line in (k) also indicate similarities (0.232 with respect to \texttt{O2Ac}).
This is not unexpected as the chemistry of these emissions is closely related
(see Sect.~\ref{sec:O}). On the other hand, the climatology for the
\chem{O_2}\mbox{(a-X)} bands in (d) shows clear discrepancies. While the tidal
features in the second half of the night are still weakly visible, the
beginning of the night reveals a strong exponential intensity decrease. As
already discussed in Sect.~\ref{sec:O}, this component is caused by the slow
decay of an a$^1\Delta_\mathrm{g}$ population produced by \chem{O_3} photolysis
at daytime \citep{noll16}.

Figure~\ref{fig:refclim_rI} shows that the climatologies of the ionospheric
red \chem{O} lines (l) and the \chem{N} doublet at 520\,\unit{nm} (n) are very
similar (uniqueness of 0.054), which demonstrates the close relation of the
chemistry of both emissions (Sects.~\ref{sec:O} and \ref{sec:N}). The highest
intensities are found between April and September after dusk. If the
minor ionospheric contribution of the green \chem{O} line (k) shows a similar
pattern, it might explain differences in comparison to the \chem{O_2}
emissions in (e) and (f) at the beginning of the night. The intensities of the
ionospheric emissions clearly decrease in the course of the night with the
exception of two local maxima in March and October at about 02:30~LT. The
general trend is expected as photoionisation is limited to daytime, which
causes a decrease of the ion density due to recombination at nighttime.
However, ion dynamics can also lead to intensity increases. Cerro Paranal is
about 15$^{\circ}$ south of the magnetic equator, where a fountain effect
exists that lifts ions to high altitudes, where recombination is slow
\citep[e.g.,][]{immel06}. At nighttime, the propagation reverses depending on
the magnetic field lines and the season-related direction of the meridional
neutral wind. Then, the airglow emission rate depends on when this ion
reservoir reaches the bottom side of the F-layer, which can lead to effects
such as the midnight brightness wave \citep[e.g.,][]{adachi10,haider22}. In
any case, this equatorial ionisation anomaly causes unusually high ionospheric
line intensities at Cerro Paranal. Interestingly, the $f_0$ climatology of the
\chem{O} recombination lines as derived from the 777\,\unit{nm} emission (m)
shows strong discrepancies. It is the pattern with the strongest variations of
all classes. A reason is the dependence on the square of the electron/ion
density \citep[e.g.,][]{makela01}. The pattern only indicates high intensities
at the beginning of the night in March and October, which brackets the high
intensity range of the red \chem{O} and \chem{N} lines. Moreover, the 02:30~LT
maxima of the latter appear to be located at the end of the nocturnal decrease
of the 777\,\unit{nm} emission. These structures seem to indicate the
descending ion propagation as the layer of the \chem{O} recombination lines is
about 50\,\unit{km} higher (Table~\ref{tab:species}).

The most unique $f_0$ climatology is related to \chem{H} (0.707 in
Table~\ref{tab:varclasses}). This is not unexpected as the radiation mechanism
is fluorescence instead of chemiluminescence (Sect.~\ref{sec:H}). This causes
a strong dependence of the emission on the shadow height as illustrated by  
Fig.~\ref{fig:refclim_rI}o, which shows high intensities after dusk and before
dawn. In addition, there also appears to be variations in the \chem{H} column
density. An example is the local maximum in June at midnight.

As discussed in Sect.~\ref{sec:climatologies}, the derived $m_\mathrm{SCE}$
climatologies have higher uncertainties due to their dependence on a
regression analysis instead of a robust average. Hence, the climatologies in
Fig.~\ref{fig:refclim_SCE} tend to look noisier than those in
Fig.~\ref{fig:refclim_rI}. Nevertheless, the existence of relatively strong
features in the plots (but not their exact shape and strength) is relatively
safe. For \chem{OH}, the average regression uncertainty is
0.05\,\unit{MJy^{-1}} (or per 100\,\unit{sfu}), whereas this value doubles for
metal-related emissions with rather small time series (see
Table~\ref{tab:varclasses}). Ionospheric emissions show the highest absolute
uncertainties (up to 0.18\,\unit{MJy^{-1}}) but the lowest uncertainties
relative to the corresponding $m_\mathrm{SCE}$. The resulting error in the
PALACE output intensity is distinctly lower than 10\% in most cases since the
deviation of the solar radio flux (SRF) from the reference of 100\,\unit{sfu}
is usually much lower than 100\,\unit{sfu}. For the whole \mbox{X-shooter}
data set that almost covers an entire solar cycle, the 27-day averages only
ranged from 67 to 166\,\unit{sfu}.

The solar cycle response is far from homogeneous in
Fig.~\ref{fig:refclim_SCE}. As discussed by \citet{noll23}, the \chem{OH}
$m_\mathrm{SCE}$ climatologies show a conspicuous maximum in the second half of
the night around July. Moreover, its location with respect to local time
depends on the effective emission height. Earlier LTs correspond to larger
altitudes, which can be explained by the upward propagation of oblique wave
fronts. As could be shown by \citet{noll23} also considering other \chem{OH}
airglow properties, the LT shifts are consistent with a period of 24\,\unit{h}
and a vertical wavelength of about 30\,\unit{km}. These properties describe
the migrating diurnal tide \citep[e.g.,][]{forbes95}, which is the most
important tidal mode at low latitudes \citep[e.g.,][]{smith12}. The diurnal
tide obviously amplifies the solar cycle effect (SCE) depending on its phase.
Tide-induced downward transport of \chem{O} radicals with enhanced production
during high solar activity to the \chem{OH}-relevant heights would be an
explanation.

An increased SCE in the middle of the year is also visible for the other
emissions in the mesopause region. Although the exact location of the centroid
of the SCE is difficult to determine, the \chem{O_2}-related emissions in (d)
to (f) and also the green \chem{O} line in (k) appear to be consistent with
the \chem{OH}-related scenario as these emissions seem to show LT centroids at
earlier times as expected for higher effective emission heights
(Table~\ref{tab:species}). The green line as highest emission would need to
peak in the evening, which is the case. For the metal-related emissions in (h)
to (j) and \chem{HO_2} in (g), the trend in LT is less clear, which might be
related to the less important role of \chem{O} radicals. Concerning changes in
the seasonal patterns, the latter emissions tend to partly show negative SCEs
in the first few months of the year. On the other hand, the \chem{O_2}-related
emissions in (d) to (f) appear to indicate additional maxima in March and
December. This trend can already be seen in the \chem{OH} hot population
climatology in (c) and a maximum in March is also visible for the green
\chem{O} line in (k). As the LTs of these features do not significantly
differ, they do not appear to be linked to the diurnal tide.

Apart from the climatological pattern, the effective SCE is interesting for a
comparison. Hence, Table~\ref{tab:varclasses} shows
$\langle m_\mathrm{SCE} \rangle$, which is the SCE averaged for the nighttime
contour area in the plots. Note that the value can slightly differ if the
effect is calculated with respect to the reference intensity of the full
climatology instead of the intensities for each grid point. For \chem{OH},
$\langle m_\mathrm{SCE} \rangle$ increases with increasing emission height
\citep{noll22}, at least for low $v^{\prime}$ and $N^{\prime}$, from about
$+0.12$ to $+0.20$\,\unit{MJy^{-1}}. According to \citet{noll23}, this trend
can partly be explained by the position of the maximum in July, which tends to
be not fully covered by the night for those emissions with the lowest SCEs as
in Fig.~\ref{fig:refclim_SCE}a. The \chem{OH}-based results are in good
agreement with earlier ones related to UVES data from 2000 to 2015
\citep{noll17}. On the other hand, the analysis of \citet{noll12} based on
FORS~1 data from 1999 to 2005 with a high average SRF of 150\,\unit{sfu} did
not show a significant effect. The result of $+0.30$\,\unit{MJy^{-1}} for
\texttt{O2a} is consistent with the SABER-based latitude-dependent study by
\citet{gao16} for the \chem{O_2}\mbox{(a-X)(0-0)} band. The same investigation
also revealed good agreement in terms of \chem{OH}. The \texttt{O2b} and
\texttt{O2Ac} classes show even higher SCEs of about $+0.40$\,\unit{MJy^{-1}}.
However, this value is still lower than $+0.63$\,\unit{MJy^{-1}} as returned
by \citet{noll12} for \chem{O_2}\mbox{(b-X)(0-1)}. The relatively strong
effects are certainly related to the fact that Reaction~(\ref{eq:O+O+M})
involves two \chem{O} radicals. In the case of the 557.7\,\unit{nm} line,
there are even three \chem{O} radicals required because of the combination of
Reactions~(\ref{eq:O+O+M}) and (\ref{eq:O2+O}). Hence, a resulting
$\langle m_\mathrm{SCE} \rangle$ of $+0.75$\,\unit{MJy^{-1}} is not unexpected.
The ESO Sky Model \citep{noll12} uses $+0.87$\,\unit{MJy^{-1}}. As reported by
\citet{zhu15}, different satellite- and ground-based studies returned typical
values of $+0.4$ to $+0.6$\,\unit{MJy^{-1}} for low latitudes (but not
covering Cerro Paranal). The same study also showed an increasing SCE with
increasing altitude for the green \chem{O} line.

As discussed by \citet{noll24}, \chem{HO_2} is very different with an SCE
close to zero. The low emission altitudes and the independence of \chem{O}
may explain this result, which was also confirmed by WACCM simulations. The
same study also revealed the weakly positive effect for \chem{FeO}
($+0.09$\,\unit{MJy^{-1}}) in Table~\ref{tab:varclasses}. Other metal-related
emissions also have rather low effects. The emission of \chem{Na} D returned a
moderate value of $+0.24$\,\unit{MJy^{-1}}, whereas \citet{noll12} measured
$+0.11$\,\unit{MJy^{-1}}. As already derived by \citet{noll19}, the mean effect
for \chem{K} 769.9\,\unit{nm} emission is even negative with
$-0.11$\,\unit{MJy^{-1}}. WACCM simulations of the \chem{Na}, \chem{Fe}, and
\chem{K} metal layers by \citet{dawkins16} indicated a similar order of the
SCEs but more negative in all cases. Hence, the metal concentrations appear
to be an important driver for the low $\langle m_\mathrm{SCE} \rangle$ values.

Very strong positive solar impacts are present for the lines originating in
the ionosphere, where photoionisation depends on extreme UV photons from the
Sun. The red \chem{O} and \chem{N} lines show about $+1.5$\,\unit{MJy^{-1}},
which is much higher than $+0.68$\,\unit{MJy^{-1}} for the red \chem{O} lines
in the ESO Sky Model. The discrepancies are most likely related to the
different sample properties. The sample of \citet{noll12} covered parts of
an earlier solar cycle with relatively high monthly SRF between 95 and
228\,\unit{sfu}. There might be a flattening of the SCE for very high SRF not
present in the \mbox{X-shooter} data set (maximum of 166\,\unit{sfu}).
Moreover, the relatively small number of 1,186 FORS\,1 spectra could have
caused sampling issues. However, the $m_\mathrm{SCE}$ climatology in
Fig.~\ref{fig:refclim_SCE}l indicates SCEs higher than $+1$\,\unit{MJy^{-1}}
for most of the year, although there are remarkable negative effects at the
beginning of the night in austral summer. As the strongest positive effects
are located around midnight, this leads to a significant change of the
intensity pattern in Fig.~\ref{fig:refclim_rI}l for high SRF. Then, the
midnight brightness wave appears to be more powerful. The dependence of
the \chem{O} recombination lines on the squared \chem{O^+} (or electron)
density, leads to the most extreme $\langle m_\mathrm{SCE} \rangle$ of
$+2.7$\,\unit{MJy^{-1}}. The pattern in Fig.~\ref{fig:refclim_SCE}m shows
some similarities with the other ionospheric lines. However, the SCE is always
positive with a minimum where also the mean intensity indicates a minimum. The
maximum SCE after dusk in May is located in a region in
Fig.~\ref{fig:refclim_rI}m with a strong intensity gradient, i.e. small
changes in the seasonal distribution can have large effects.

In contrast to the ionospheric lines, only a moderately positive effect is
present for \chem{H\alpha} emission ($+0.17$\,\unit{MJy^{-1}}). This is about
3 times lower than found by \citet{nossal12}. However, their data were from
Wisconsin in the United States and only covered 22 nights between 1997 and
2008. The $m_\mathrm{SCE}$ climatology in Fig.~\ref{fig:refclim_SCE}o
indicates the largest solar impact in the second half of the night.

The residual variation $\sigma_{f,0}$ as derived after the subtraction of the
SCE-corrected reference climatological model is shown in
Fig.~\ref{fig:refclim_rdI} for the different variability classes. As
$\sigma_{f,0}$ corresponds to the standard deviation, which enhances the
impact of outliers, the uncertainties are higher than in the case of the
$f_0$ climatologies. Measurement errors could cause significant deviations.
Nevertheless, the main structures should be reliable in most cases.

\citet{noll23} analysed the \chem{OH}-related $\sigma_{f,0}$ climatologies in
detail. As shown in (a) to (c), the climatologies reveal a remarkable seasonal
pattern with maxima in January and June/July. In a similar way as discussed
for the SCE, the centroids of the variability in the maximum months depend on
LT. The migrating diurnal tide also affects the perturbations causing these
variations. For this reason, the features in (c) occur earlier than in (a). A
separation of 4\,\unit{h} would correspond to 5\,\unit{km} in altitude. In
order to better understand the spectrum of periods that causes these
structures, \citet{noll23} studied the residual variability as a function of
time difference. This approach revealed that periods shorter than 1 day
dominate in the middle of the year in austral winter, which suggests that the
emission variations are either directly related to (especially long-period)
gravity waves \citep[GWs, e.g.,][]{fritts03} and/or the tidal modes are
modified by interaction with GWs. A survey of GW-related literature with a
focus on South America \citep[e.g.,][]{alexander15,ern18,cao22} indicated that
the main origin of the waves is at higher latitudes along the Andes, which is
a global GW hot spot in austral winter. The waves then reached Cerro Paranal
either directly or indirectly via secondary waves. The $\sigma_{f,0}$ maximum
in January also indicates GW contributions, but especially from deep
convection across the Andes. However, the main reason for the austral summer
maximum is the occurrence of quasi-two-day waves (Q2DWs), the strongest waves
at low southern latitudes \citep[e.g.,][]{ern13,gu19,tunbridge11}. They are
usually active only a few weeks in austral summer. \citet{noll22} used the
strong event in 2017 to study \chem{OH} effective emission heights (see
Sect.~\ref{sec:climatologies}). The same approach was applied by
\citet{noll24} for the emissions of \chem{HO_2} and \chem{FeO}.

We therefore expect to see GW and Q2DW features in the $\sigma_{f,0}$
climatologies for different emissions in the mesopause region. Indeed, the
seasonal structures show similarities in Fig.~\ref{fig:refclim_rdI}. Good
examples are \chem{O_2}\mbox{(b-X)} (e) and \chem{HO_2} (g). In contrast,
\chem{K} (j) shows very different features, which might be related to the very
small sample size for this weak emission. In the case of
\chem{O_2}\mbox{(a-X)} (d), the situation in the first half of the night is
probably complicated by the contribution of emissions from relatively low
heights related to \chem{O_3} photolysis (see Sect.~\ref{sec:O2}). In the same
way as \citet{noll24}, we checked the period-dependent contributions to winter
and summer, which confirmed the \chem{OH}-related findings. In principle, the
centroid LT should match the height of the corresponding emission layer in all
seasons. Although this is not well constrained in many cases, the trends are
promising. \chem{HO_2} peaks later than, e.g., \chem{FeO} and
\chem{O_2}\mbox{(b-X)}. In this context, the result for the green \chem{O}
line (k) is most difficult to interpret as the winter and summer maxima are at
very different LTs. As the quality of the data of this line is relatively good
and the reference height is close to the top of the mesopause region, these
discrepancies might point to height-dependent changes in the importance of GWs
and Q2DWs. 

In Table~\ref{tab:varclasses}, we also list the night averages of
$\sigma_{f,0}$ for each class. Here, the \chem{OH} hot population class
\texttt{OH6d} indicates the lowest relative residual variability of 0.25. On
the other hand, the intermediate-$N^{\prime}$ class \texttt{OH3b} shows the
maximum for \chem{OH} of 0.35. This result reflects the variable mixing of
cold and hot rotational populations, which causes additional variability for
lines in classes like \texttt{OH3b} \citep{noll23}. As discussed in
Sect.~\ref{sec:climatologies}, this effect had an influence on the
classification scheme for \chem{OH}. The other emissions of the mesopause
region tend to have similar or somewhat higher $\langle \sigma_{f,0} \rangle$
as/than \texttt{OH3b}. The maximum of 0.46 for \chem{K} might be partly caused
by noise, although the other alkali metal \chem{Na} indicates the second
highest value of 0.42. The results for \chem{OH}, \chem{O_2}\mbox{(b-X)},
\chem{Na}, and green \chem{O} line are roughly consistent with those of
\citet{noll12} based on FORS\,1 data. 

The ionospheric lines indicate the largest $\langle \sigma_{f,0} \rangle$ with
a maximum of 0.97 for the \chem{O} recombination lines, which is comparable to
the results for the SCE. For the red \chem{O} lines, \citet{noll12} obtained
values of about 0.9, which is higher than the \mbox{X-shooter}-based mean
standard deviation of about 0.68. As indicated by a comparison of
Figs.~\ref{fig:refclim_rI} and \ref{fig:refclim_rdI}, the high variability
mainly originates from combinations of month and LT with low average
intensity. The maximum in $\sigma_{f,0}$ at midnight in December for the red
\chem{O} (l) and \chem{N} (n) lines corresponds to a minimum in the
corresponding $f_0$ climatologies. Typical reasons for variability in
ionospheric airglow emissions are plasma instabilities that cause effects like
plasma bubbles \citep[e.g.,][]{makela06} and wave-like disturbances with and
without links to lower atmospheric layers \citep[e.g.,][]{paulino16}.

Fluorescent \chem{H} emission only shows moderate residual variability (0.35).
The highest $\sigma_{f,0}$ are located near the minima in $f_0$
(Figs.~\ref{fig:refclim_rI}o and \ref{fig:refclim_rdI}o). Hence, the \chem{H}
density could be more variable at higher altitudes. However, the larger
variation in the shadow height due to changes in the line of sight (which was
not corrected, see Sect.~\ref{sec:H}) could also play a role.

\section{Evaluation}
\label{sec:eval}

\begin{figure*}[tp]
\includegraphics[width=15.9cm]{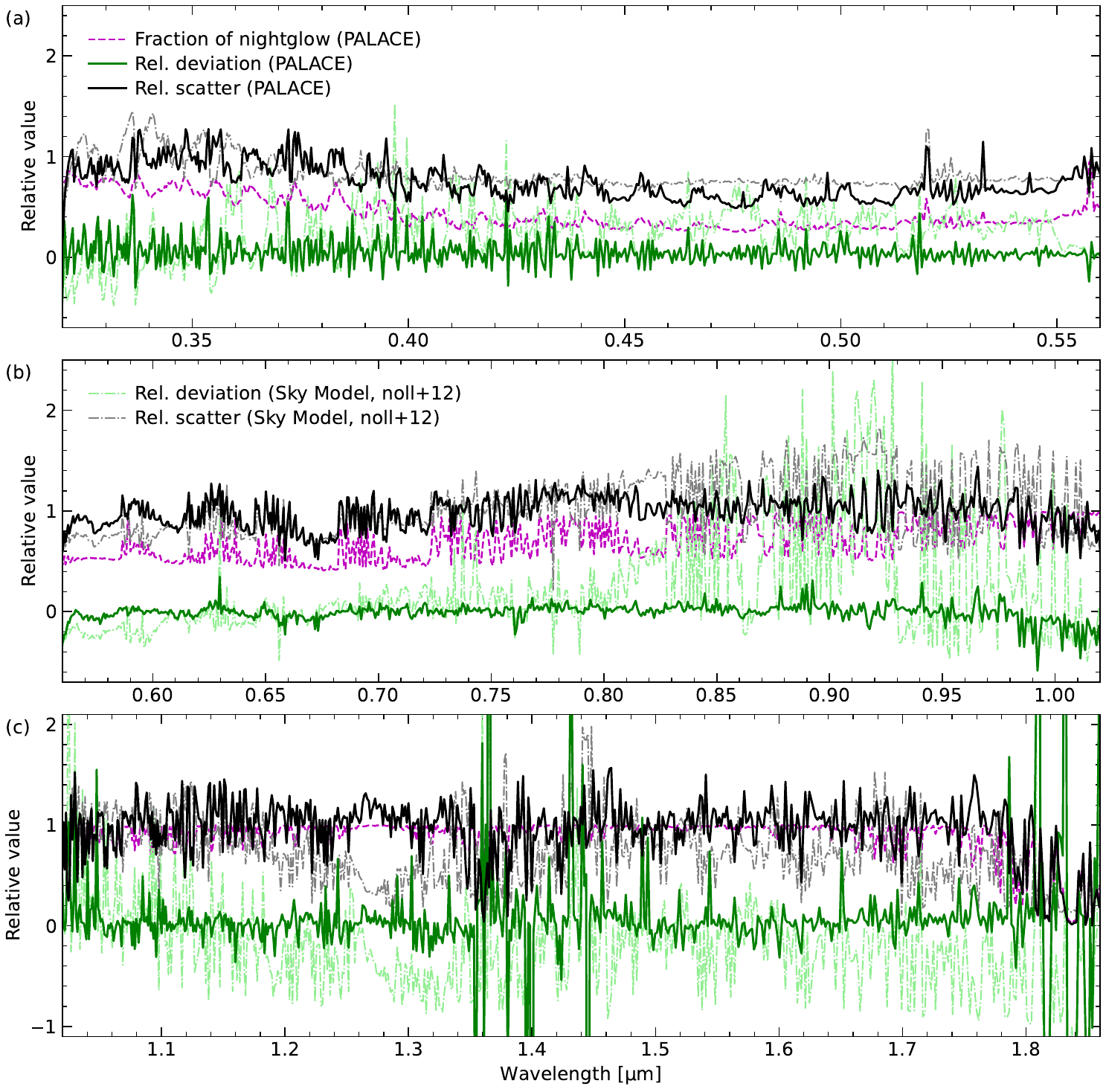}
\caption{Evaluation of PALACE (solid lines) and the airglow component of the
  ESO Sky Model \citep{noll12} (dash-dotted lines) based on a comparison to
  6,874 \mbox{X-shooter} spectra after the subtraction of non-airglow
  components calculated by means of the ESO Sky Model. Results are provided
  for the wavelength range from 0.32 to 1.86\,\unit{\mu{}m} divided into three
  ranges related to the different \mbox{X-shooter} arms. The data points
  correspond to mean values in bins with a width of about 0.001 times the
  central wavelength. The green and light green curves show the mean deviation
  of the two models calculated for the different observing conditions from the
  \mbox{X-shooter} airglow emission relative to the average of the latter. The
  black and grey curves display the ratio of the mean residual variability of
  the models and the standard deviation of the difference between model and
  measured data. The dashed magenta curve indicates the average fraction of
  the PALACE airglow emission compared to the full sky brightness that also
  considers the Sky Model non-airglow components.}
\label{fig:reseval}
\end{figure*}

The model development described in Sects.~\ref{sec:lines} to
\ref{sec:variability} was complex. Although many checks were performed for
the different steps of the procedure, it is not clear how well the final
PALACE model represents the observed data. Therefore, we carried out an
evaluation of the model based on high-quality \mbox{X-shooter} spectra with
very low noise level and without obvious systematic errors in the central
parts of the different arms. \citet{noll24} selected such a sample (consisting
of 10,633 spectra) for the investigation of the nightglow continuum. We
further reduced this by excluding spectra where the exposures were split
depending on the \mbox{X-shooter} arm. In this way, we avoided the averaging
of spectra before the analysis. The resulting sample contained 7,195 spectra
in all arms. We also neglected spectra taken before 5 February 2010 as they
showed systematic effects in certain near-UV wavelength ranges connected with
the different echelle orders. These calibration issues would have increased
the scatter with respect to PALACE. As a consequence, the final sample
comprised 6,874 spectra. As the night-sky radiation also involves scattered
moonlight, scattered starlight, and zodiacal light and the thermal radiation
of telescope and instrument also disturbs, non-airglow contributions had to be
subtracted. Here, we used the corresponding spectra of the ESO Sky Model
\citep{noll12,jones13} that were already calculated by \citet{noll24} for the
continuum study. For the resulting \mbox{X-shooter} airglow spectra (divided
into three arm-dependent parts), we then calculated PALACE model spectra in
air wavelengths under the consideration of the observing conditions described
by the input parameters \texttt{z}, \texttt{mbin}, \texttt{tbin},
\texttt{srf}, and \texttt{pwv} (Table~\ref{tab:parameters}).

For the evaluation, we binned the PALACE and \mbox{X-shooter} spectra by
using wavelength-dependent bins with a size of about 0.001 times the
wavelength. This step is important as the modelled simple Gaussian line
profiles depending on the parameter \texttt{resol} cannot reproduce the real
line-spread function, which can cause significant residuals. This is even an
issue for more sophisticated kernels due to the pixelation of the data and
wavelength calibration uncertainties. Our choice of the bin size corresponds
to a resolving power of 1,000, which is sufficiently low compared to values
of 3,200 to 18,400 for the \mbox{X-shooter} data \citep{vernet11}. The
difference of the averaged fluxes of the PALACE and \mbox{X-shooter}
spectra in each wavelength bin was then analysed by the calculation of mean
and standard deviation for the whole sample. The results were refined by the
application of a $\sigma$-clipping approach for the flux data of each bin.
This was important as systematic outliers in the observed data are possible.
In particular, strong lines of astrophysical origin (that were not an issue
for the continuum study of \citet{noll24}) can have a significant effect.
However, the wavelength-dependent fraction of rejections was always lower than
3\%. In most cases, only a few spectra were excluded for a given bin.
Nevertheless, the sample was about 22\% smaller for wavelengths beyond
2.1\,\unit{\mu{}m} due to a special set-up for background reduction in the
instrument at shorter wavelengths that blocked this range \citep{vernet11}. 

The results of the analysis are shown in Fig.~\ref{fig:reseval} for the
wavelength range from 0.32 to 1.86\,\unit{\mu{}m}. Longer wavelengths than
about 1.8\,\unit{\mu{}m} are difficult to study because of increasing thermal
emissions from the instrument and the atmosphere. The situation further
deteriorates due to the presence of relatively strong absorption bands, which
affect the airglow, whereas the emission of the instrument is not
significantly absorbed. First, we discuss the relative deviation, which is the
mean difference between PALACE and \mbox{X-shooter} fluxes divided by the
mean \mbox{X-shooter} flux. As desired, the green solid curve is close to 0,
although with some scatter. For the wavelength ranges from 0.32 to
0.56\,\unit{\mu{}m} (UVB arm), 0.56 to 1.02\,\unit{\mu{}m} (VIS arm), and
1.02 to 1.78\,\unit{\mu{}m} (NIR arm), we obtained mean relative deviations of
only $+0.053$, $-0.008$, and $+0.032$, respectively. The latter value is not
significantly affected by the few outliers near 1.4\,\unit{\mu{}m}, which are
related to strong water vapour absorption and hence very low fluxes. The
impact of the instrument-related systematic discrepancies at wavelengths above
0.99\,\unit{\mu{}m} in the VIS arm (overlap with NIR arm) can also be
neglected ($+0.002$ without this range). 

The relative scatter is also shown in the figure (black solid curve). This
quantity represents the ratio of the sample-averaged residual variability from
the model (Sect.~\ref{sec:overview}) and the mean standard deviation from the
difference of the PALACE and \mbox{X-shooter} fluxes. The values should
ideally be close to 1 for our high-quality \mbox{X-shooter} test sample. For
the three wavelength ranges defined above, we calculated mean ratios of 0.769,
0.983, and 1.002, respectively. The former value for the UVB arm is distinctly
below 1. Figure~\ref{fig:reseval} shows particularly low values down to about
0.5 in the range between 420 and 520\,\unit{nm}. The main reason for these
discrepancies is the fact that other sky radiance components had to be
subtracted from the \mbox{X-shooter} spectra. As these components also have
uncertainties, the modelled airglow variability is lower than the measured
variability. This effect is particularly strong where the fraction of airglow
emission compared to the total emission is relatively low. For illustration,
this ratio is also plotted in Fig.~\ref{fig:reseval} (magenta dashed curve).
Between 420 and 520\,\unit{nm}, its mean value is just 0.32 (with a minimum of
0.25), i.e. airglow is a minor contribution. If we only select the 34\% of
bins with an airglow fraction of at least 0.5, the mean relative scatter
already increases to 0.961, which is more convincing. The VIS-arm range is
also partly affected. There, a focus on a minimum airglow fraction of 0.5
causes a slight increase of the mean relative scatter to 1.024. While
scattered moonlight and zodiacal light can have strong impacts in the visual
range, airglow is the dominating nocturnal radiation source in the NIR-arm
range up to about 1.78\,\unit{\mu{}m} as demonstrated by the magenta dashed
curve in (c). For longer wavelengths (where no nightglow continuum could be
measured), the model accuracy remains uncertain, although the only relevant
emission lines are related to the well-modelled \chem{OH} radical
(Fig.~\ref{fig:refspec}).

In the ranges where an evaluation was possible, the performance of PALACE is
convincing as the typical deviations for the two investigated properties are
only of the order of a few per cent. In principle, a relative scatter slightly 
above 1 could even be expected as the model standard deviations for all lines
and continua were just averaged for simplicity (see Sect.~\ref{sec:overview}).
In particular, wavelength regions with similar contributions from variability
classes with very different climatologies should be affected. An example
would be ranges in the VIS arm where the \chem{OH} and \chem{FeO}-like
classes (see Sect.~\ref{sec:variability}) are important. Indeed, the
potentially affected regions indicate a slightly increased relative scatter in
Fig.~\ref{fig:reseval}b. Concerning the relative deviation, the UVB-arm regime
in (a) shows a small positive offset, which tends to increase with decreasing
wavelength ($+0.075$ between 320 and 360\,\unit{nm}). This result seems to
reflect the difficulty to measure the nightglow in a range where other sky
brightness components are relatively bright and atmospheric scattering is
relatively strong but uncertain due to variable aerosol properties. There
might be additional bias in the model which is not visible in
Fig.~\ref{fig:reseval} as PALACE is based on the same \mbox{X-shooter} data
that were used for the evaluation. However, a complex flux calibration
approach was applied \citep{noll22} that should limit the mean uncertainties
to a few per cent. Moreover, the comparison of \chem{OH} populations from
\mbox{X-shooter} data and the UVES mean spectrum from \citet{noll20} based on
data from 2000 to 2015 \citep{noll17} showed good agreement (see
Sect.~\ref{sec:OH}). The deviations were only of the order of a few per cent
and mainly related to real population differences as indicated by their
dependence on the upper vibrational level. Finally, it is possbile that
significant long-term changes in the airglow mean intensity and variability
occur. For recognising such trends, the covered period of the \mbox{X-shooter}
data of 10 years is too short.  

PALACE was developed for the same location as the ESO Sky Model, which is
briefly described in Sect.~\ref{sec:intro}. Hence, it is interesting to
compare both models. In the previous sections, we have already made
comparisons with respect to the reference intensities and climatologies of
the green and red \chem{O} lines, the \chem{Na} D doublet, \chem{OH}, and
\chem{O_2}\mbox{(b-X)(0-1)} that were measured by \citet{noll12} in
low-resolution FORS\,1 data. There are clear similarities, although some
discrepancies are also observed (see Sects.~\ref{sec:lines} and
\ref{sec:comparison}). Some uncertainties are related to the very different
solar radio flux ranges for the \mbox{X-shooter} and FORS\,1 spectra.
Moreover, the ESO Sky Model climatologies are too coarse (18 compared to 134
useful grid points) for a detailed comparison. An evaluation of the mean
nightglow continuum of the ESO Sky Model (derived from the FORS\,1 data and a
few early \mbox{X-shooter} NIR-arm spectra) was already performed by
\citet{noll24}, which indicated good agreement up to about 800\,\unit{nm} and
major disrepancies at longer wavelengths, although the \chem{HO_2}
1.51\,\unit{\mu{}m} peak (Fig.~\ref{fig:refcont}) was visible
\citep[see also][]{jones19}. \citet{noll12} measured the continuum variability
only at 543\,\unit{nm}, which would best fit to the \chem{FeO} class. However,
an SCE of $+0.61$\,\unit{MJy^{-1}} and different climatological patterns as for
\chem{Na} do not match the expectations (see Table~\ref{tab:varclasses}).
There might also be significant contributions from \chem{O_2} and other
emissions. The low resolving power of the FORS\,1 spectra and the lack of
near-IR data also affected the line model, which is only based on simple
theoretical calculations beyond 925\,\unit{nm}.

In order to better understand the overall performance of the widely used ESO
Sky Model, we carried out the same comparison to the selected
\mbox{X-shooter} data set as for PALACE. As the removal of the non-airglow
components in the \mbox{X-shooter} spectra already required the calculation of
the Sky Model, we could directly use these spectra. Their line-spread function
is a combination of boxcar and Gaussian optimised for the different arms and
slit widths. However, this difference in comparison to PALACE (only Gaussian)
should not matter due to the binning of the wavelength grid. The results of
the analysis are also presented in Fig.~\ref{fig:reseval} (dash-dotted
curves).

For the UVB-arm range, relative deviation (light green) and scatter (grey) are
$+0.239$ and 0.842 on average. If only bins with an airglow fraction of at
least 0.5 are considered, these values change to $+0.067$ and 0.989. Thus, the
relative scatter appears to be realistic and similar to PALACE, whereas the
relative deviation is clearly too high. As illustrated by
Fig.~\ref{fig:reseval}a, the largest positive deviations correlate with low
airglow fractions. Hence, there could have been issues with the separation of
airglow and other radiance components in the FORS\,1 spectra. Systematic
uncertainties in the flux calibration of the data that were collected by
\citet{patat08} could contribute here.

In the VIS-arm range, the average relative deviation amounts to $+0.198$,
which is mostly driven by the wavelength range between 800 and 930\,\unit{nm},
where even deviations up to about 2.0 are visible (b). As wavelengths with
strong line emission (see Fig.~\ref{fig:refspec}) show values near 0, this is
caused by the unrealistic FORS\,1-based airglow continuum in this range
\citep{noll24}, where the spectral resolution and quality of the flux
calibration were obviously insufficient. The average relative scatter of 1.018
looks convincing. However, there are significant variations of this quantity.
At wavelengths where the continuum error is high, the scatter is too high as
well. On the other hand, the variation of the \chem{OH} bands modelled by a
single climatology (without significant solar cycle effect) appears to be
underestimated.

This issue seems to extend to the NIR-arm range (c), where even the average
scatter decreases to 0.762 for the range up to 1.78\,\unit{\mu{}m}. Apart from
the \chem{OH} bands, the \chem{O_2}\mbox{(a-X)(0-0)} band near
1.27\,\unit{\mu{}m} is affected. Note that \citet{noll12} did not have the
data to measure the variability of this band and therefore used the results
for \chem{O_2}\mbox{(b-X)(0-1)}, which is quite different as (d) and (e) in
Fig.~\ref{fig:refclim_rI} illustrate. The \chem{HO_2} variability in (g) is
also very different from the situation near 543\,\unit{nm}. The mean relative
deviation also indicates an underestimation between 1.02 and
1.78\,\unit{\mu{}m} ($-0.182$). As the structures in Fig.~\ref{fig:reseval}c
are relatively similar to those of the relative scatter, the same emissions
are affected. The deviations are probably related to the extrapolation of the
reference intensities derived from the FORS\,1 data, which relied on the
simple population model of \citet{rousselot00} for \chem{OH} (see
Sect.~\ref{sec:intro}). As this model only assumes cold rotational
populations, the hot population lines are far too weak. Moreover, the use of
old $A$ coefficients from HITRAN2008 \citep{rothman09} could have had an
impact. As the modelled \chem{O_2}\mbox{(a-X)} emissions were scaled to the
\chem{OH} emissions based on the small sample of \mbox{X-shooter} spectra
\citep{noll14}, significant deviations are also present, at least for the
strong 1.27\,\unit{\mu{}m} band. At the position of the weak
1.58\,\unit{\mu{}m} band, there is no clear offset.

Sky-brightness models are crucial for an efficient scheduling of the
telescope operations at large astronomical observatories \citep{noll12}. Time
estimates are obtained via exposure time calculators (ETCs). For the ESO
sites in Chile, see \texttt{https://etc.eso.org/}. In the case of
overestimations, precious observing time can be wasted, whereas the opposite
case can lead to data of too low quality for the scientific goals. As the time
predictions are usually made in advance (especially during the application
process for the observing time), there are unavoidable uncertainties for
individual observations as the actual atmospheric conditions are not
considered. However, the overall performance of an ETC will clearly depend on
the quality of the sky-brightness model. Hence, the discussed shortcomings of
the current ESO Sky Model with respect to airglow emission can have a
significant impact on the efficiency of the telescope operations.

In the near-IR range shown in Fig.~\ref{fig:reseval}c, airglow is clearly
dominating the night-sky brightness. Even scattered light from the Moon is
only a minor contribution in this wavelength regime \citep{jones19}. Hence,
the underestimations of the airglow radiance and its variability by the ESO
Sky Model should significantly affect the performance of the ETC. In the
wavelength range from 1.02 to 1.78\,\unit{\mu{}m}, 23\% of the bins show
deviations that are larger than 50\% of the \mbox{X-shooter}-related flux.
Assuming an astronomical object that is much fainter than the airglow
emission, the signal-to-noise ratio is approximately proportional to the ratio
of the detector electron counts from the object and the square root of those
from the sky ($N_\mathrm{obj} / \sqrt{N_\mathrm{sky}}$) if read-out noise and
dark currents can be neglected. Hence, a relative deviation of $-0.5$ in
Fig.~\ref{fig:reseval}c would correspond to an overestimation of the
signal-to-noise ratio by a factor of about 1.4. In order to achieve the same
real quality of the data, the exposure time would need to be doubled, which
would have a major impact on the scheduling of the observations. The opposite
case (i.e. an overestimation of the required exposure time) is expected for
the airglow continuum at wavelengths of about 920\,\unit{nm}, where
Fig.~\ref{fig:reseval}b indicates a flux of the ESO Sky Model that is about 3
times higher than expected. As the airglow continuum only provides about 50\%
of the total sky brightness at these wavelengths for the analysed
\mbox{X-shooter} data on average, the exposure time could be decreased by a
factor of about 1.5 for our simplified scenario in dark nights without a
bright Moon.

In conclusion, the distinctly better performance of PALACE with respect to
the prediction of the airglow brightness should also have a significant impact
on the quality of astronomical exposure time calculations and the planning of
telescope operations. Moreover, we expect an influence on the development of
new astronomical instruments as the sky brightness is also a basis for the
design of instruments in the context of the corresponding scientific goals.
The brightness and variability of airglow also matters for the processing of
astronomical data. In particular, the ESO Sky Model is used for the removal of
airglow lines in one-dimensional spectra \citep{noll14}. Although the
subtraction is based on an observed spectrum scaled by wavelength-dependent
factors, the model is required for the identification of lines and the initial
weights of different variability groups to each pixel. Here, the PALACE model
data should also allow for improvements.

\conclusions[Conclusions]  
\label{sec:conclusions}

The pioneering ESO Sky Model for Cerro Paranal in Chile has been very
important with respect to astronomical observing proposal evaluation,
observation scheduling, data processing, and the design of new instruments.
As discussed in the previous section, the crucial airglow component is,
however, affected by systematic issues due to the lack of suitable data for
the development of the model by \citet{noll12}. The data situation has
improved a lot in the past decade, especially by the availability of a
well-calibrated \mbox{X-shooter} data set of the order of $10^5$ spectra
covering the wide wavelength from 0.3 to 1.8\,\unit{\mu{}m} and a period of
10 years. Moreover, the authors of this study carried out several studies that
significantly improved the understanding of airglow line and pseudo-continuum
emissions at Cerro Paranal. Therefore, the preconditions to build a
significantly better model are fulfilled. As a consequence, we created the
Paranal Airglow Line And Continuum Emission (PALACE) model. It consists of a
comprehensive line list with 26,541 entries, 3 continuum components, and
two-dimensional climatologies of relative intensity, solar cycle effect, and
residual variability for 23 variability classes. The model is provided in
connection with a Python/Cython code for the calculation of airglow spectra
(with uncertainty estimates) depending on different input parameters. The
comparison of the model and the observed \mbox{X-shooter} spectra (where the
other sky radiance components were still removed by means of the ESO Sky
Model) showed a convincing agreement with average uncertainties in terms of
the mean difference and the scatter of only a few per cent. This performance
should be appealing for the use of PALACE in similar contexts as the current
ESO Sky Model. With respect to astronomical observatories, such applications
are the evaluation of observing proposals, the scheduling of telescope time,
the design of instruments, and algorithms for data processing.

The development of PALACE also resulted in new insights in airglow physics
and chemistry. For the \chem{OH} line list, a comprehensive population model
with one to three temperature components for the vibrational levels $v$ from 2
to 10 was created based on fits of \mbox{X-shooter} data as well as UVES data
from previous studies \citep{cosby06,noll20}. In this context, also the input
Einstein-$A$ coefficients were further improved compared to \citet{noll20}.
Population models were also created for the excited
a$^1\Delta_\mathrm{g}$($v \leq 1$) and b$^1\Sigma^+_\mathrm{g}$($v \leq 2$)
states of \chem{O_2} based on new \mbox{X-shooter} measurements near 0.865 and
1.27\,\unit{\mu{}m} and UVES data from \citet{cosby06}. A model for
intensities of \chem{O} recombination lines (e.g. at 777\,\unit{nm}) was
derived from the combination of \mbox{X-shooter}-based intensities,
theoretical recombination coefficients, and $A$ coefficients. The measurements
of atomic lines also include green \chem{O}, red \chem{O}, \chem{Na} D,
\chem{N} 520\,\unit{nm}, and several \chem{H} Balmer lines. From
\citet{noll19}, the UVES-based time series of the \chem{K} 769.9\,\unit{nm}
line was taken. For the derivation of the continuum model consisting of the
components of \chem{HO_2}, \chem{FeO} (plus other molecules), and unresolved
\chem{O_2} at short wavelengths, we used the \mbox{X-shooter}-based results of
\citet{noll24}.

Climatologies from the time series data of 392 different emissions were
considered for the PALACE variability model consisting of 23 classes.
Extending the studies of \citet{noll19,noll23,noll24}, the relative intensity
climatologies of emissions in the mesopause region can be grouped depending on
the most crucial chemical species for the variations, i.e. either \chem{H},
\chem{O}, or \chem{O_3}. The intensities of ionospheric lines mainly depend on
the electron/ion density, whereas \chem{H} fluorescence emissions are strongly
influenced by the height of the Earth's shadow. The comparison of
two-dimensional climatologies of month and local time for the solar cycle
effect and residual variability (including the effective nighttime values)
also revealed interesting similarities and differences which were not
investigated before. The results show the height-dependent impact of wave-like
perturbations such as tides (especially the migrating diurnal tide), gravity
waves (especially long periods), and planetary waves (especially the
quasi-two-day wave) on the climatological pattern. For ionospheric lines, the
change of the vertical ionisation distribution is important, where waves and
instabilities can lead to additional variability. As the main goal of this
study was the creation of PALACE, a more detailed study of the time series and
climatological data was out of the focus. Hence, the large data set should be
valuable for further investigations.




\codedataavailability{The Python/Cython code of PALACE v1.0, the necessary
  model files in FITS table format, the reference test output, and different
  supplementary ASCII tables for a better understanding of the model and the
  reproduction of the figures are provided at
  \texttt{https://zenodo.org/records/14064023} \citep{noll24ds} under a
  \mbox{CC-BY-4.0} licence for the data as well as a GNU GPLv3 licence for the
  code. In the case of future versions of PALACE, the paper-related release
  files will remain untouched, but a link to the revised model will be
  provided in the metadata. Apart from own measurements, the study also used
  \mbox{X-shooter}-based data of previous publications \citep{noll23,noll24}
  released via Zenodo \citep{noll23ds1,noll23ds2}. The basic \mbox{X-shooter}
  data for this project originate from the ESO Science Archive Facility at
  \texttt{http://archive.eso.org} and are related to various observing
  programmes that were carried out between October 2009 and September 2019.
  Moreover, UVES-related data sets jointly published with the articles by
  \citet{cosby06}, \citet{noll19}, and \citet{noll20} were considered for the
  model development. The raw UVES data and more processed Phase~3 spectra are
  also provided by the ESO Science Archive Facility.}












\authorcontribution{SN designed and organised the project, performed the
  the analysis of the data, derived and tested the model, wrote the code,
  created figures and tables, and is the main author of the paper text. The
  co-authors contributed to the paper content (especially CS) and the
  flowchart of PALACE (PH and CS). Moreover, PH and WK tested the code. WK
  performed the basic processing of the \mbox{X-shooter} spectra. SK managed
  the infrastructure for the processing and storage of the \mbox{X-shooter}
  data.}

\competinginterests{Competing interests are not present.} 


\begin{acknowledgements}
  The work of Stefan Noll was partly financed by the project
  \mbox{NO\,1328/1-3} of the German Research Foundation (DFG). We are grateful
  to Michael Bittner from DLR for his support in terms of the infrastructure
  for the project and the funding of this publication. Moreover, we thank
  Sabine M\"ohler from ESO for her support with respect to the
  \mbox{X-shooter} calibration data. Finally, we are grateful to the two
  anonymous reviewers, who appreciated our extensive work, which was a major
  effort. Their specific recommendations further improved this manuscript.
\end{acknowledgements}







\bibliographystyle{copernicus}
\bibliography{Nolletal2025.bib}

\end{document}